\documentclass[english, aps, prl, reprint, superscriptaddress]{revtex4-1}
\usepackage[T1]{fontenc}
\usepackage[latin9]{inputenc}
\usepackage{amsmath}
\usepackage{amssymb}
\usepackage{graphicx}
\usepackage{babel}
\usepackage{color}
\usepackage[caption=false]{subfig}

\begin{document}

\title{Exploring the Function Space of Deep-Learning Machines}

\author{Bo Li}
\email{bliaf@connect.ust.hk}

\selectlanguage{english}

\affiliation{Department of Physics, The Hong Kong University of Science and Technology,
Hong Kong}

\author{David Saad}
\email{d.saad@aston.ac.uk}

\selectlanguage{english}

\affiliation{Non-linearity and Complexity Research Group, Aston University, Birmingham
B4 7ET, United Kingdom}
\begin{abstract}
The function space of deep-learning machines is investigated by studying growth in the entropy of functions of a given error with respect to a reference function, realized by a deep-learning machine. Using physics-inspired methods we study both sparsely and densely connected architectures to discover a layerwise convergence of candidate functions, marked by a corresponding reduction in entropy when approaching the reference function, gain insight into the importance of having a large number of layers, and observe phase transitions as the error increases.
\end{abstract}
\maketitle
Deep-learning machines (DLMs) have both fascinated and bewildered the scientific community and have given rise to an active and ongoing debate~\cite{Elad_SIAM2017}. They are carefully structured layered networks of nonlinear elements, trained on data to perform complex tasks such as speech recognition, image classification, and natural language processing. While their phenomenal engineering successes~\cite{LeCun2015} have been broadly recognized, their scientific foundations remain poorly understood, 
particularly their good ability to generalize well from a limited number of examples with respect to the degrees of freedom~\cite{Chiyuan2017, Chaudhari2017, Neyshabur2017}
and the nature of the layerwise internal representations~\cite{Zeiler2014, Yosinski2015}.

Supervised learning in DLMs is based on the introduction of example pairs of input and output patterns, which serve as constraints on space of candidate functions. As more examples are introduced, the function space monotonically decreases. Statistical physics methods have been successful in gaining insight into both pattern-storage \cite{Gardner1988} and learning scenarios, mostly in single-layer machines~\cite{Hertz1991} but also in simple two-layer scenarios~\cite{Watkin1993,Saad95}. However, extending these methods to DLMs is difficult due to the recursive application of nonlinear functions in successive layers and the undetermined degrees of freedom in intermediate layers. While training examples determine both input and output patterns, the constraints imposed on hidden-layer representations are difficult to pin down. These constitute the main difficulties for a better understanding of DLMs.

In this Letter, we propose a general framework for analyzing DLMs by mapping them onto a dynamical system and by employing the generating functional (GF) approach to analyze their typical behavior. More specifically, we investigate the landscape in function space around a reference function by perturbing its parameters (weights in the DLM setting) and quantifying the entropy of the corresponding functions space for a given level of error with respect to the reference function. 
This provides a measure for the abundance of nearly perfect solutions and hence an indication for the ability to obtain good approximations using DLMs. The function error measure is defined as the expected difference (Hamming distance in the discrete case) between the perturbed and reference functions' outputs given the same input (additional explanation is provided in Ref. ~\cite{Li2017sup}).\nocite{Schwartz1990, Dominicis1978, Lee2018, Shang2016, Springenberg2015} This setup is reminiscent of the teacher-student scenario, commonly used in the neural networks literature~\cite{Saad98a} where the average error serves as a measure of distance between the perturbed and reference network in function space. For certain classes of reference networks, we obtain closed form solutions of the error as a function of perturbation on each layer, and consequently the weight-space volume for a given level of function error. By virtue of supervised learning and constraints imposed by the examples provided, high-error functions will be ruled out faster than those with low errors, such that the candidate function space is reduced and the concentration of low-error functions increases.
A somewhat similar approach, albeit based on recursive mean field relations between each two consecutive layers separately, has been used to probe the expressivity of DLMs~\cite{Poole2016}. 

Through the GF framework and entropy maximization, we analyze the typical behavior of different classes of models including networks with continuous and binary parameters (weights) and different topologies, both  fully and sparsely connected. We find that as one lowers the error level, typical functions gradually better match the reference network starting from earlier layers to later ones. More drastically, for fully connected binary networks, weights in earlier layers of the perturbed functions will perfectly match those of the reference function, implying a possible successive layer by layer learning behavior. Sparsely connected topologies exhibit phase transitions with respect to the number of layers, by varying the magnitude of perturbation, similar to the phase transitions in noisy Boolean computation~\cite{Mozeika2009}, which support the need of deep networks for improving generalization.

\begin{figure}
\includegraphics[scale=0.4]{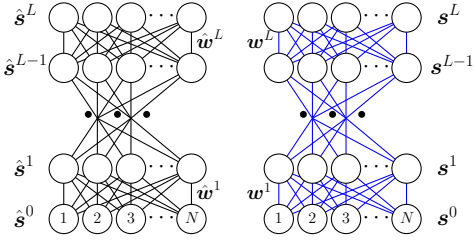}
\caption{The model of two coupled DLMs. The reference and perturbed 
functions are denoted by $\{\hat{\boldsymbol{w}}^{l}\}$ (black edges) and $\{\boldsymbol{w}^{l}\}$
 (blue edges), respectively.  \label{fig:two_systems}}
\end{figure}

\textit{Densely connected network models.}\textendash The model considered here comprises two coupled feed-forward DLMs as illustrated in Fig.~\ref{fig:two_systems}, one of which serves as the reference function and the other is obtained by perturbing the reference network parameters.
We first consider the densely connected networks. Each network is composed of $L\!+\!1$ layers of $N$ neurons each. The reference function is parametrized by $N^{2}\!\times\! L$ weight variables $\hat{w}_{ij}^{l}, \forall ~l\!=\!1,2,...,L, ~i,j\!=\!1,2,...,N$, and maps an $N$-dimensional input $\hat{\boldsymbol{s}}^{0}\!\in\!\{-1,1\}^{N}$ to an $N$-dimensional output $\hat{\boldsymbol{s}}^{L}\!\in\!\{-1,1\}^{N}$ , through intermediate-layer internal representations and according to the
stochastic rule:
\begin{align}
P(\hat{\boldsymbol{s}}^{L}|\hat{\boldsymbol{w}},\hat{\boldsymbol{s}}^{0}) & =\prod_{l=1}^{L}P(\hat{\boldsymbol{s}}^{l}|\hat{\boldsymbol{w}}^{l},\hat{\boldsymbol{s}}^{l-1}).\label{eq:dynamic_rule}
\end{align}
The $i$th neuron in the $l$th layer experiences a local field
$\hat{h}_{i}^{l}(\hat{\boldsymbol{w}}^{l},\hat{\boldsymbol{s}}^{l-1})\!=\!\frac{1}{\sqrt{N}}\sum_{j}\hat{w}_{ij}^{l}\hat{s}_{j}^{l-1},$
and its state is determined by the conditional probability 
\begin{equation}
P(\hat{s}_{i}^{l}|\hat{\boldsymbol{w}}^{l},\hat{\boldsymbol{s}}^{l-1})=\frac{e^{\beta\hat{s}_{i}^{l}\hat{h}_{i}^{l}(\hat{\boldsymbol{w}}^{l},\hat{\boldsymbol{s}}^{l-1})}}{2\cosh\left[\beta\hat{h}_{i}^{l}(\hat{\boldsymbol{w}}^{l},\hat{\boldsymbol{s}}^{l-1})\right]},\label{eq:node_conditional_prob}
\end{equation}
where the temperature $\beta$ quantifies the strength of thermal noise. In the noiseless limit $\beta\!\to\!\infty$, node $i$ represents a perceptron $\hat{s}_{i}^{l}\!=\!\text{sgn}(\hat{h}_{i}^{l})$ and Eq.~(\ref{eq:dynamic_rule}) corresponds to a deterministic neural network with a \emph{sign} activation function. The perturbed network operates in the same manner, but the weights $w_{ij}^{l}$ are obtained by applying independent perturbation to each of the reference weights; the perturbed weights $w_{ij}^{l}$ give rise to a function that is correlated with the reference function. 

We focus on the similarity between reference and perturbed functions outputs for randomly sampled input patterns $\boldsymbol{s}^{0}\!=\!\hat{\boldsymbol{s}}^{0}$,
drawn from some distribution $P(\hat{\boldsymbol{s}}^{0})$. Considering the joint probability  of the two systems, 
\begin{eqnarray}
\label{eq:joint_density}
P\left[\{\hat{\boldsymbol{s}}^{l}\},\{\boldsymbol{s}^{l}\}\right] & = & P(\hat{\boldsymbol{s}}^{0})\prod_{i=1}^{N}\delta_{s_{i}^{0},\hat{s}_{i}^{0}} \\
 &  & \prod_{l=1}^{L}P(\hat{\boldsymbol{s}}^{l}|\hat{\boldsymbol{w}}^{l},\hat{\boldsymbol{s}}^{l-1})P(\boldsymbol{s}^{l}|\boldsymbol{w}^{l},\boldsymbol{s}^{l-1}), \nonumber
\end{eqnarray}
where the weight parameters $\{\hat{w}_{ij}^{l}\}$ and $\{w_{ij}^{l}\}$ are quenched disordered variables. We consider two cases, where the weights are continuous or discrete variables drawn from the Gaussian and Bernoulli distributions, respectively. The quantities of interest are the overlaps between the two functions at the different layers $q^{l}(\hat{\boldsymbol{w}},\boldsymbol{w})\!\equiv\!\frac{1}{N}\sum_{i}\langle\hat{s}_{i}^{l}s_{i}^{l}\rangle$, where angle brackets $\langle\cdots\rangle$ denote the average over the joint probability $P[\{\hat{\boldsymbol{s}}^{l}\},\{\boldsymbol{s}^{l}\}]$. The $N$ outputs represent $N$ weakly coupled Boolean functions of the same form of disorder, and thus share the same average behavior.

The form of probability distribution Eq. (\ref{eq:joint_density}) is analogous to the dynamical evolution of disordered Ising spin systems~\cite{Hatchett2004} if the layers are viewed as discrete time steps of parallel dynamics. We therefore apply the GF formulation from statistical physics to these deep feed-forward functions similarly to the approach used to investigate random Boolean formulas~\cite{Mozeika2009}. We compute the GF
$\Gamma[\hat{\boldsymbol{\psi}},\boldsymbol{\psi}]\!=\!\left\langle e^{-\mathrm{i}\sum_{l,i}(\hat{\psi}_{i}^{l}\hat{s}_{i}^{l}+\psi_{i}^{l}s_{i}^{l})}\right\rangle,$
from which the moments can be calculated; e.g., $q^{l}(\hat{\boldsymbol{w}},\boldsymbol{w})\!=\!-\!\frac{1}{N}\sum_{i}\lim_{\hat{\boldsymbol{\psi}},\boldsymbol{\psi}\to0}\frac{\partial^{2}}{\partial\hat{\psi}_{i}^{l}\partial\psi_{i}^{l}}\Gamma[\hat{\boldsymbol{\psi}},\boldsymbol{\psi}]$.
Assuming the systems are self-averaging for $N\!\to\!\infty$ and computing the disorder average (denoted by the upper line) 
$\overline{\Gamma[\hat{\boldsymbol{\psi}},\boldsymbol{\psi}]}$, the disorder-averaged overlaps can be obtained, $q^{l}\!=\!\frac{1}{N}\sum_{i=1}\overline{\langle\hat{s}_{i}^{l}s_{i}^{l}\rangle}.$ 
For convenience, we introduce the field doublet $H^{l}\!\equiv\![\hat{h}^{l},h^{l}]^{T}$. Expressing the GF $\Gamma[\hat{\boldsymbol{\psi}},\boldsymbol{\psi}]$ by macroscopic order parameters and averaging over the disorder yields the saddle-point integral $\overline{\Gamma}\!=\!\int\{d\boldsymbol{q}d\boldsymbol{\mathcal{Q}}\}e^{N\Psi[\boldsymbol{q},\boldsymbol{\mathcal{Q}}]}$, where $\Psi[...]$ is \cite{Li2017sup}
\begin{align}
\Psi & =\mathrm{i}\sum_{l=0}^{L}\mathcal{Q}^{l}q^{l}+\log\int\prod_{l=1}^{L}d\hat{h}^{l}dh^{l}\sum_{\{\hat{s}^{l},s^{l}\}}M[\hat{s},s,\hat{h},h],\label{eq:saddle_surface}
\end{align}
and the effective single site measure $M[...]$ has the following form for both
continuous and binary weights:
\begin{align}
& M[\hat{s},s,\hat{h},h]  =P(\hat{s}^{0})\delta_{\hat{s}^{0},s^{0}}e^{-\mathrm{i}\sum_{l=0}^{L}\mathcal{Q}^{l}\hat{s}^{l}s^{l}}\nonumber \\
 & \times\prod_{l=1}^{L}\left\{ \frac{e^{\beta\hat{s}^{l}\hat{h}^{l}}}{2\cosh\beta\hat{h}^{l}}\frac{e^{\beta s^{l}h^{l}}}{2\cosh\beta h^{l}}~
\frac{e^{-\frac{1}{2}(H^{l})^{T}\cdot\Sigma_{l}^{-1}\cdot H^{l}}}{\sqrt{(2\pi)^{2}|\Sigma_{l}(q^{l-1})|}}
\right\} .\label{eq:effective_single_site_measure}
\end{align}
The Gaussian density of the local field $\{\hat{h}^{l},h^{l}\}$ in Eq. (\ref{eq:effective_single_site_measure}) comes from summing a large number of random variables in $\hat{h}^{l}$ and $h^{l}$. The precision matrix $\Sigma_{l}^{-1}$, linking the effective field $\hat{h}^{l}$ and $h^{l}$, measures the correlation between internal fields of 
the two systems and depends on the overlap $q^{l-1}$ of the previous layer. In the limit $N\!\to\!\infty$, the GF $\overline{\Gamma}$ is dominated by the extremum of $\Psi$. Variation with respect to $\mathcal{Q}^{l}$ gives rise to saddle-point equations of the order parameters $q^{l}\!=\!\langle\hat{s}^{l}s^{l}\rangle_{M[...]}$,
where the average is taken over the measure $M[...]$ of Eq. (\ref{eq:effective_single_site_measure}).
The conjugate order parameter $\mathcal{Q}^{l}$, ensuring the normalization
of the measure, vanishes identically. It leads to the evolution equation
\cite{Li2017sup} 
\begin{equation}
q^{l}  \!=\!\int d\hat{h}^{l}dh^{l}\tanh(\beta\hat{h}^{l})\tanh(\beta h^{l})
 \frac{e^{ -\frac{1}{2}(H^{l})^{T}\cdot\Sigma_{l}^{-1}\cdot H^{l}} }{\sqrt{(2\pi)^{2}|\Sigma_{l}|}}  .
\end{equation}

The overlap evolution is somewhat similar to the dynamical mean field relation in Ref.~\cite{Poole2016}, but the objects investigated and the remainder of the study are different. We focus on the function-space landscape rather than the sensitivity of function to input perturbations.

\textit{Densely connected continuous weights.}\textendash In the first
scenario, we assume weight variables $\hat{w}_{ij}^{l}$ to be independently drawn from
a Gaussian density $\mathcal{N}(0,\sigma^{2})$ and the perturbed weights to have the form $w_{ij}^{l}\!=\!\sqrt{1-(\eta^{l})^{2}}\hat{w}_{ij}^{l}\!+\!\eta^{l}\delta w_{ij}^{l}$, where $\delta w_{ij}^{l}$ are drawn from $\mathcal{N}(0,\sigma^{2})$ independently of $\hat{w}_{ij}^{l}$. It ensures that $w_{ij}^{l}$  has the same variance $\sigma^{2}$. The parameter $\eta^{l}$ quantifies the strength of perturbation introduced in layer $l$. In this case the
covariance matrix between the local fields $\hat{h}^{l}$ and $h^{l}$
takes the form
\begin{equation}
\Sigma_{l}(\eta^{l},q^{l-1})=\sigma^{2}\begin{bmatrix}1 & \sqrt{1-(\eta^{l})^{2}}q^{l-1}\\
\sqrt{1-(\eta^{l})^{2}}q^{l-1} & 1
\end{bmatrix},
\end{equation}
leading to the closed form solution of the overlap as $\beta\!\to\!\infty$, 
\begin{equation}
q^{l}=\frac{2}{\pi}\sin^{-1}\left(\sqrt{1-(\eta^{l})^{2}}q^{l-1}\right).
\end{equation}
Of particular interest is the final-layer overlap given the same input for the two systems under  specific perturbations
$q^{L}(\{\eta^{l}\},q^{0}\!=\!1)$. The average error $\varepsilon\!=\!\frac{1}{2}(1-q^{L})$
measures the typical distance between the two mappings.

\begin{figure}
	\includegraphics[scale=0.22]{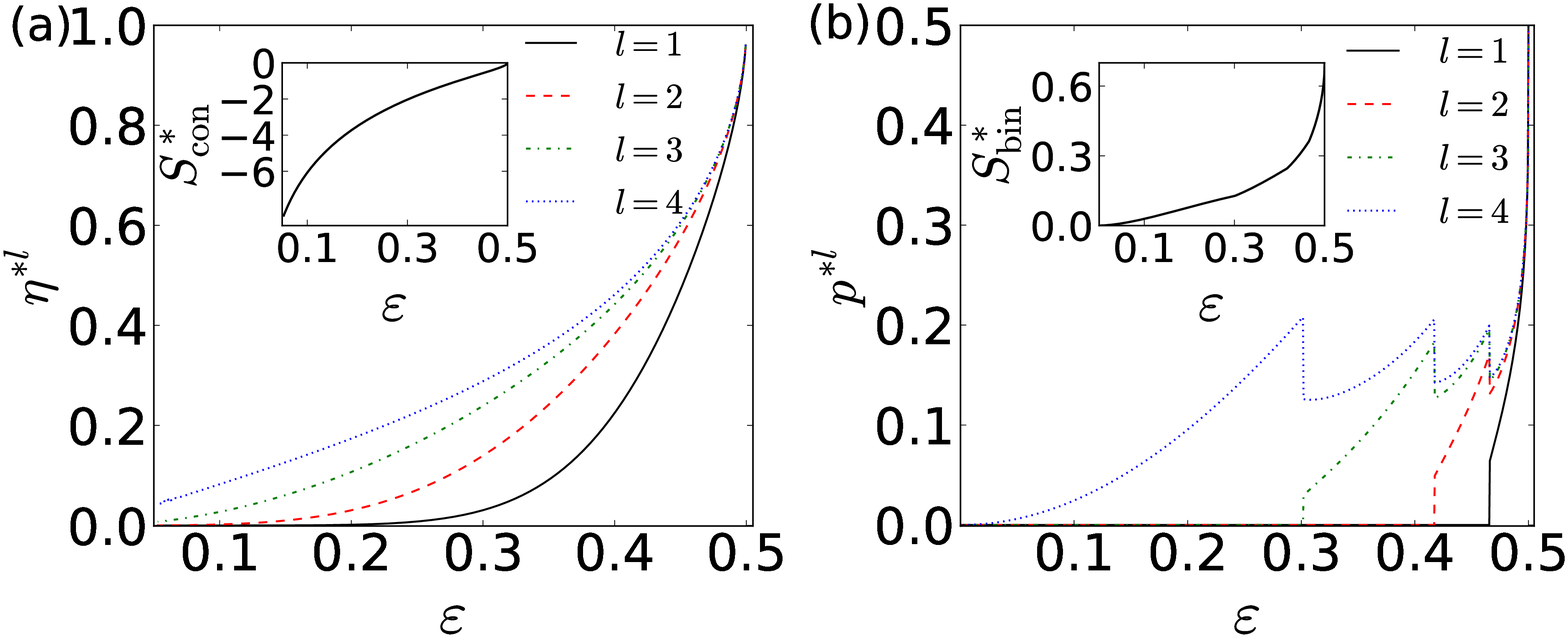}
	\includegraphics[scale=0.22]{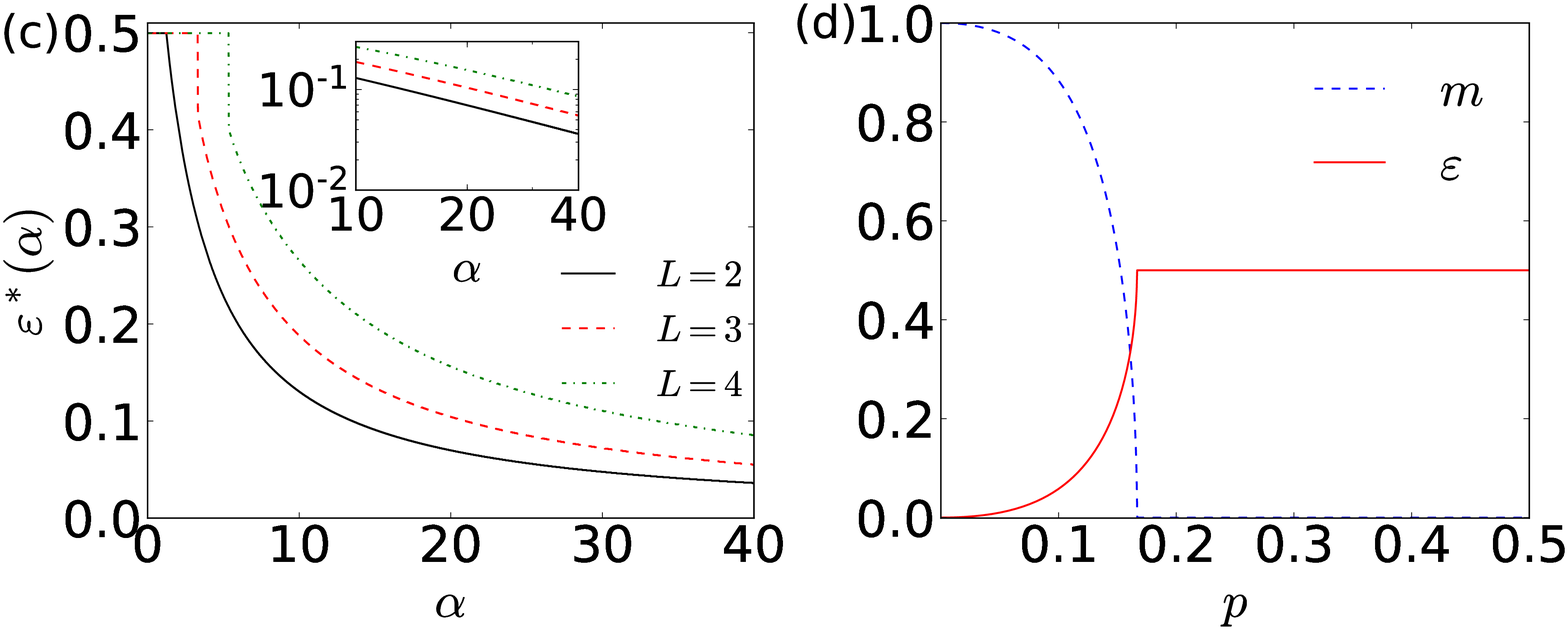}
	\includegraphics[scale=0.22]{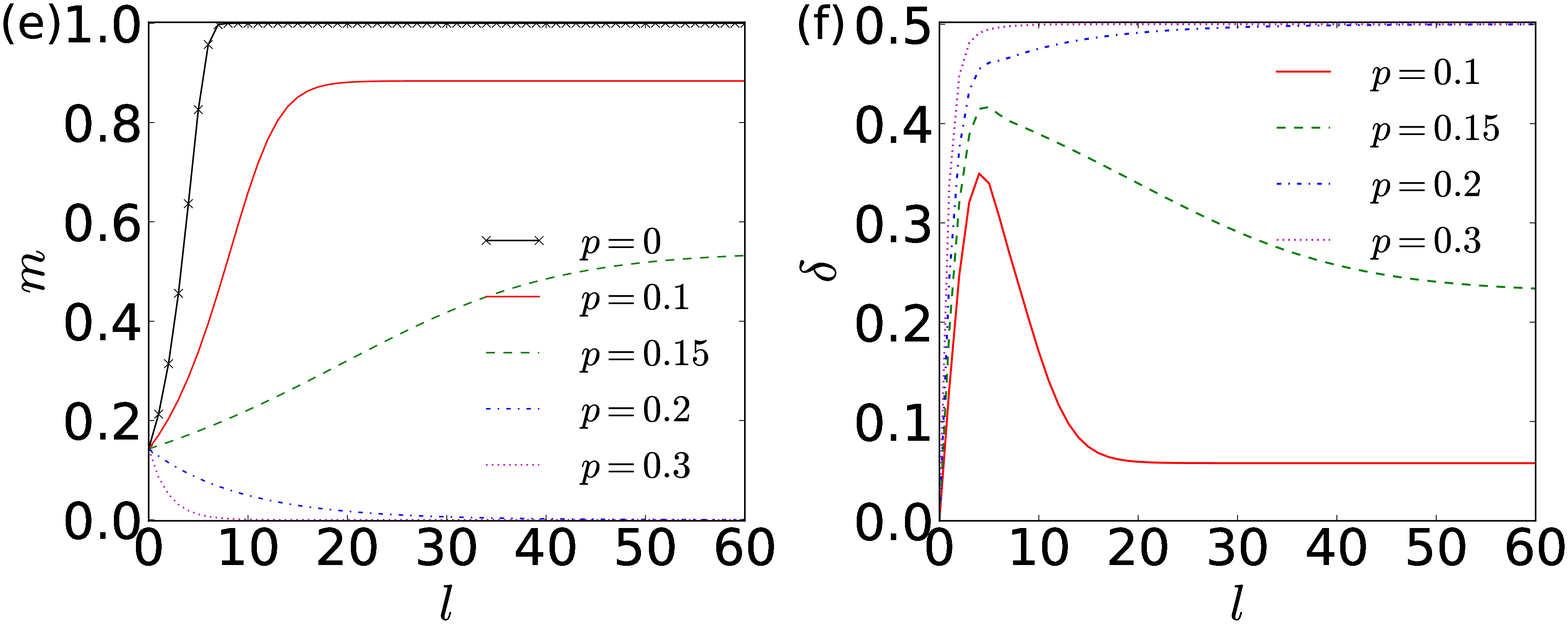}
\caption{Maximal-entropy perturbations as a function of output error $\varepsilon$ for a four-layer  densely connected networks with (a) continuous weights and (b) binary weights. Inset represents the growth in entropy with respect to $\varepsilon$. (c) Generalization curves of densely connected networks with continuous weights by using the annealed approximation. The inset demonstrates the classical asymptotic behavior of $\varepsilon^{*}\!\sim\! \alpha^{-1}$ in the large $\alpha$ limit~\cite{Engel2001}.  (d) Stationary magnetization $m$ and function error $\varepsilon$ for sparsely connected MAJ-$3$-based DLMs as a function of perturbation probability $p$ in networks with binary weights. We show the evolution of (e) magnetization and (f) internal activation error $\delta$ over layers. Note that $p\!=\!0$ corresponds to the reference network. 
All results are obtained in the deterministic limit $\beta \to \infty$.  \label{fig:results}}
\end{figure}

The number of solutions at a given distance (error) $\varepsilon$ away from the reference function is indicative of how difficult it is to obtain this level of approximation at the vicinity of the exact function. Let the $N$-dimensional vectors $\hat{\boldsymbol{w}}^{l,i}$ and $\boldsymbol{w}^{l,i}$
denote the weights of the $i$th perceptron of the reference and perturbed systems
at layer $l$, respectively; the expected angle between them is $\theta^{l}\!=\!\sin^{-1}\eta^{l}$.
Then the perceptron $\boldsymbol{w}^{l,i}$ occupies on average 
an angular volume around $\hat{\boldsymbol{w}}^{l,i}$  as $\Omega(\eta^{l})\sim\sin^{N-2}\theta^{l}\!=\!(\eta^{l})^{N-2}$~\cite{Seung1992,Engel2001}. The total weight-space volume of the  perturbed system is $\Omega_{\text{tot}}(\{\eta^{l}\})\!=\!\prod_{l=1}^{L}\prod_{i}(\eta^{l})^{N-2}$,
and the corresponding entropy density is
\begin{equation}
S_{\text{con}}(\{\eta^{l}\})=\frac{1}{LN^{2}}\log\Omega_{\text{tot}}(\{\eta^{l}\})\approx\frac{1}{L}\sum_{l=1}^{L}\log\eta^{l}.\label{eq:continuous_entropy}
\end{equation}
In the thermodynamic limit $N\to\infty$, the set of perturbed functions at distance $\varepsilon$ away from the reference function is dominated by those with perturbation vector $\{\eta^{*l}\}$, which maximizes the entropy $S_{\text{con}}(\{\eta^{l}\})$ subject to the constraint $q^{L}(\{\eta^{l}\})\!=\!1\!-\!2\varepsilon$. The result of $\{\eta^{*l}\}$ for a four-layer network, shown in Fig.~\ref{fig:results}(a), reveals that the dominant perturbation $\eta^{*l}$ to the reference network decays faster for smaller $l$ values; this indicates that closer to the reference function, solutions are dominated by functions where early-layer weights match better the reference network.
Consequently, high-$\varepsilon$ functions are ruled out faster during training through the successful alignment of earlier layers, resulting in the increasing concentration of low-$\varepsilon$ functions and better generalization.
We denote the maximal weight-space volume at distance $\varepsilon$ away from the reference function as $\Omega_0(\varepsilon)\equiv \Omega_{\text{tot}}(\{ \eta^{*l} \})$.

Supervised learning is based on the introduction of input-output example pairs. Introducing constraints, in the form of $P \!\equiv\! \alpha LN^{2}$ examples provided by the reference function, the weight-space volume at small distance $\varepsilon$ away from the reference function is reshaped as $\Omega_{\alpha}(\varepsilon) \!=\! \Omega_0(\varepsilon)(1-\varepsilon)^{P}$  in the annealed approximation~\cite{Seung1992, Engel2001}; details of the derivation can be found in Ref.~\cite{Li2017sup}. The typical distance $\varepsilon^{*}(\alpha)\!=\!\text{argmax}_{\varepsilon} \Omega_{\alpha}(\varepsilon)$ can be interpreted as the generalization error in the presence of $P$ examples, giving rise to an approximate generalization curve shown in Fig.~\ref{fig:results}(c).  
These are expected to be valid in the small $\varepsilon$ (large $\alpha$) limit on which the perturbation analysis is based. 
It is observed that typically a large number of examples ($\alpha \!\gg\! 10$) are needed for good generalization. This may imply that DLMs trained on realistic data sets (usually $\alpha\! \ll \!1$) occupy a small, highly biased subspace, different from the typical function space analyzed here (e.g., the handwritten digit MNIST database~\cite{LeCun1998} represents highly biased inputs that occupy a very small fraction of the input space). Note that the results correspond to a typical generalization performance under the assumption of self-averaging, potentially with unlimited computational resources and independently of the training rule used.

\textit{Densely connected binary weights.}\textendash Once trained, networks with binary weights are highly efficient computationally, which is especially useful in devices with limited memory or computational resources~\cite{Courbariaux2016, Rastegari2016}. Here we consider a reference network with binary weight variables
drawn from the distribution $P(\hat{w}_{ij}^{l})\!=\!\frac{1}{2}\delta_{\hat{w}_{ij}^{l},1}\!+\!\frac{1}{2}\delta_{\hat{w}_{ij}^{l},\!-\!1}$, while
the perturbed network weights follow the distribution
$P(w_{ij}^{l})\!=\!(1\!-\!p^{l})\delta_{w_{ij}^{l},\hat{w}_{ij}^{l}}\!+\!p^{l}\delta_{w_{ij}^{l},\!-\!\hat{w}_{ij}^{l}}$,
where $p^{l}$ is the flipping probability at layer $l$. The covariance matrix 
\begin{equation}
\Sigma_{l}(p^{l},q^{l-1})=\begin{bmatrix}1 & (1-2p^{l})q^{l-1}\\
(1-2p^{l})q^{l-1} & 1
\end{bmatrix},
\end{equation}
gives rise to overlaps $q^{l}$ as $\beta\!\to\!\infty$  of the form
\begin{equation}
q^{l}=\frac{2}{\pi}\sin^{-1}\left((1-2p^{l})q^{l-1}\right).
\end{equation}
The entropy density of the perturbed system is given by 
\begin{equation}
S_{\text{bin}}(\{p^{l}\})=\frac{1}{L}\sum_{l=1}^{L}-p^{l}\log p^{l}-(1-p^{l})\log(1-p^{l}).
\end{equation}
Similarly, the entropy $S_{\text{bin}}(\{p^{l}\})$ is maximized by the perturbation vector $\{p^{*l}\}$ 
subject to $q^{L}(\{p^{l}\})\!=\!1\!-\!2\varepsilon$  at a distance $\varepsilon$ away from the reference function. The result of $\{p^{*l}\}$ for a four-layer binary neural network is shown in Fig.~\ref{fig:results}(b). Surprisingly, as $\varepsilon$ decreases, the first-layer weights are first to align perfectly with those of the reference function followed by the second-layer weights and so on. The discontinuities come from the nonconvex nature of the entropy landscape $S_{\text{bin}}(\{p^{l}\})$ when one restricts the perturbed system to the nonlinear $\varepsilon$-error surface satisfying $q^{L}(\{p^{l}\})\!=\!1\!-\!2\varepsilon$.
Nevertheless, there exist many more high-$\varepsilon$ than low-$\varepsilon$ functions for densely connected binary networks [as indicated by the entropy shown in the inset of Fig.~\ref{fig:results}(b)], and it remains to explore how low generalization error functions could be identified.

\textit{Sparsely connected binary weights.}\textendash Lastly, we consider the sparsely connected DLM with binary weights; these topologies are of interest to practitioners due to the reduction in degrees of freedom and their computational and energy efficiency.
The layered setup is similar to the previous case, except that unit $i$ at layer $l$ is randomly connected to a small number $k$ of units in layer $(l\!-\!1)$ and its local field is given by $\hat{h}_{i}^{l}(\hat{\boldsymbol{w}}^{l},\hat{\boldsymbol{s}}^{l-1})\!=\!\frac{1}{\sqrt{k}}\sum_{j}A_{ij}^{l}\hat{w}_{ij}^{l}\hat{s}_{j}^{l-1}$, where the adjacency matrix $A^{l}$ represents the connectivity between the two layers. The perturbed network has the same topology but its weights are randomly flipped, $P(w_{ij}^{l})\!=\!(1\!-\!p^{l})\delta_{w_{ij}^{l},\hat{w}_{ij}^{l}}\!+\!p^{l}\delta_{w_{ij}^{l},\!-\!\hat{w}_{ij}^{l}}$;
the activation and the joint probability of the two systems follow from Eqs. (\ref{eq:node_conditional_prob}) and (\ref{eq:joint_density}).
Unlike the case of densely connected networks, the magnetization $m^{l}\!\equiv\!\frac{1}{N}\sum_{i}s_{i}^{l}$ also plays an important role in the evolution of sparse networks. The GF approach gives rise to the order parameter $\mathcal{P}^{l}(\hat{s},s)\!\equiv\!\frac{1}{N}\sum_{i}\delta_{\hat{s}_{i}^{l},\hat{s}}\delta_{s_{i}^{l},s}$
relating to the magnetization and overlap by $\mathcal{P}^{l}(\hat{s},s)\!=\!\frac{1}{4}(1+\hat{s}\hat{m}^{l}+sm^{l}+\hat{s}sq^{l})$. 

The random topology provides an additional disorder to average over. For simplicity, we assign the reference weights to $\hat{w}_{ij}^{l}\!=\!1$,
which in the limit $\beta\!\to\!\infty$ relate to the $k$-majority gate (MAJ-$k$)-based Boolean
formulas that provide all Boolean functions with uniform probability at the large $L$ limit~\cite{Savicky1990,Mozeika2010}. For a uniform perturbation over layers $p^{l}\!=\!p$, we focus on  functions generated in the deep regime $L\!\to\!\infty$, where the order parameters take the form
\begin{eqnarray}
m^{l}&=&\sum_{\{s_{j}\}}\prod_{j=1}^{k}\frac{1}{2}\left[1+s_{j}m^{l-1}(1-2p)\right]\text{sgn}\left[\sum^{k}_{j=1}s_{j}\right],\label{eq:sparse_m}
\\
q^{l} & =&\sum_{\{s_{j},\hat{s}_{j}\}}\prod_{j=1}^{k}\frac{1}{4}\left[1+\hat{s}_{j}\hat{m}^{l-1}+s_{j}m^{l-1}(1-2p)\right.  \label{eq:sparse_q} \\
 && \quad\left.+s_{j}\hat{s}_{j}q^{l-1}(1-2p)\right]\text{sgn}\left[\sum^{k}_{j=1}\hat{s}_{j}\right]\text{sgn}\left[\sum^{k}_{j=1}s_{j}\right]. \nonumber
\end{eqnarray}

For finite $k$, the macroscopic observables at layer $l$ are polynomially dependent on the observables at layer $(l\!-\!1)$ up to order $k$.
In the limit $L\!\to\!\infty$, the Boolean functions generated depend on the initial magnetization $m^{0}\!=\!\frac{1}{N}\sum_{i}s_{i}^{0}$.
Here, we consider a biased case with initial conditions $\hat{m}^{0}\!=\!m^{0}\!>\!0$ and $q^{0}\!=\!1$. The reference function admits a stationary solution $\hat{m}^{\infty}\!=\!1$, computing a 1-bit information-preserving majority function~\cite{Mozeika2010}. Both magnetization of the perturbed function $m^{\infty}$ and the function error $\varepsilon\!=\!\frac{1}{2}(1\!-\!q^{\infty})$ exhibit a transition from the ordered phase to the paramagnetic phase at some critical perturbation level $p_{c}$, below which the perturbed network computes the reference function with error $\varepsilon\!<\!\frac{1}{2}$.
The results for $k\!=\!3$ are shown in Fig.~\ref{fig:results}(d). Interestingly, the critical perturbation $p_{c}$ coincides with the location of the critical thermal noise $\epsilon_{c}\!=\!\frac{1}{2}(1\!-\!\tanh\beta_{c})$ for noisy $k$-majority gate-based Boolean formulas; for $k\!=\!3$, the critical perturbation  $p_{c}\!=\!\frac{1}{6}$~\cite{Mozeika2009}. Below $p_{c}$, there exist two ordered states with $m^{\infty}\!=\!\pm\sqrt{(1\!-\!6p)/(1\!-\!2p)^{3}}$, and the overlap satisfies $q^{\infty}\!=\!m^{\infty}$~\cite{Li2017sup}, 
which is also reminiscent of the thermal noise-induced solutions~\cite{Mozeika2009}.
However, the underlying physical implications are drastically different. Here it indicates that even in the deep network regime there exists a large number ${Nk \choose Nkp}^{L}$ of networks that can reliably represent the reference function when $p\!<\!p_{c}$. This function landscape is important for learning tasks to achieve a similar rule to the reference function. The propagation of internal error $\delta(l)\!\equiv\!\frac{1}{2}(1-q^{l})$, shown in Fig.~\ref{fig:results}(f), exhibits a stage of error increase followed by a stage of error decrease for $p\!<\!p_{c}$.  Consequently a successful sparse DLM requires more layers to reduce errors and provide a higher similarity to the reference function when we approach $p_{c}$, indicating the need of deep networks in such models.

In summary, we propose a GF analysis to probe the function landscapes of DLMs, focusing on the entropy of functions, given their error with respect to a reference function. The entropy maximization of densely connected networks at fixed error to the reference function indicates that weights of earlier layers are the first to align with reference function parameters when the error decreases. It highlights the importance of early-layer weights for reliable computation~\cite{Raghu2017} and sheds light on the parameter learning dynamics in function space during the learning process. We also investigate the phase transitions behavior in sparsely connected networks, which advocate the use of deeper machines for suppressing errors with respect to the reference function in these models. 
The suggested GF framework is very general and can accommodate other structures and computing elements, e.g., continuous variables, other activation functions (such as the commonly used ReLU activation function~\cite{Li2017sup}) and more complicated weight ensembles. 
In Ref.~\cite{Li2017sup}, we also demonstrate the effect of negatively or positively correlated weight variables on the expressive power of networks with ReLU activation and their impact on the function space, and investigate the behavior of simple convolutional DLMs.
Moreover, the GF framework allows one to investigate other aspects as well, including finite size effects and the use of perturbative expansion to provide a systematic analysis of the interactions between network elements. This is a step towards a principled investigation of the typical behavior of DLMs and we envisage follow-up work on various aspects of the learning process. 
\begin{acknowledgments}

We thank K. Y. Michael Wong for helpful discussions. Support from The Leverhulme Trust (RPG-2013-48) (D.S.) and Research Grants Council of Hong Kong (Grants No. 605813 and No. 16322616) (B.L.) is acknowledged.
\end{acknowledgments}


\widetext
\clearpage
\begin{center}
\textbf{\large Exploring the Function Space of Deep-Learning Machines\\
Supplemental Material}
\end{center}
\setcounter{equation}{0}
\setcounter{figure}{0}
\setcounter{table}{0}
\setcounter{page}{1}
\makeatletter
\renewcommand{\theequation}{S\arabic{equation}}
\renewcommand{\thefigure}{S\arabic{figure}}
\renewcommand{\bibnumfmt}[1]{[S#1]}
\renewcommand{\citenumfont}[1]{S#1}

\subsection{Function Space and the Entropy of Solutions}

\begin{figure}[h]
\includegraphics{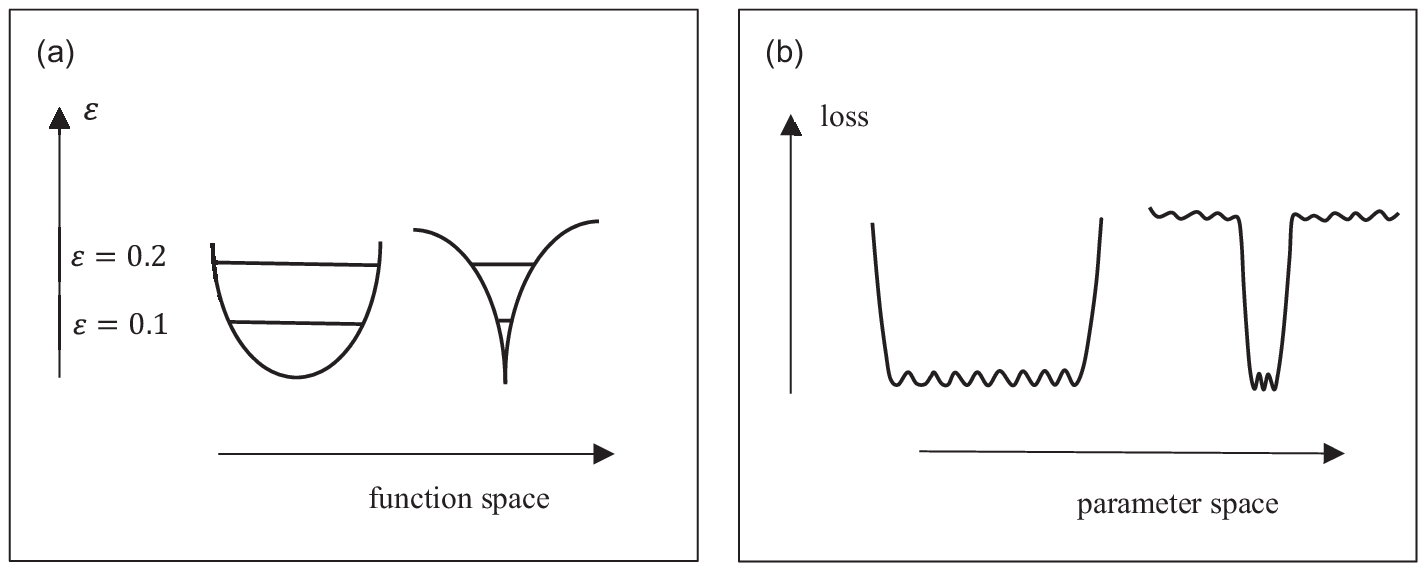} \caption{A pictorial illustration of two type of landscape of the (a) function
space and (b) parameter space. (a) the y-axis represents the distance-measure
from the reference function and the x-axis the entropy at that distance
(error) level. (b) the y-axis stands for the loss (lowest value -
closest to the reference function) and the x-axis for the change in
parameters. \label{fig:functionspace}}
\end{figure}

Our study explores the function space and especially the number of
solutions in the vicinity of a reference function. It relies on the
assumption, in the absence of any other detailed information, that
the bigger the volume of solutions is, the easier it will be to find
one. For instance, if there exist many low-error functions in function
space (Fig.~\ref{fig:functionspace}(a) left and \ref{fig:functionspace}(b)
left), then it is in general easier to achieve good alignment with
the reference function (generalization) by some generic unspecified
learning algorithm. On the other hand, if the volume of low-error
functions is very small compared to high-error solutions (Fig.~\ref{fig:functionspace}(a)
right and \ref{fig:functionspace}(b) right), then it is generally
harder to achieve good alignment (generalization) since the function
space is dominated by the high-error functions. This concept has been
introduced, albeit for simpler frameworks~\cite{Schwartz1990_sm}.

\subsection{Disorder Averaged Generating Functional for Densely-Connected Networks}

Here we give detailed derivations of the generating functional for
densely connected networks

\begin{align}
\Gamma[\hat{\boldsymbol{\psi}},\boldsymbol{\psi}] & =\sum_{\{s_{i}^{l},\hat{s}_{i}^{l}\}_{\forall l,i}}P(\hat{\boldsymbol{s}}^{0})\prod_{i=1}^{N}\delta_{s_{i}^{0},\hat{s}_{i}^{0}}\prod_{l=1}^{L}P(\hat{\boldsymbol{s}}^{l}|\hat{\boldsymbol{w}}^{l},\hat{\boldsymbol{s}}^{l-1})P(\boldsymbol{s}^{l}|\boldsymbol{w}^{l},\boldsymbol{s}^{l-1})e^{-\mathrm{i}\sum_{l,i}(\hat{\psi}_{i}^{l}\hat{s}_{i}^{l}+\psi_{i}^{l}s_{i}^{l})}\nonumber \\
 & =\sum_{\{s_{i}^{l},\hat{s}_{i}^{l}\}_{\forall l,i}}P(\boldsymbol{\hat{s}}^{0})\prod_{i}\delta_{\hat{s}_{i}^{0},s_{i}^{0}}\prod_{l=1}^{L}\prod_{i}\frac{e^{\beta\hat{s}_{i}^{l}\hat{h}_{i}^{l}(\hat{\boldsymbol{w}}^{l},\hat{\boldsymbol{s}}^{l-1})}}{2\cosh\left[\beta\hat{h}_{i}^{l}(\hat{\boldsymbol{w}}^{l},\hat{\boldsymbol{s}}^{l-1})\right]}\frac{e^{\beta s_{i}^{l}h_{i}^{l}(\boldsymbol{w}^{l},\boldsymbol{s}^{l-1})}}{2\cosh\left[\beta s_{i}^{l}h_{i}^{l}(\boldsymbol{w}^{l},\boldsymbol{s}^{l-1})\right]},
\end{align}
with the local field $\hat{h}_{i}^{l}(\hat{\boldsymbol{w}}^{l},\hat{\boldsymbol{s}}^{l-1})=\frac{1}{\sqrt{N}}\sum_{j}\hat{w}_{ij}^{l}\hat{s}_{j}^{l-1}$.
To deal with the non-linearity of the local fields $\hat{h}_{i}^{l}(\hat{\boldsymbol{w}}^{l},\hat{\boldsymbol{s}}^{l-1})$
in the conditional probability, we introduce auxiliary fields through
the integral representation of $\delta$-function

\begin{equation}
1=\int_{-\infty}^{\infty}\frac{d\hat{h}_{i}^{l}d\hat{x}_{i}^{l}}{2\pi}e^{\mathrm{i}\hat{x}_{i}^{l}\left(\hat{h}_{i}^{l}-\frac{1}{\sqrt{N}}\sum_{j}\hat{w}_{ij}^{l}\hat{s}_{j}^{l-1}\right)},\quad1=\int_{-\infty}^{\infty}\frac{dh_{i}^{l}dx_{i}^{l}}{2\pi}e^{\mathrm{i}x_{i}^{l}\left(h_{i}^{l}-\frac{1}{\sqrt{N}}\sum_{j}w_{ij}^{l}s_{j}^{l-1}\right)},
\end{equation}
which allows us to express the quench random variables $\hat{w}_{ij}^{l}$
and $w_{ij}^{l}$ linearly in the exponents, leading to 
\begin{align}
\Gamma[\hat{\boldsymbol{\psi}},\boldsymbol{\psi}] & =\int\prod_{l=1}^{L}\prod_{i}\frac{d\hat{h}_{i}^{l}d\hat{x}_{i}^{l}}{2\pi}\frac{dh_{i}^{l}dx_{i}^{l}}{2\pi}\sum_{\{\hat{s}_{i}^{l},s_{i}^{l}\}_{\forall l,i}}P(\boldsymbol{\hat{s}}^{0})\prod_{i}\delta_{\hat{s}_{i}^{0},s_{i}^{0}}e^{-\mathrm{i}\sum_{l,i}(\hat{\psi}_{i}^{l}\hat{s}_{i}^{l}+\psi_{i}^{l}s_{i}^{l})}\nonumber \\
 & \times\exp\left\{ \frac{-\mathrm{i}}{\sqrt{N}}\sum_{l=1}^{L}\left[\sum_{ij}\hat{w}_{ij}^{l}\hat{x}_{i}^{l}\hat{s}_{j}^{l-1}+\sum_{ij}w_{ij}^{l}x_{i}^{l}s_{j}^{l-1}\right]\right\} \label{eq:gen_func_linear_w_dw}\\
 & \times\exp\left\{ \sum_{l=1}^{L}\sum_{i}\left[\beta\hat{s}_{i}^{l}\hat{h}_{i}^{l}+\beta s_{i}^{l}h_{i}^{l}-\log2\cosh\beta\hat{h}_{i}^{l}-\log2\cosh\beta h_{i}^{l}+\mathrm{i}\hat{x}_{i}^{l}\hat{h}_{i}^{l}+\mathrm{i}x_{i}^{l}h_{i}^{l}\right]\right\} .\label{eq:gen_func_fields}
\end{align}

Assuming the system is self-averaging, the disorder average can be
traced over \textit{ab initio} \cite{Dominicis1978_sm}, leaving the
disorder averaged generating functional $\overline{\Gamma[\hat{\boldsymbol{\psi}},\boldsymbol{\psi}]}$.
We consider two types of networks, one with continuous weight variable
following a Gaussian distribution and the other with binary weight
variables. In both cases, we consider input distribution of the form
\begin{equation}
P\left(\hat{\boldsymbol{s}}^{0}\right)=\prod_{i}P(\hat{s}_{i}^{0})=\prod_{i}\left(\frac{1}{2}\delta_{\hat{s}_{i}^{0},1}+\frac{1}{2}\delta_{\hat{s}_{i}^{0},-1}\right).
\end{equation}

\subsubsection{Continuous weight variables with Gaussian disorder}

In this case, we assume that weight variables are independent and
follow a Gaussian density $\hat{w}_{ij}^{l}\sim\mathcal{N}(0,\sigma^{2})$,
and that the perturbed weight has the form $w_{ij}^{l}=\sqrt{1-(\eta^{l})^{2}}\hat{w}_{ij}^{l}+\eta^{l}\delta w_{ij}^{l}$
where $\delta w_{ij}^{l}$ also have density $\mathcal{N}(0,\sigma^{2})$
but is independent of $\hat{w}_{ij}^{l}$. For weight
variables that are independent and identically distributed, an alternative
derivation based on central limit theorem can be employed~\cite{Poole2016_sm}.
Nevertheless, the proposed GF  framework  is more principled and general,
and can accommodate many possible extensions.

Averaging Eq.~(\ref{eq:gen_func_linear_w_dw}) over the weight $\{\hat{w}_{ij}^{l}\}$
and perturbation $\{\delta w_{ij}^{l}\}$ gives 
\begin{align}
 & \int P(\hat{\boldsymbol{w}})P(\delta\boldsymbol{w})d\hat{\boldsymbol{w}}d\delta\boldsymbol{w}\exp\left\{ \frac{-\mathrm{i}}{\sqrt{N}}\sum_{l=1}^{L}\left[\sum_{ij}\hat{w}_{ij}^{l}\left(\hat{x}_{i}^{l}\hat{s}_{j}^{l-1}+\sqrt{1-(\eta^{l})^{2}}x_{i}^{l}s_{j}^{l-1}\right)+\eta^{l}\sum_{ij}\delta w_{ij}^{l}x_{i}^{l}s_{j}^{l-1}\right]\right\} \nonumber \\
= & \exp\left\{ -\sigma^{2}\sum_{l=1}^{L}\left[\frac{1}{2}\sum_{i}(\hat{x}_{i}^{l})^{2}+\frac{1}{2}\sum_{i}(x_{i}^{l})^{2}+\sqrt{1-(\eta^{l})^{2}}\sum_{i}\hat{x}_{i}^{l}x_{i}^{l}\left(\sum_{j}\frac{1}{N}\hat{s}_{j}^{l-1}s_{j}^{l-1}\right)\right]\right\} ,\label{eq:x_density_from_int}
\end{align}
leading to the disorder-averaged generating functional 
\begin{align}
\overline{\Gamma[\hat{\boldsymbol{\psi}},\boldsymbol{\psi}]} & =\int\prod_{l=1}^{L}\prod_{i}\frac{d\hat{h}_{i}^{l}d\hat{x}_{i}^{l}}{2\pi}\frac{dh_{i}^{l}dx_{i}^{l}}{2\pi}\sum_{\{\hat{s}_{i}^{l},s_{i}^{l}\}_{\forall l,i}}P(\boldsymbol{\hat{s}}^{0})\prod_{i}\delta_{\hat{s}_{i}^{0},s_{i}^{0}}e^{-\mathrm{i}\sum_{l,i}(\hat{\psi}_{i}^{l}\hat{s}_{i}^{l}+\psi_{i}^{l}s_{i}^{l})}\nonumber \\
 & \times\exp\left\{ -\sigma^{2}\sum_{l=1}^{L}\left[\frac{1}{2}\sum_{i}(\hat{x}_{i}^{l})^{2}+\frac{1}{2}\sum_{i}(x_{i}^{l})^{2}+\sqrt{1-(\eta^{l})^{2}}\sum_{i}\hat{x}_{i}^{l}x_{i}^{l}\left(\sum_{j}\frac{1}{N}\hat{s}_{j}^{l-1}s_{j}^{l-1}\right)\right]\right\} \nonumber \\
 & \times\exp\left\{ \sum_{l=1}^{L}\sum_{i}\left[\beta\hat{s}_{i}^{l}\hat{h}_{i}^{l}+\beta s_{i}^{l}h_{i}^{l}-\log2\cosh\beta\hat{h}_{i}^{l}-\log2\cosh\beta h_{i}^{l}+\mathrm{i}\hat{x}_{i}^{l}\hat{h}_{i}^{l}+\mathrm{i}x_{i}^{l}h_{i}^{l}\right]\right\} .
\end{align}

Site factorization can be achieved by defining the macroscopic order
parameter $q^{l}:=\sum_{j}\frac{1}{N}\hat{s}_{j}^{l}s_{j}^{l}$ through
the integral representation of the $\delta$-function 
\begin{equation}
1=\int\frac{d\mathcal{Q}^{l}dq^{l}}{2\pi/N}\exp\left\{ \mathrm{i}N\mathcal{Q}^{l}\left[q^{l}-\frac{1}{N}\sum_{i}s_{i}^{l}\hat{s}_{i}^{l}\right]\right\} ,
\end{equation}
which leads to 
\begin{align}
\overline{\Gamma[\hat{\boldsymbol{\psi}},\boldsymbol{\psi}]} & =\int\prod_{l=0}^{L}\frac{d\mathcal{Q}^{l}dq^{l}}{2\pi/N}\exp\left\{ \mathrm{i}N\sum_{l=0}^{L}\mathcal{Q}^{l}q^{l}\right\} \nonumber \\
 & \times\int\prod_{l=1}^{L}\prod_{i}\frac{d\hat{h}_{i}^{l}d\hat{x}_{i}^{l}}{2\pi}\frac{dh_{i}^{l}dx_{i}^{l}}{2\pi}\sum_{\{\hat{s}_{i}^{l},s_{i}^{l}\}_{\forall l,i}}P(\boldsymbol{\hat{s}}^{0})\prod_{i}\delta_{\hat{s}_{i}^{0},s_{i}^{0}}e^{-\mathrm{i}\sum_{l,i}(\hat{\psi}_{i}^{l}\hat{s}_{i}^{l}+\psi_{i}^{l}s_{i}^{l})}\nonumber \\
 & \times\exp\left\{ -\sigma^{2}\sum_{l=1}^{L}\left[\frac{1}{2}\sum_{i}(\hat{x}_{i}^{l})^{2}+\frac{1}{2}\sum_{i}(x_{i}^{l})^{2}+\sqrt{1-(\eta^{l})^{2}}\sum_{i}\hat{x}_{i}^{l}x_{i}^{l}q^{l-1}\right]\right\} \nonumber \\
 & \times\exp\left\{ \sum_{l=1}^{L}\sum_{i}\left[\beta\hat{s}_{i}^{l}\hat{h}_{i}^{l}+\beta s_{i}^{l}h_{i}^{l}-\log2\cosh\beta\hat{h}_{i}^{l}-\log2\cosh\beta h_{i}^{l}+\mathrm{i}\hat{x}_{i}^{l}\hat{h}_{i}^{l}+\mathrm{i}x_{i}^{l}h_{i}^{l}\right]\right\} \nonumber \\
 & \times\exp\left(-\mathrm{i}\sum_{l=0}^{L}\sum_{i}\mathcal{Q}^{l}\hat{s}_{i}^{l}s_{i}^{l}\right).
\end{align}
Now the spin and field variables are the same for any site $i$; we
therefore consider the generating functional as a function of site-independent
conjugate fields, i.e., $\hat{\psi}_{i}^{l}=\hat{\psi}^{l}$ and $\psi_{i}^{l}=\psi^{l}$,
which takes the form 
\begin{align}
\overline{\Gamma[\hat{\psi},\psi]} & =\int\prod_{l=0}^{L}\frac{d\mathcal{Q}^{l}dq^{l}}{2\pi/N}\exp\left\{ \mathrm{i}N\sum_{l=0}^{L}\mathcal{Q}^{l}q^{l}\right\} \nonumber \\
 & \times\left\{ \int\prod_{l=1}^{L}\frac{d\hat{h}^{l}d\hat{x}^{l}}{2\pi}\frac{dh^{l}dx^{l}}{2\pi}\sum_{\{\hat{s}^{l},s^{l}\}_{\forall l}}P(\hat{s}^{0})\delta_{\hat{s}^{0},s^{0}}e^{-\mathrm{i}\sum_{l}(\hat{\psi}^{l}\hat{s}^{l}+\psi^{l}s^{l})}\right.\nonumber \\
 & \times\exp\left(-\sigma^{2}\sum_{l=1}^{L}\left[\frac{1}{2}(\hat{x}^{l})^{2}+\frac{1}{2}(x^{l})^{2}+\sqrt{1-(\eta^{l})^{2}}\hat{x}^{l}x^{l}q^{l-1}\right]+\mathrm{i}\sum_{l=1}^{L}\left(\hat{x}^{l}\hat{h}^{l}+x^{l}h^{l}\right)\right)\label{eq:x_gaussian_density}\\
 & \times\exp\left(\sum_{l=1}^{L}\left[\beta\hat{s}^{l}\hat{h}^{l}+\beta s^{l}h^{l}-\log2\cosh\beta\hat{h}^{l}-\log2\cosh\beta h^{l}\right]\right)\nonumber \\
 & \left.\times\exp\left(-\mathrm{i}\sum_{l=0}^{L}\mathcal{Q}^{l}\hat{s}^{l}s^{l}\right)\right\} ^{N}.\label{eq:factorized_gen_func}
\end{align}

For convenience, we define the fields doublet $X^{l}:=[\hat{x}^{l},x^{l}]^{T}$,
$H^{l}:=[\hat{h}^{l},h^{l}]^{T}$ and the covariance matrix 
\begin{equation}
\Sigma_{l}(\eta^{l},q^{l-1})=\sigma^{2}\begin{bmatrix}1 & \sqrt{1-(\eta^{l})^{2}}q^{l-1}\\
\sqrt{1-(\eta^{l})^{2}}q^{l-1} & 1
\end{bmatrix}.
\end{equation}
The density of $\{\hat{x}^{l},x^{l}\}$ in Eq.~(\ref{eq:x_gaussian_density})
has the form $\exp\{-\sum_{l}[\frac{1}{2}(X^{l})^{T}\cdot\Sigma_{l}\cdot X^{l}+\mathrm{i}(X^{l})^{T}\cdot H^{l}]\}$,
which can be integrated over $\{X^{l}\}$, yielding the joint Gaussian
density of $H^{l}$ with precision matrix $\Sigma_{l}^{-1}$ 
\begin{equation}
\frac{1}{\sqrt{(2\pi)^{2}|\Sigma_{l}|}}\exp\left\{ -\sum_{l}\frac{1}{2}(H^{l})^{T}\cdot\Sigma_{l}(\eta^{l},q^{l-1})^{-1}\cdot H^{l}\right\} .\label{eq:h_gaussian_density}
\end{equation}

\subsubsection{Binary weight variables}

For the binary weight variables, we assume a disorder of the form
$P(\hat{w}_{ij}^{l})=\frac{1}{2}\delta_{\hat{w}_{ij}^{l},1}+\frac{1}{2}\delta_{\hat{w}_{ij}^{l},-1}$,
$P(w_{ij}^{l})=(1-p^{l})\delta_{w_{ij}^{l},\hat{w}_{ij}^{l}}+p^{l}\delta_{w_{ij}^{l},-\hat{w}_{ij}^{l}}$,
where $p^{l}$ is the flipping probability at layer $l$. Averaging
Eq.~(\ref{eq:gen_func_linear_w_dw}) over the weight $\{\hat{w}_{ij}^{l}\}$
and the perturbation $\{\delta w_{ij}^{l}\}$ gives 
\begin{align}
 & \prod_{l=1}^{L}\prod_{ij}\left[(1-p^{l})\cos\left(\frac{1}{\sqrt{N}}(\hat{x}_{i}^{l}\hat{s}_{j}^{l-1}+x_{i}^{l}s_{j}^{l-1})\right)+p^{l}\cos\left(\frac{1}{\sqrt{N}}(\hat{x}_{i}^{l}\hat{s}_{j}^{l-1}-x_{i}^{l}s_{j}^{l-1})\right)\right]\nonumber \\
\approx & \prod_{l=1}^{L}\prod_{ij}\left[(1-p^{l})\left[1-\frac{1}{2N}(\hat{x}_{i}^{l}\hat{s}_{j}^{l-1}+x_{i}^{l}s_{j}^{l-1})^{2}\right]+p^{l}\left[1-\frac{1}{2N}(\hat{x}_{i}^{l}\hat{s}_{j}^{l-1}-x_{i}^{l}s_{j}^{l-1})^{2}\right]\right]\nonumber \\
\approx & \prod_{l=1}^{L}\prod_{ij}\exp\left\{ -\frac{1}{2N}\left[(\hat{x}_{i}^{l})^{2}+(x_{i}^{l})^{2}+2(1-2p^{l})\hat{x}_{i}^{l}x_{i}^{l}\hat{s}_{j}^{l-1}s_{j}^{l-1}\right]\right\} \nonumber \\
= & \exp\left\{ -\sum_{l=1}^{L}\sum_{i}\left[\frac{1}{2}(\hat{x}_{i}^{l})^{2}+\frac{1}{2}(x_{i}^{l})^{2}+(1-2p^{l})\hat{x}_{i}^{l}x_{i}^{l}\left(\sum_{j}\frac{1}{N}\hat{s}_{j}^{l-1}s_{j}^{l-1}\right)\right]\right\} ,
\end{align}
where the large $N$ property is used in the second and third line.
The resulting density is similar to the density of the continuous
weight variables in Eq.~(\ref{eq:x_density_from_int}), and therefore
the same derivations can be applied. By identifying the covariance
matrix 
\begin{equation}
\Sigma_{l}(p^{l},q^{l-1})=\begin{bmatrix}1 & (1-2p^{l})q^{l-1}\\
(1-2p^{l})q^{l-1} & 1
\end{bmatrix},
\end{equation}
the density for the site independent local field $H^{l}$ also has
the Gaussian density in the form of 
\begin{equation}
\frac{1}{\sqrt{(2\pi)^{2}|\Sigma_{l}|}}\exp\left\{ -\sum_{l}\frac{1}{2}(H^{l})^{T}\cdot\Sigma_{l}(p^{l},q^{l-1})^{-1}\cdot H^{l}\right\} .
\end{equation}

\subsubsection{Effective single site measure and saddle point equations}

In summary, the cases of continuous weight variables and binary weight
variables have the unified expressions by properly identifying the
covariance matrix for the local fields $\{H^{l}\}$. The generating
functional in both cases has the form 
\begin{align}
\overline{\Gamma} & =\int\prod_{l=0}^{L}\frac{d\mathcal{Q}^{l}dq^{l}}{2\pi/N}e^{N\Psi[\boldsymbol{q},\boldsymbol{\mathcal{Q}}]},
\end{align}
with the exponent

\begin{align}
\Psi[\boldsymbol{q},\boldsymbol{\mathcal{Q}}] & =\mathrm{i}\sum_{l=0}^{L}\mathcal{Q}^{l}q^{l}+\log\int\prod_{l=1}^{L}d\hat{h}^{l}dh^{l}\sum_{\{\hat{s}^{l},s^{l}\}}M[\hat{s},s,\hat{h},h],\label{eq:saddle_potential}
\end{align}
and the effective single site measure $M[...]$ has the form of 
\begin{align}
M\left[\{\hat{s}^{l},s^{l},\hat{h}^{l},h^{l}\}\right] & =P(\hat{s}^{0})\delta_{\hat{s}^{0},s^{0}}e^{-\mathrm{i}\sum_{l=0}^{L}\mathcal{Q}^{l}\hat{s}^{l}s^{l}}e^{-i\sum_{l}(\hat{\psi}^{l}\hat{s}^{l}+\psi^{l}s^{l})}\nonumber \\
 & \times\prod_{l=1}^{L}\left\{ \frac{e^{\beta\hat{s}^{l}\hat{h}^{l}}}{2\cosh\beta\hat{h}^{l}}\frac{e^{\beta s^{l}h^{l}}}{2\cosh\beta h^{l}}\frac{1}{\sqrt{(2\pi)^{2}|\Sigma_{l}(q^{l-1})|}}\exp\left[-\frac{1}{2}(H^{l})^{T}\cdot\Sigma_{l}(q^{l-1})^{-1}\cdot H^{l}\right]\right\} \label{eq:effective_single_site_measure_dense}
\end{align}

Now that the potential $\Psi[...]$ is expressed by macroscopic order
parameters $\{\mathcal{Q}^{l},q^{l}\}$, the conjugate fields $\{\hat{\psi}^{l},\psi^{l}\}$
in Eq.~(\ref{eq:effective_single_site_measure_dense}) can be omitted.
For $N\to\infty$, $\overline{\Gamma}$ is dominated by the extremum
of $\Psi[\boldsymbol{q},\boldsymbol{\mathcal{Q}}]$ given by $\partial\Psi/\partial\mathcal{Q}^{l}=0$
and $\partial\Psi/\partial q^{l}=0$ 
\begin{align}
q^{l} & =\langle\hat{s}^{l}s^{l}\rangle_{M[...]}=\frac{\int\prod_{l=1}^{L}d\hat{h}^{l}dh^{l}\sum_{\{\hat{s}^{l},s^{l}\}}\hat{s}^{l}s^{l}M\left[\{\hat{s}^{l},s^{l},\hat{h}^{l},h^{l}\}\right]}{\int\prod_{l=1}^{L}d\hat{h}^{l}dh^{l}\sum_{\{\hat{s}^{l},s^{l}\}}M\left[\{\hat{s}^{l},s^{l},\hat{h}^{l},h^{l}\}\right]},\label{eq:saddle_eq_q}\\
\mathrm{i}\mathcal{Q}^{l} & =-\frac{\int\prod_{l=1}^{L}d\hat{h}^{l}dh^{l}\sum_{\{\hat{s}^{l},s^{l}\}}\frac{\partial}{\partial q^{l}}M\left[\{\hat{s}^{l},s^{l},\hat{h}^{l},h^{l}\}\right]}{\int\prod_{l=1}^{L}d\hat{h}^{l}dh^{l}\sum_{\{\hat{s}^{l},s^{l}\}}M\left[\{\hat{s}^{l},s^{l},\hat{h}^{l},h^{l}\}\right]}.\label{eq:saddle_eq_Q}
\end{align}

Since the two systems are interlinked through the input vectors $\delta_{\hat{s}^{0},s^{0}}$
and the final overlap $q^{L}$ does not show up in the measure $M[...]$,
the saddle point equations Eq.~(\ref{eq:saddle_eq_q}) and Eq.~(\ref{eq:saddle_eq_Q})
have the boundary conditions 
\begin{equation}
q^{0}=1,\qquad\mathrm{i}\mathcal{Q}^{L}=0.\label{eq:boundary_condition_QL}
\end{equation}

\subsubsection{Simplifications of the saddle point equations}

The saddle point of $i\mathcal{Q}^{l}$ can be further simplified
by iterating backward from the boundary condition Eq.~(\ref{eq:boundary_condition_QL}).
We start from computing $i\mathcal{Q}^{L-1}$ by substituting Eq.
(\ref{eq:boundary_condition_QL}) into Eq.~(\ref{eq:saddle_eq_Q})
\begin{align}
\mathrm{i}\mathcal{Q}^{L-1} & =-\frac{\int d\hat{h}^{L}dh^{L}\frac{\partial}{\partial q^{L-1}}\left\{ \frac{1}{\sqrt{(2\pi)^{2}|\Sigma_{L}(q^{L-1})|}}\exp\left[-\frac{1}{2}(H^{L})^{T}\cdot\Sigma_{L}(q^{L-1})^{-1}\cdot H^{L}\right]\right\} }{\int d\hat{h}^{L}dh^{L}\frac{1}{\sqrt{(2\pi)^{2}|\Sigma_{L}(q^{L-1})|}}\exp\left[-\frac{1}{2}(H^{L})^{T}\cdot\Sigma_{L}(q^{L-1})^{-1}\cdot H^{L}\right]}\nonumber \\
 & =-\int dH^{L}\frac{\partial}{\partial q^{L-1}}\left\{ \int\frac{dX}{(2\pi)^{2}}\exp\left[-\frac{1}{2}X^{T}\cdot\Sigma_{L}(q^{L-1})\cdot X+iX\cdot H^{L}\right]\right\} \nonumber \\
 & =\frac{1}{(2\pi)^{2}}\frac{\partial\Sigma_{L,12}(q^{L-1})}{\partial q^{L-1}}\int dH^{L}dX(\hat{x}x)\exp\left[-\frac{1}{2}X^{T}\cdot\Sigma_{L}(q^{L-1})\cdot X+iX\cdot H^{L}\right],
\end{align}
where $\partial\Sigma_{L,12}(q^{L-1})/\partial q^{L-1}=\sigma^{2}\sqrt{1-(\eta^{L})^{2}}$
for continuous weights and $\partial\Sigma_{L,12}(q^{L-1})/\partial q^{L-1}=1-2p^{l}$
for binary weights. Further notice that 
\begin{align}
 & \int dH^{L}dX(\hat{x}x)\exp\left[-\frac{1}{2}X^{T}\cdot\Sigma_{L}(q^{L-1})\cdot X+iX\cdot H^{L}\right]\nonumber \\
= & (2\pi)^{2}\int dX(\hat{x}x)\delta(\hat{x})\delta(x)\exp\left[-\frac{1}{2}X^{T}\cdot\Sigma_{L}(q^{L-1})\cdot X\right]=0,
\end{align}
giving $\mathrm{i}\mathcal{Q}^{L-1}=0$. This can be used to show
that $\mathrm{i}\mathcal{Q}^{L-2}=0$ etc and finally we conclude
that in the saddle point 
\begin{equation}
\mathrm{i}\mathcal{Q}^{l}=0,\quad\forall l.
\end{equation}

Therefore the overlap $q^{l}$ can be simplified as 
\begin{equation}
q^{l}=\int d\hat{h}^{l}dh^{l}\tanh(\beta\hat{h}^{l})\tanh(\beta h^{l})\frac{1}{\sqrt{(2\pi)^{2}|\Sigma_{l}(q^{l-1})|}}\exp\left[-\frac{1}{2}(H^{l})^{T}\cdot\Sigma_{l}(q^{l-1})^{-1}\cdot H^{l}\right].\label{eq:ql_integration}
\end{equation}
Consider the deterministic limit $\beta\to\infty$, $\tanh(\beta h^{l})\to\text{sgn}(h^{l})$,
the double integration in Eq.~(\ref{eq:ql_integration}) can be carried
out analytically, yielding 
\begin{align}
q^{l} & =\frac{2}{\pi}\tan^{-1}\left(\frac{\Sigma_{l,12}(q^{l-1})}{\sqrt{|\Sigma_{l}(q^{l-1})|}}\right)=\begin{cases}
\frac{2}{\pi}\sin^{-1}\left(\sqrt{1-(\eta^{l})^{2}}q^{l-1}\right), & \text{continuous weights,}\\
\\
\frac{2}{\pi}\sin^{-1}\left((1-2p^{l})q^{l-1}\right), & \text{binary weights.}
\end{cases}\label{eq:overlap_dynamics_dense}
\end{align}

\begin{figure}
\includegraphics[scale=0.42]{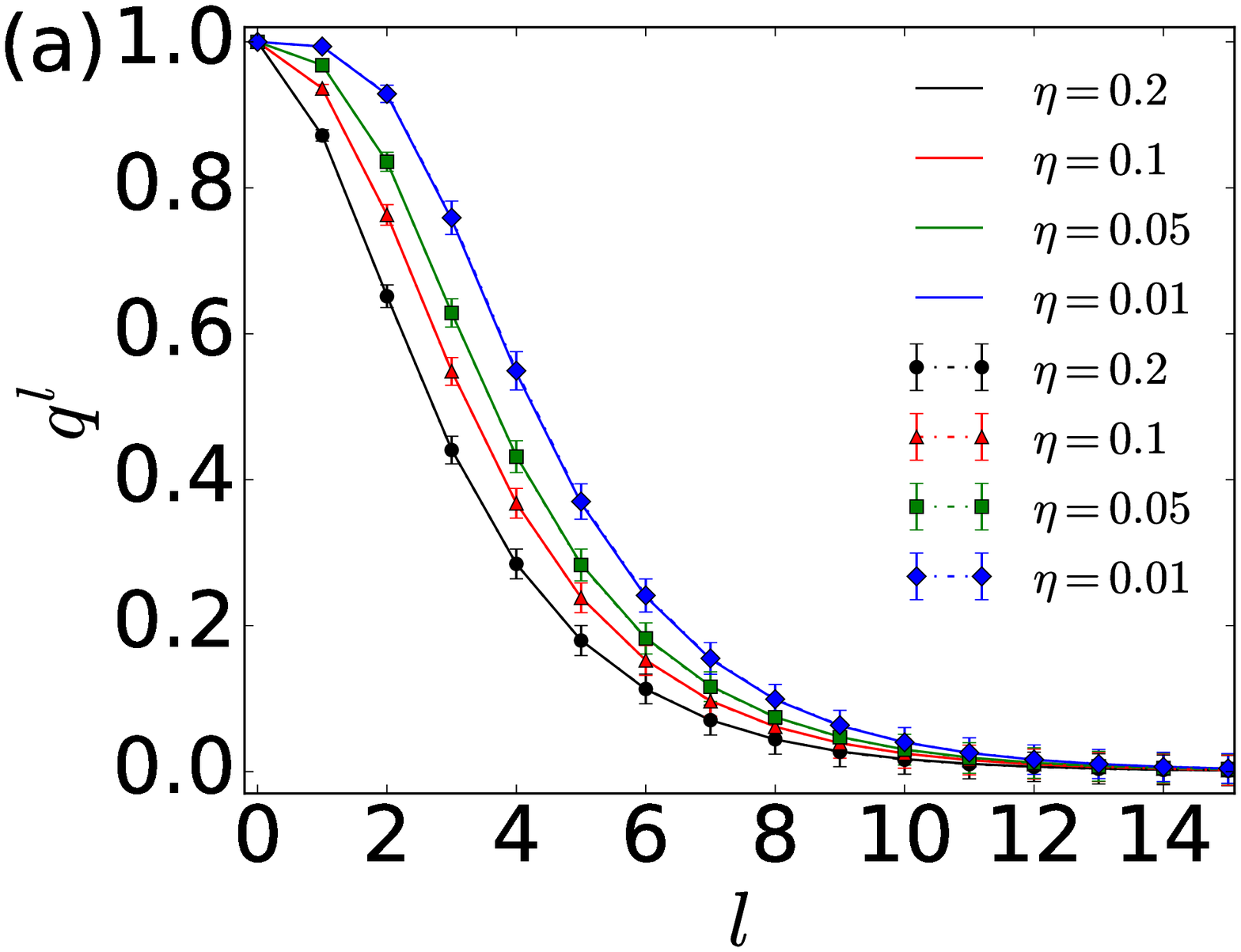}\includegraphics[scale=0.42]{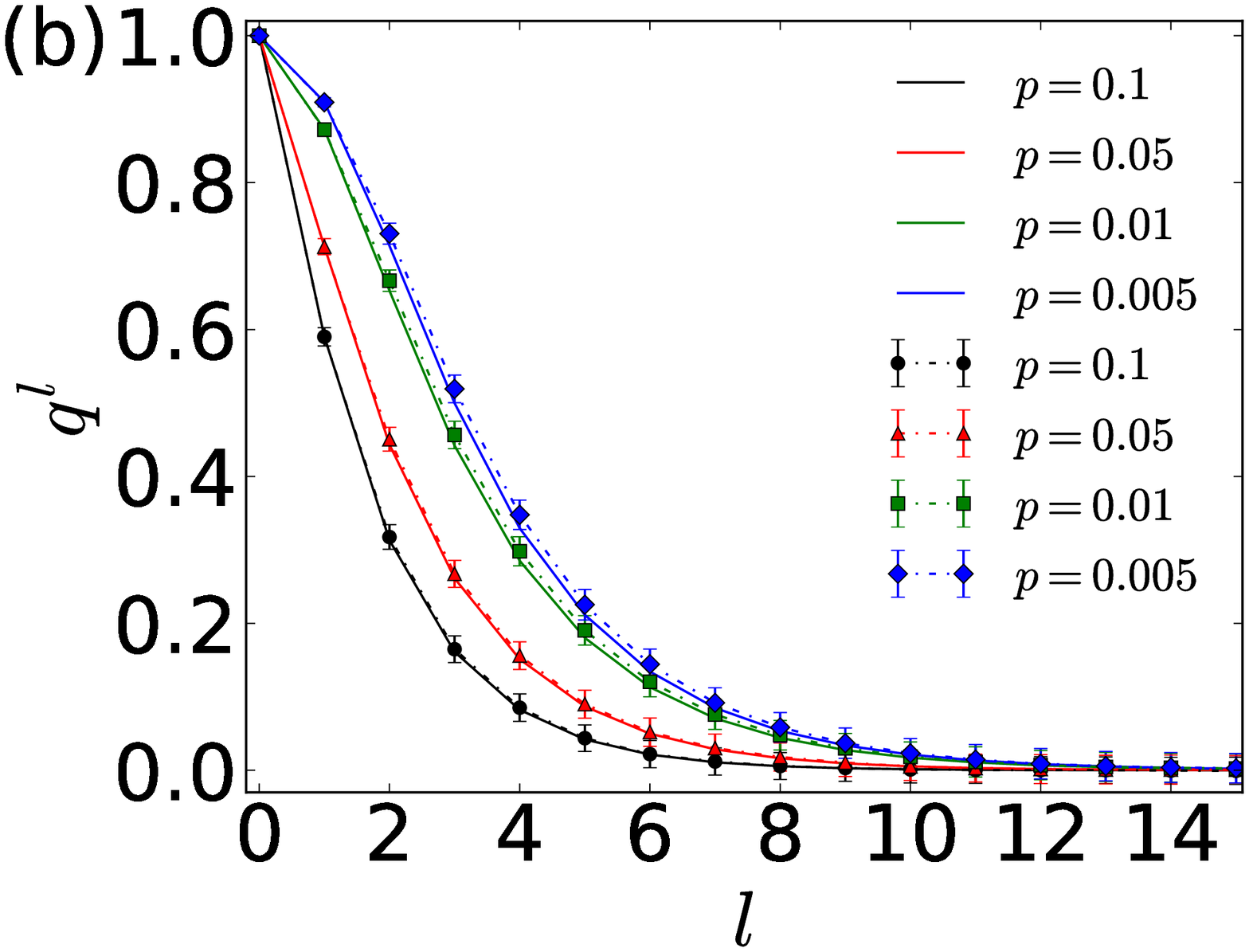}

\caption{Comparison of the overlap evolution in layers in densely-connected networks (Eq.~(\ref{eq:overlap_dynamics_dense}), solid lines) to Monte Carlo simulations (dashed-dotted lines with
markers), which shows a perfect match between the two approaches.
The perturbations are homogeneous in both cases, i.e., $\eta^{l}=\eta$
and $p^{l}=p$. The systems are of size $N=4000$ and the results
are averaged over $100$ disorder realizations. (a) Continuous weight
variables. (b) Binary weight variables. \label{fig:compare_to_MC_dense} }
\end{figure}

The mean field overlap evolution in layers of Eq.~(\ref{eq:overlap_dynamics_dense})
are perfectly confirmed by Monte Carlo simulations as shown in Fig.
\ref{fig:compare_to_MC_dense}.

\subsection{Maximum-Entropy Perturbation}

In this section we describe the procedure for finding the dominating
entropy spread for a given generalization error in the continuous-weight
scenario

\begin{align}
\max_{\{\eta^{l}\}}\quad & S_{\text{con}}(\{\eta^{l}\})=\frac{1}{L}\sum_{l=1}^{L}\log\eta^{l},\label{eq:max_ent_obj_func}\\
\text{s.t.}\quad & q^{L}(\{\eta^{l}\})=1-2\varepsilon.\label{eq:max_ent_eta_constraint}
\end{align}
To ease the nonlinear function-composition constraint in Eq.~(\ref{eq:max_ent_eta_constraint}),
we first notice in Eq.~(\ref{eq:overlap_dynamics_dense}) that $q^{l}$
is a monotonic function of $\eta^{l}$ which allows us to express
$\eta^{l}$ as 
\begin{equation}
\eta^{l}(q^{l-1},q^{l})=\sqrt{1-\frac{\sin^{2}(\frac{\pi}{2}q^{l})}{(q^{l-1})^{2}}},
\end{equation}
with $\sin(\frac{\pi}{2}q^{l})\leq q^{l-1}$. Then the maximization
problem can be recast as 
\begin{align}
\max_{\{q^{l}\}}\quad & S_{\text{con}}(\{\eta^{l}(q^{l-1},q^{l})\})=\frac{1}{L}\sum_{l=1}^{L}\log\eta^{l}(q^{l-1},q^{l}),\\
\text{s.t.}\quad & \sin(\frac{\pi}{2}q^{l})\leq q^{l-1},\quad l=1,...,L\nonumber \\
 & q^{0}=1,\nonumber \\
 & q^{L}=1-2\varepsilon,\label{eq:max_ent_q_constraints}
\end{align}
which can be solved numerically. The inequality constraints in Eq.
(\ref{eq:max_ent_q_constraints}) are non-convex and the minimizers
$\{q^{*l}\}$ found are possibly local minima. We start from multiple
initial guesses and select the best solution obtained. Then the maximum-entropy
perturbations are given by $\eta^{*l}=\eta^{l}(q^{*l-1},q^{*l})$.
We denote the corresponding optimal entropy at distance-$\varepsilon$
from the reference function as $S_{\text{con}}^{*}(\varepsilon)$.

\begin{figure}
\includegraphics[scale=0.45]{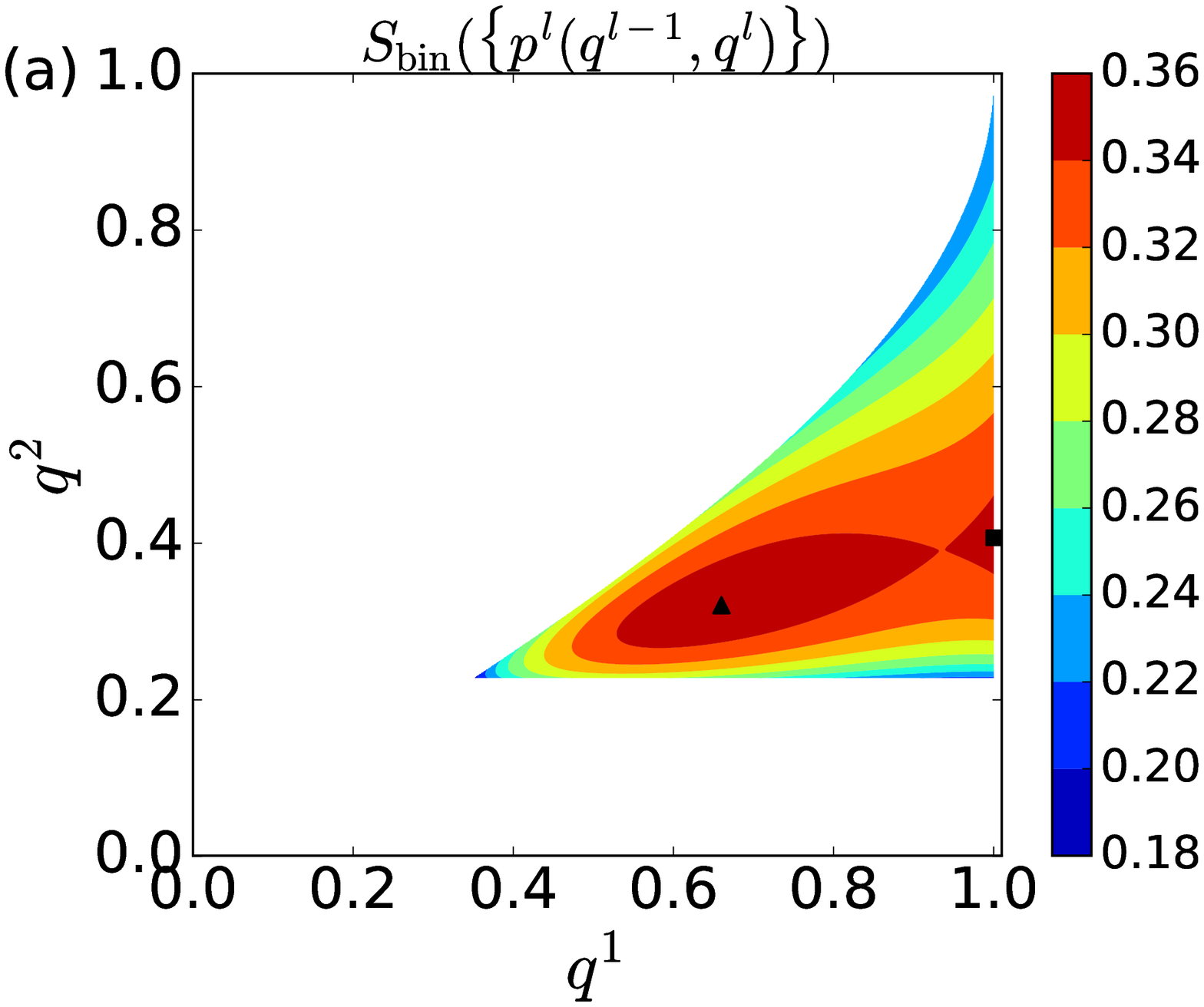}\includegraphics[scale=0.45]{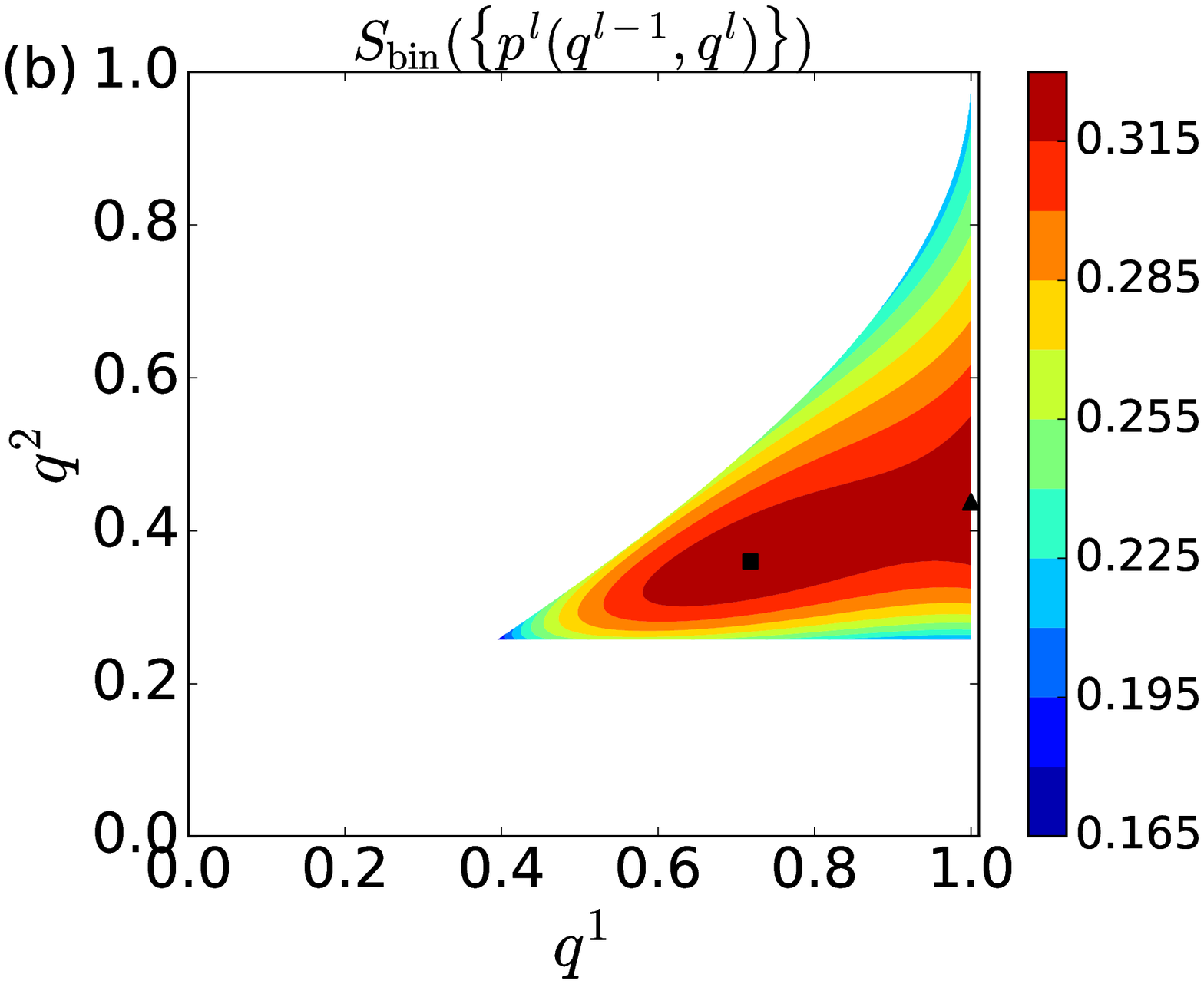}

\caption{Entropy of $L=3$ binary-weight networks $S_{\text{bin}}(\{p^{l}(q^{l-1},q^{l})\})$
as a function of the hidden-layer overlap $q^{1}$ and $q^{2}$. The
input and output overlaps are fixed as $q^{0}=1$ and $q^{L=3}=1-2\varepsilon$.
The global maximum is marked as a triangle and the local but not global
maximum is marked as a square. (a) $\varepsilon=0.427$; there exist
two local maxima and the global maximum is located in the interior
of the feasible region. (b) $\varepsilon=0.417$; the global maximum
jumps from the interior to the boundary of the feasible region with
$q^{1}=1$; it indicates $p^{*1}=0$, i.e., the first-layer weights
match with that of the reference network perfectly $\boldsymbol{w}^{1}=\hat{\boldsymbol{w}}^{1}$.
\label{fig:entropy_landscape_L3} }
\end{figure}

The procedures for the binary-weight problem are similar, with the
perturbation expressed in overlaps as 
\begin{equation}
p^{l}(q^{l-1},q^{l})=\frac{1}{2}\left(1-\frac{\sin(\frac{\pi}{2}q^{l})}{q^{l-1}}\right).
\end{equation}
The abrupt transition of the maximum-entropy perturbations $\{p^{*l}\}$
comes from the transitions of the global maximum of $S_{\text{bin}}(\{p^{l}(q^{l-1},q^{l})\})$
from the interior of the feasible region to its boundary. An example
of such jump in the maximum for a system with $L=3$ is illustrated
in Fig \ref{fig:entropy_landscape_L3}. This jump in entropy maximum
is reminiscent of the shift of free energy minimum occurred in first
order phase transitions.

\subsection{The Annealed Approximation of Learning}

In this section we give more details of the annealed approximation
of learning~\cite{Seung1992_sm,Engel2001_sm}. The $(L+1)$-layer
network constructed here maps an $N$-dimensional input $\{s_{i}^{0}\}_{i=1,...,N}$
to an $N$-dimensional output $\{s_{i}^{L}\}_{i=1,...,N}$, where
each component $s_{i}^{L}$ is the output of a certain Boolean function
with $N$ dimensional input, e.g., a single instance of a binary classifier.
Therefore all the $N$ components of the $\boldsymbol{s}^{L}$ are
the outputs of $N$ weakly coupled Boolean functions that share the
same typical properties.

\begin{figure}
\includegraphics[scale=0.4]{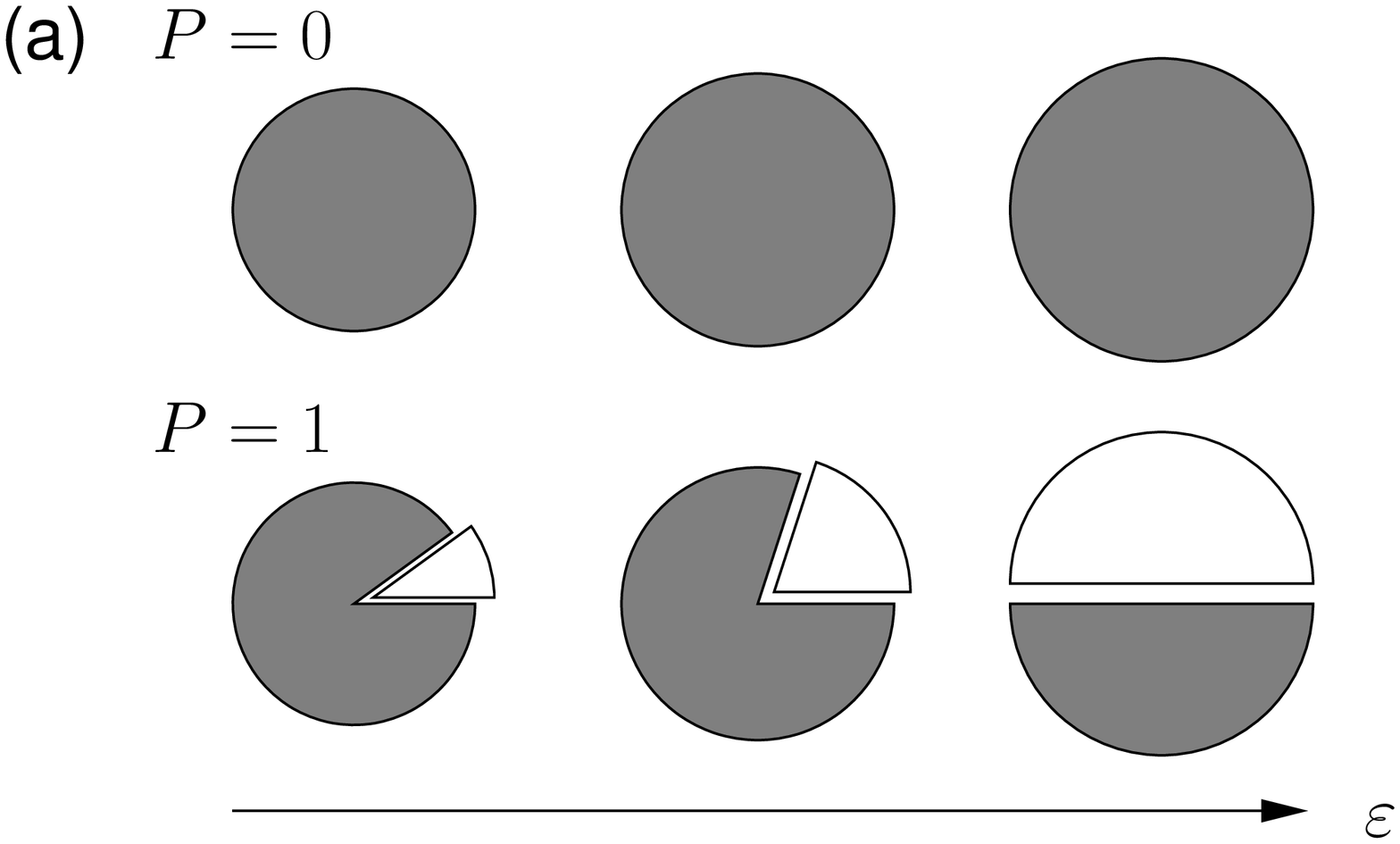} \includegraphics[scale=0.4]{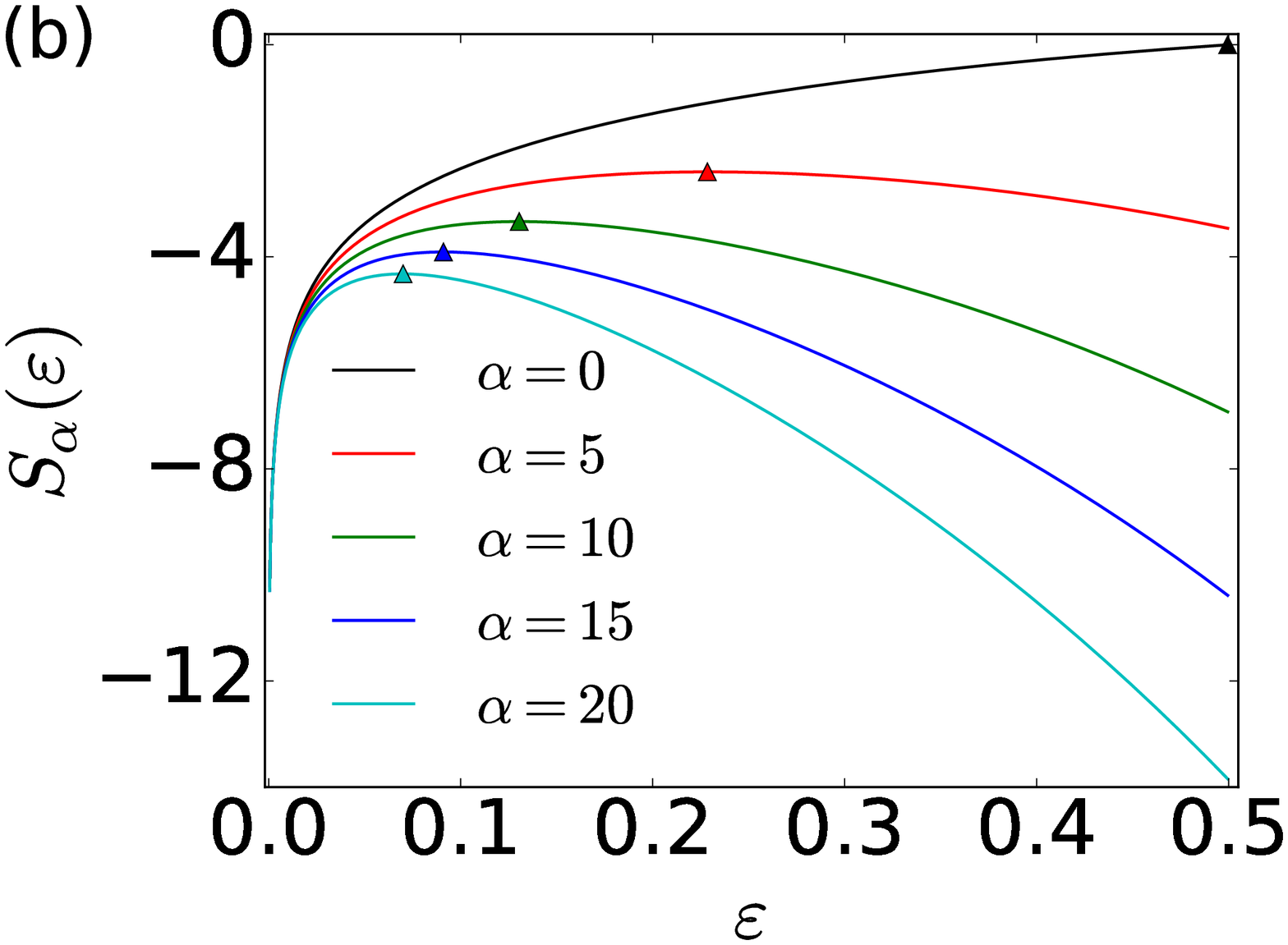}
\caption{(a) Schematic illustration of the annealed learning process. The dark
area represents the phase volume compatible with the examples at a
given level of generalization error $\varepsilon$. Upper row corresponds
to the case where no example is present. In the lower row, a single
example is introduced and the phase volume is re-shaped as $\Omega_{1}(\varepsilon)\!=\!\Omega_{0}(\varepsilon)(1-\varepsilon)$;
on average more high-$\varepsilon$ functions are incompatible with
the example than the low-$\varepsilon$ functions, which will be ruled
out in the learning process (marked as white regions removed). It
results in the increased concentration of low-$\varepsilon$ functions.
(b) The annealed entropy density $S_{\alpha}(\varepsilon)$ vs $\varepsilon$
for different $\alpha$ values. The two layer network ($L=2$) is
densely-connected with continuous-weight variables. The space of candidate
functions is dominated by functions that maximize the entropy $S_{\alpha}(\varepsilon)$
for a particular $\alpha$ value, marked by a triangle. \label{fig:anneal_learning_process} }
\end{figure}

Consider the case of densely-connected networks with continuous weights.
We start by showing how the profile of the typical phase volume at
a distance $\varepsilon$ from the reference function $\Omega_{0}(\varepsilon)\equiv\Omega_{\text{tot}}(\{\eta^{*l}\})$,
is re-shaped by the constraints represented by the introduction of
examples. The function distance (error) $\varepsilon$ we defined
is closely related to the generalization error in supervised learning.
Suppose a randomly chosen example (input-output pair provided by the
reference function) is introduced, the distance-$\varepsilon$ functions
on average has probability $\varepsilon$ of providing the wrong output
based on the input of the introduced example. On average, $(1-\varepsilon)$-fraction
of the $\Omega_{0}(\varepsilon)$ solutions are compatible with the
present example; therefore the remaining volume of compatible solutions
is $\Omega_{0}(\varepsilon)(1-\varepsilon)$. High-$\varepsilon$
functions are ruled out faster, resulting in increasing concentration
of the low-$\varepsilon$ functions. This learning process is illustrated
in Fig.~\ref{fig:anneal_learning_process}(a).

Assuming the examples are weakly correlated, the phase volume at the
presence of $P=\alpha LN^{2}$ examples using the annealed approximation
is given by $\Omega_{\alpha}(\varepsilon)\!=\!\Omega_{0}(\varepsilon)(1-\varepsilon)^{P}$.
The corresponding annealed entropy density is $S_{\alpha}(\varepsilon)=\log\Omega_{\alpha}(\varepsilon)/(LN^{2})=S_{\text{con}}^{*}(\varepsilon)+\alpha\log(1-\varepsilon)$.
In the large $N$ limit, the generalization error can be obtained
by 
\begin{align}
\varepsilon^{*}(\alpha) & =\text{argmax}_{\varepsilon}\Omega_{\alpha}(\varepsilon)\nonumber \\
 & =\text{argmax}_{\varepsilon}S_{\alpha}(\varepsilon).
\end{align}
The learning process in a network of $L=2$ is illustrated in Fig.~\ref{fig:anneal_learning_process}(b).

Although the generalization curve obtained by the annealed learning
theory is only an approximation, it produces the correct qualitative
behaviors in many cases, e.g., the large $\alpha$ scaling~\cite{Engel2001_sm}.
We also note that our derivation is based on the small perturbation
limit. Therefore, we expect the obtained generalization curves are
more relevant in the small $\varepsilon$ (large $\alpha$) limit.
While the annealed approximation is qualitatively correct in
the asymptotic limit of large $\alpha$ for realizable rules (where
the target functions can be completely realized by at least one of
the networks), it may fail for unrealizable rules \cite{Seung1992_sm}. Since our framework is based on exploring the function space in the neighborhood of an existing reference function, we expect the annealed approximation to provide qualitatively correct results. We also remark that such analysis of the learning task (which is called Gibbs learning~\cite{Engel2001_sm}) describes the expected typical generalization behavior, irrespective of the training rule used and with potentially unlimited computing resources. The performance of specific learning rules are subject to computational limitations and may provide different results, depending on the rule used and the energy surface of the specific problem. 

\subsection{Saddle Point Equations for Sparsely-Connected Networks}

The calculations of the sparsely connected binary networks are similar
to the densely-connected case, except that the local fields are only
contributed by finite number of elements $\hat{h}_{i}^{l}(A^{l},\hat{\boldsymbol{w}}^{l},\hat{\boldsymbol{s}}^{l-1})=\frac{1}{\sqrt{k}}\sum_{j}A_{ij}^{l}\hat{w}_{ij}^{l}\hat{s}_{j}^{l-1}$
where $A_{ij}^{l}$ is the connectivity matrix between layer $l-1$
and layer $l$ satisfying $\sum_{j}A_{ij}^{l}=k$; this induces additional
disorder of a different type. Such network topology is used in previous
studies of random Boolean formula~\cite{Mozeika2010_sm}. To make
use of the steps of calculations therein, we adapt the notion of adjacency
matrix $A_{ij}^{l}$ to the connectivity tensor as $A_{i_{1},...,i_{k}}^{l,i}$,
indicating the $i$-th neuron at layer $l$ is connected to the neurons
at $(l-1)$-th layer with indices $i_{1},...,i_{k}$. The local field
is expressed as 
\begin{align}
\hat{h}_{i}^{l}(A^{l,i},\hat{\boldsymbol{w}}^{l},\hat{\boldsymbol{s}}^{l-1}) & =\frac{1}{\sqrt{k}}\left(A_{j_{1},j_{2},...,j_{k}}^{l,i}\hat{w}_{ij_{1}}^{l}\hat{s}_{j_{1}}^{l-1}+A_{j_{1},j_{2},...,j_{k}}^{l,i}\hat{w}_{ij_{2}}^{l}\hat{s}_{j_{2}}^{l-1}+\cdots+A_{j_{1},j_{2},...,j_{k}}^{l,i}\hat{w}_{ij_{k}}^{l}\hat{s}_{j_{k}}^{l-1}\right),\nonumber \\
 & :=\sum_{j_{1},j_{2}...,j_{k}}^{N}A_{j_{1},j_{2},...,j_{k}}^{l,i}\hat{\alpha}_{\hat{w}}^{l,i}\left(\hat{s}_{j_{1}}^{l-1},\hat{s}_{j_{2}}^{l-1},...,\hat{s}_{j_{k}}^{l-1}\right),
\end{align}
where we have introduced the notation $\hat{\alpha}_{\hat{w}}^{l,i}$
to mimic the gate output of Boolean formula as in Ref. \cite{Mozeika2010_sm}.

The connectivity tensor follows the probability distribution 
\begin{equation}
P(\{A_{i_{1},...,i_{k}}^{l,i}\})=\frac{1}{Z_{A}}\prod_{l=1}^{L}\prod_{i}\left\{ \delta\left(\sum_{j_{1},...,j_{k}}^{N}A_{j_{1},...,j_{k}}^{l,i},1\right)\prod_{i_{1},...,i_{k}}^{N}\left[\frac{1}{N^{k}}\delta_{A_{i_{1},...,i_{k}}^{l,i},1}+(1-\frac{1}{N^{k}})\delta_{A_{i_{1},...,i_{k}}^{l,i},0}\right]\right\} ,\label{eq:connectivity_tensor_prob}
\end{equation}
where $Z_{A}$ is the normalization factor. In Eq.~(\ref{eq:connectivity_tensor_prob})
we have made use of the fact that the probability of having the same
indices in $\{i_{1},i_{2},...,i_{k}\}$ of $A_{i_{1},...,i_{k}}^{l,i}$
vanishes in the limit $N\to\infty$, or the fact that $\frac{1}{N(N-1)\cdots(N-k+1)}\approx\frac{1}{N^{k}}$
for finite $k$.

The average over the connectivity follows exactly the same procedures
as in Ref. \cite{Mozeika2010_sm} and the average over weight disorder
$P(\hat{\boldsymbol{w}})$ and $P(\boldsymbol{w})$ can be viewed
as an average over the gate disorder $\hat{\alpha}_{\hat{w}}$ and
$\alpha_{w}$. With the help of the functional order parameter $\mathcal{P}^{l}(\hat{s},s):=\frac{1}{N}\sum_{i}\delta_{\hat{s}_{i}^{l},\hat{s}}\delta_{s_{i}^{l},s}$
relating the magnetization $m^{l}:=\frac{1}{N}\sum_{i}s_{i}^{l}$
and overlap $q^{l}$ through 
\begin{equation}
\mathcal{P}^{l}(\hat{s},s)=\frac{1}{4}(1+\hat{s}\hat{m}^{l}+sm^{l}+\hat{s}sq^{l}),\label{eq:Pl_with_m_q}
\end{equation}
we are eventually led to the self-consistent saddle point equation
\cite{Mozeika2010_sm} 
\begin{equation}
\mathcal{P}^{l}(\hat{s},s)=\left\langle \delta_{\hat{s}^{l},\hat{s}}\delta_{s^{l},s}\right\rangle _{M[...]},\label{eq:Pl_average_over_M}
\end{equation}
with the effective single-site measure 
\begin{align}
M\left[\{\hat{s}^{l},s^{l}\}\right] & =P(\hat{s}^{0})\delta_{\hat{s}^{0},s^{0}}\prod_{l=1}^{L}\left\{ \sum_{\{\hat{s}_{j},s_{j}\}}\prod_{j=1}^{k}\mathcal{P}^{l}(\hat{s}_{j},s_{j})\right.\nonumber \\
 & \qquad\left.\times\left\langle \frac{\exp\left[\beta\hat{s}^{l}\hat{\alpha}_{\hat{w}}\left(\hat{s}_{1},\hat{s}_{2},...,\hat{s}_{k}\right)\right]}{2\cosh\left[\beta\hat{\alpha}_{\hat{w}}\left(\hat{s}_{1},\hat{s}_{2},...,\hat{s}_{k}\right)\right]}\frac{\exp\left[\beta s^{l}\alpha_{w}\left(s_{1},s_{2},...,s_{k}\right)\right]}{2\cosh\left[\beta s^{l}\alpha_{w}\left(s_{1},s_{2},...,s_{k}\right)\right]}\right\rangle _{\hat{\alpha}_{\hat{w}},\alpha_{w}}\right\} .\label{eq:effective_single_site_measure_sparse}
\end{align}

Note the spin variables $s_{j}$, $s^{l}$ and $s$ in this section
are different objects. From a physical point of view, the dynamical
variable $s^{l}$ at layer $l$ experiences a local field $\alpha_{w}\left(s_{1},s_{2},...,s_{k}\right)$
contributed by ``dummy spins'' $\{s_{1},s_{2},...,s_{k}\}$ which
reflects the mean field interactions from $k$ spins from layer $l-1$.
The macroscopic variables $s,\hat{s}$ are defined to express the
magnetizations and overlaps.

By using Eqs.~(\ref{eq:Pl_with_m_q}),(\ref{eq:Pl_average_over_M})
and (\ref{eq:effective_single_site_measure_sparse}), the evolution
of magnetization and overlap reads 
\begin{align}
\hat{m}^{l} & =\sum_{\{\hat{s}_{j}\}}\prod_{j=1}^{k}\left[\frac{1+\hat{s}_{j}\hat{m}^{l-1}}{2}\right]\left\langle \tanh\left[\beta\hat{\alpha}_{\hat{w}}\left(\hat{s}_{1},\hat{s}_{2},...,\hat{s}_{k}\right)\right]\right\rangle _{\hat{\alpha}_{\hat{w}}},\label{eq:mhat_dyn_alpha}\\
m^{l} & =\sum_{\{s_{j}\}}\prod_{j=1}^{k}\left[\frac{1+s_{j}m^{l-1}}{2}\right]\left\langle \tanh\left[\beta\alpha_{w}\left(s_{1},s_{2},...,s_{k}\right)\right]\right\rangle _{\hat{\alpha}_{\hat{w}},\alpha_{w}},\label{eq:m_dyn_alpha}\\
q^{l} & =\sum_{\{s_{j},\hat{s}_{j}\}}\prod_{j=1}^{k}\left[\frac{1+\hat{s}_{j}\hat{m}^{l-1}+s_{j}m^{l-1}+\hat{s}_{j}s_{j}q^{l-1}}{4}\right]\nonumber \\
 & \qquad\times\left\langle \tanh\left[\beta\hat{\alpha}_{\hat{w}}\left(\hat{s}_{1},\hat{s}_{2},...,\hat{s}_{k}\right)\right]\tanh\left[\beta\alpha_{w}\left(s_{1},s_{2},...,s_{k}\right)\right]\right\rangle _{\hat{\alpha}_{\hat{w}},\alpha_{w}}.\label{eq:q_dyn_alpha}
\end{align}

In the following we assume $k$ is odd.

\subsubsection{Average over weight disorder}

We consider the weights to be independent quench random variables
drawn from the distribution 
\begin{equation}
P_{\hat{w}}(\hat{w}_{ij}^{l})=\delta_{\hat{w}_{ij}^{l},1},\label{eq:sparse_wh_distribution}
\end{equation}
\begin{align}
P_{w}(w_{ij}^{l}) & =(1-p)\delta_{w_{ij}^{l},\hat{w}_{ij}^{l}}+p\delta_{w_{ij}^{l},-\hat{w}_{ij}^{l}}\nonumber \\
 & =(1-p)\delta_{w_{ij}^{l},1}+p\delta_{w_{ij}^{l},-1}.\label{eq:sparse_w_distribution}
\end{align}

In the limit $\beta\to\infty$, we have $\tanh\left[\beta\alpha_{w}\left(s_{1},s_{2},...,s_{k}\right)\right]=\text{sgn}\left(\sum_{j=1}^{k}w_{j}s_{j}\right)$
where $w_{j}$ follows the distribution $P_{w}(w_{j})$ in Eq.~(\ref{eq:sparse_w_distribution}).
Including the disorder distributions Eq.~(\ref{eq:sparse_wh_distribution})
and Eq.~(\ref{eq:sparse_w_distribution}) in the saddle point equations
gives 
\begin{equation}
m^{l}=\sum_{\{s_{j}\}}\prod_{j=1}^{k}\left[\frac{1+s_{j}m^{l-1}}{2}\right]\sum_{\{w_{j}\}}P_{w}(\{w_{j}\})\text{sgn}\biggr(\sum_{j=1}^{k}w_{j}s_{j}\biggr),\label{eq:m_dyn_pw}
\end{equation}
\begin{equation}
q^{l}=\sum_{\{\hat{s}_{j},s_{j}\}}\prod_{j=1}^{k}\left[\frac{1+\hat{s}_{j}\hat{m}^{l-1}+s_{j}m^{l-1}+\hat{s}_{j}s_{j}q^{l-1}}{4}\right]\text{sgn}\biggr(\sum_{j=1}^{k}\hat{s}_{j}\biggr)\sum_{\{w_{j}\}}P_{w}(\{w_{j}\})\text{sgn}\biggr(\sum_{j=1}^{k}w_{j}s_{j}\biggr).\label{eq:q_dyn_pw}
\end{equation}
These expressions can be further simplified as follows. We start with
Eq.~(\ref{eq:m_dyn_alpha}) by defining a new spin variable $\sigma_{j}:=w_{j}s_{j}$,
through the integral representation of the Kronecker delta 
\begin{equation}
\delta(\sigma_{j},w_{j}s_{j})=\int_{-\pi}^{\pi}\frac{d\theta_{j}}{2\pi}e^{\mathrm{i}\theta_{j}(\sigma_{j}-w_{j}s_{j})}=\begin{cases}
1 & \text{if }\sigma_{j}=w_{j}s_{j}\\
0 & \text{others}
\end{cases}
\end{equation}
\begin{equation}
\sum_{\sigma_{j}=\pm1}\delta(\sigma_{j},w_{j}s_{j})=1.
\end{equation}
It allows us to express the dynamics of the magnetization as 
\begin{eqnarray}
m^{l} & = & \sum_{\{s_{j}\}}\prod_{j=1}^{k}\left[\frac{1+s_{j}m^{l-1}}{2}\right]\sum_{\{w_{j}\}}P(\{w_{j}\})\sum_{\{\sigma_{j}\}}\prod_{j}\delta(\sigma_{j},w_{j}s_{j})\text{sgn}\biggr(\sum_{j=1}^{k}w_{j}s_{j}\biggr)\nonumber \\
 & = & \sum_{\{s_{j}\}}\prod_{j=1}^{k}\left[\frac{1+s_{j}m^{l-1}}{2}\right]\sum_{\{w_{j}\}}P(\{w_{j}\})\sum_{\{\sigma_{j}\}}\prod_{j}\int_{-\pi}^{\pi}\frac{d\theta_{j}}{2\pi}e^{\mathrm{i}\theta_{j}(\sigma_{j}-w_{j}s_{j})}\text{sgn}\biggr(\sum_{j=1}^{k}\sigma_{j}\biggr).
\end{eqnarray}
We then notice that 
\begin{align}
 & \sum_{\{w_{j}\}}P(\{w_{j}\})e^{-\mathrm{i}\sum_{j=1}^{k}\theta_{j}w_{j}s_{j}}\nonumber \\
= & \sum_{\{w_{j}\}}\prod_{j}\left\{ \left[p\delta_{w_{j},-1}+(1-p)\delta_{w_{j},1}\right]e^{-\mathrm{i}\theta_{j}w_{j}s_{j}}\right\} \nonumber \\
= & \prod_{j}\left[pe^{\mathrm{i}\theta_{j}s_{j}}+(1-p)e^{-\mathrm{i}\theta_{j}s_{j}}\right],
\end{align}
which leads to 
\begin{eqnarray}
m^{l} & = & \sum_{\{\sigma_{j}\}}\sum_{\{s_{j}\}}\prod_{j=1}^{k}\left\{ \left[\frac{1+s_{j}m^{l-1}}{2}\right]\int_{-\pi}^{\pi}\frac{d\theta_{j}}{2\pi}e^{\mathrm{i}\theta_{j}\sigma_{j}}\left[pe^{\mathrm{i}\theta_{j}s_{j}}+(1-p)e^{-\mathrm{i}\theta_{j}s_{j}}\right]\right\} \text{sgn}\biggr(\sum_{j=1}^{k}\sigma_{j}\biggr)\nonumber \\
 & = & \sum_{\{\sigma_{j}\}}\sum_{\{s_{j}\}}\prod_{j=1}^{k}\left\{ \left[\frac{1+s_{j}m^{l-1}}{2}\right]\left[p\delta(s_{j},-\sigma_{j})+(1-p)\delta(s_{j},\sigma_{j})\right]\right\} \text{sgn}\biggr(\sum_{j=1}^{k}\sigma_{j}\biggr)\nonumber \\
 & = & \sum_{\{s_{j}\}}\prod_{j=1}^{k}\left[\frac{1+s_{j}m^{l-1}(1-2p)}{2}\right]\text{sgn}\biggr(\sum_{j=1}^{k}s_{j}\biggr),\label{eq:m_dyn_1m2p}
\end{eqnarray}
where in the last step we trace over $\{s_{j}\}$ and replace the
dummy spin variables $\{\sigma_{j}\}$ by $\{s_{j}\}$.

Performing the same derivation for the dynamics of the overlap in
Eq.~(\ref{eq:q_dyn_pw}), we have 
\begin{eqnarray}
q^{l} & = & \sum_{\{\hat{s}_{j},s_{j}\}}\prod_{j=1}^{k}\left[\frac{1+\hat{s}_{j}\hat{m}^{l-1}+s_{j}m^{l-1}+\hat{s}_{j}s_{j}q^{l-1}}{4}\right]\text{sgn}\biggr(\sum_{j=1}^{k}\hat{s}_{j}\biggr)\sum_{\{w_{j}\}}P(w_{j})\sum_{\{\sigma_{j}\}}\prod_{j}\delta(\sigma_{j},w_{j}s_{j})\text{sgn}\biggr(\sum_{j=1}^{k}\sigma_{j}\biggr)\nonumber \\
 & = & \sum_{\{\hat{s}_{j},s_{j}\}}\prod_{j=1}^{k}\left[\frac{1+\hat{s}_{j}\hat{m}^{l-1}+s_{j}m^{l-1}+\hat{s}_{j}s_{j}q^{l-1}}{4}\right]\text{sgn}\biggr(\sum_{j=1}^{k}\hat{s}_{j}\biggr)\sum_{\{\sigma_{j}\}}\prod_{j}\left[p\delta(s_{j},-\sigma_{j})+(1-p)\delta(s_{j},\sigma_{j})\right]\text{sgn}\biggr(\sum_{j=1}^{k}\sigma_{j}\biggr)\nonumber \\
 & = & \sum_{\{\hat{s}_{j},s_{j}\}}\prod_{j=1}^{k}\left[\frac{1+\hat{s}_{j}\hat{m}^{l-1}+s_{j}m^{l-1}(1-2p)+\hat{s}_{j}s_{j}q^{l-1}(1-2p)}{4}\right]\text{sgn}\biggr(\sum_{j=1}^{k}\hat{s}_{j}\biggr)\text{sgn}\biggr(\sum_{j=1}^{k}s_{j}\biggr),\label{eq:q_dyn_1m2p}
\end{eqnarray}
where in the last step we trace again over $\{s_{j}\}$ and replace
the dummy spin variables $\{\sigma_{j}\}$ by $\{s_{j}\}$.

\subsubsection{Expressions for $k=3$}

For the particular case of $k=3$, we have equations of the form 
\begin{equation}
\hat{m}^{l}=\frac{1}{2}\left[3\hat{m}^{l-1}-(\hat{m}^{l-1})^{3}\right],
\end{equation}
\begin{equation}
m^{l}=\frac{1}{2}\left[3m^{l-1}(1-2p)-(m^{l-1})^{3}(1-2p)^{3}\right],\label{eq:m_dyn_k3}
\end{equation}
\begin{align}
q^{l} & =\frac{3}{2}\hat{m}^{l-1}m^{l-1}(1-2p)-\frac{3}{4}q^{l-1}(m^{l-1})^{2}(1-2p)^{3}-\frac{3}{4}q^{l-1}(\hat{m}^{l-1})^{2}(1-2p)\nonumber \\
 & \quad+\frac{3}{4}q^{l-1}(1-2p)+\frac{1}{4}(q^{l-1})^{3}(1-2p)^{3}.\label{eq:q_dyn_k3}
\end{align}

\begin{figure}
\includegraphics[scale=0.42]{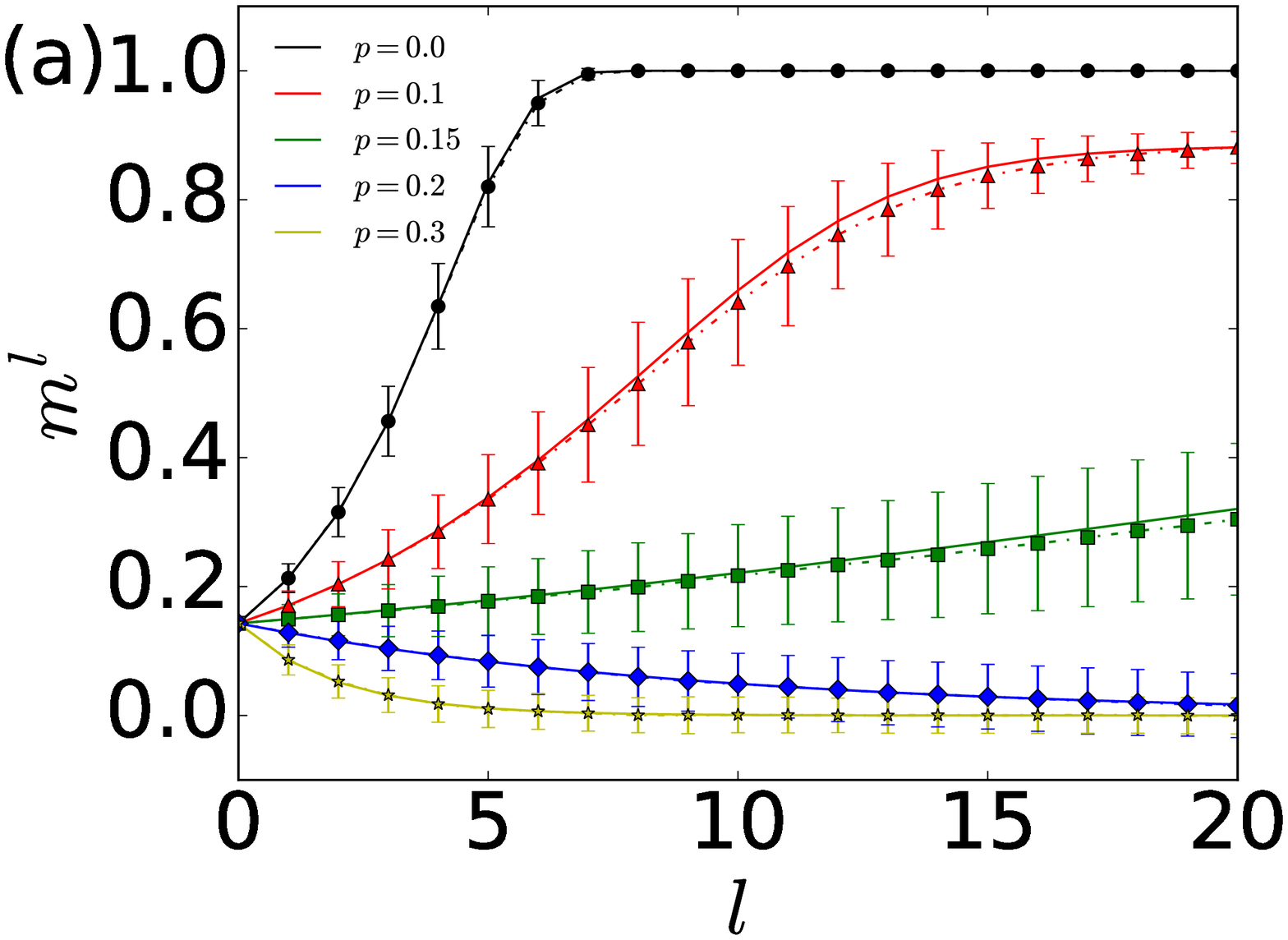}\includegraphics[scale=0.42]{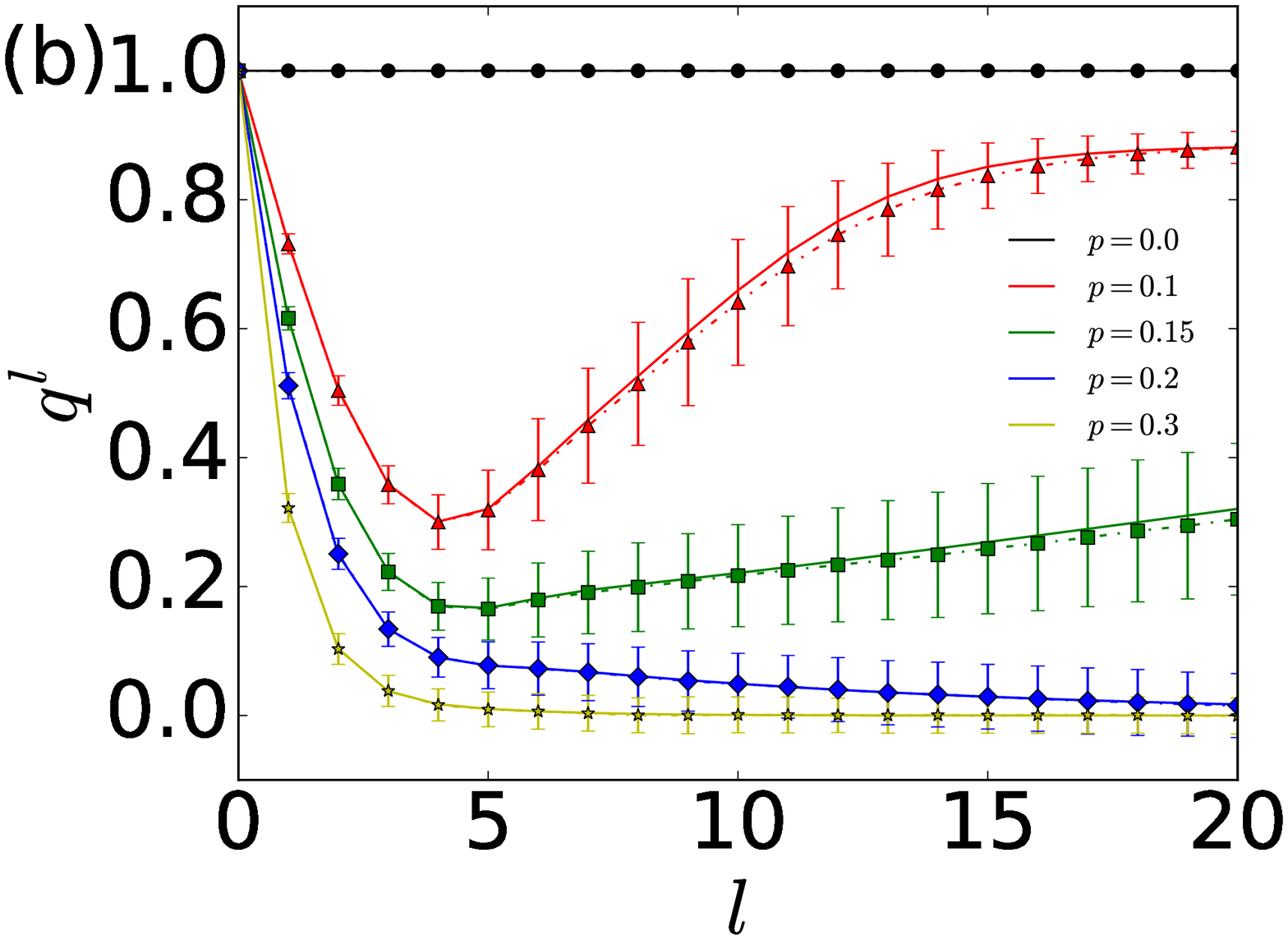}\caption{Comparing the mean field order parameter evolution in layers, in sparsely-connected networks, following
Eq.~(\ref{eq:m_dyn_k3}) in (a) and Eq.~(\ref{eq:q_dyn_k3}) in
(b) (solid lines) to Monte Carlo simulations (dashed-dotted lines
with markers). The results show a perfect match between analysis and
numerical results. The systems are of size $N=2000$ and the results
are averaged over $100$ disorder realizations. \label{fig:compare_to_MC_sparse}}
\end{figure}

The evolutions of $m^{l}$ and $q^{l}$ are validated by Monte Carlo
simulations as shown in Fig.~\ref{fig:compare_to_MC_sparse}.

In the deep network limit $l\to\infty$, $m^{\infty}=0$ is always
a solution of Eq.~(\ref{eq:m_dyn_k3}), which is stable for $p>p_{c}=1/6$.
For $p<1/6$, the solution $m^{\infty}=0$ becomes unstable and two
stable solutions emerge $m^{\infty}=\pm\sqrt{\frac{1-6p}{(1-2p)^{3}}}$.

Consider the initial condition $\hat{m}^{0}=m^{0}>0$ and $q^{0}=1$,
then the reference function admits a stationary solution $\hat{m}^{\infty}\!=\!1$,
which implies $\hat{s}_{i}^{\infty}\!=\!1$, $\forall i$. So we have
the relation $q^{\infty}\!=\!\frac{1}{N}\sum_{i}\hat{s}_{i}^{\infty}s_{i}^{\infty}\!=\!\frac{1}{N}\sum_{i}s_{i}^{\infty}\!=\!m^{\infty}$.

\subsubsection{Critical Perturbation $p_{c}$ for general $k$}

By examining the stability of the steady solution $m^{\infty}=0$
in Eq.~(\ref{eq:m_dyn_1m2p}), we can derive the the critical perturbation
$p_{c}$ for general $k$. We first notice that Eq.~(\ref{eq:m_dyn_1m2p})
can be expressed as 
\begin{equation}
m^{l}=\sum_{n=0}^{k}{k \choose n}\left[\frac{1+m^{l-1}(1-2p)}{2}\right]^{n}\left[\frac{1-m^{l-1}(1-2p)}{2}\right]^{k-n}\text{sgn}(2n-k).
\end{equation}
Perturbing $m^{l-1}$ and $m^{l}$ around $0$ gives

\begin{eqnarray}
\delta m^{l} & = & \sum_{n=0}^{k}{k \choose n}\left\{ (k-n)\left(\frac{1}{2}\right)^{k-1}\left(-\frac{1-2p}{2}\delta m^{l-1}\right)+n\left(\frac{1}{2}\right)^{k-1}\left(\frac{1-2p}{2}\delta m^{l-1}\right)\right\} \text{sgn}(2n-k)\nonumber \\
 &  & \quad+\mathcal{O}\left((\delta m^{l-1})^{2}\right)\\
 & = & \delta m^{l-1}(1-2p)\left(\frac{1}{2}\right)^{k}\sum_{n=0}^{k}{k \choose n}(2n-k)\text{sgn}(2n-k)+\mathcal{O}\left((\delta m^{l-1})^{2}\right).
\end{eqnarray}
By using the identity $\sum_{n=0}^{k}{k \choose n}(2n-k)\text{sgn}(2n-k)=2k{k-1 \choose (k-1)/2}$
for odd integer $k$, one obtains the instability condition 
\begin{equation}
\frac{\delta m^{l}}{\delta m^{l-1}}=(1-2p)\left(\frac{1}{2}\right)^{k-1}k{k-1 \choose (k-1)/2}>1,
\end{equation}
\begin{align}
p & <p_{c}(k):=\frac{1}{2}-2^{k-2}/\left[k{k-1 \choose (k-1)/2}\right].\label{eq:critical_flip_noise}
\end{align}
The obtained critical flip noise $p_{c}(k)$ in Eq.~(\ref{eq:critical_flip_noise})
has exactly the same form as the critical thermal noise $\epsilon_{c}(k)$
\cite{Mozeika2010_sm}. Therefore, they show the same locations for
the corresponding phase transitions.

\subsection{The Case of Continuous Spin Variables with ReLU Activation Functions\label{subsec:relu_activiation}}

In this section we consider the case of continuous spin variable $\hat{s}_{i}^{l},s_{i}^{l}\in\mathbb{R}$
and the rectified linear unit (ReLU) activation function 
\begin{equation}
\phi(x)=\max(0,x).
\end{equation}
We assume a fully connected network structure with the local field
on $\hat{s}_{i}^{l}$ given by $\hat{h}_{i}^{l}(\hat{\boldsymbol{w}}^{l},\hat{\boldsymbol{s}}^{l-1})=\frac{1}{\sqrt{N}}\sum_{j}\hat{w}_{ij}^{l}\hat{s}_{j}^{l-1}$
and the state of $\hat{s}_{i}^{l}$ determined by the conditional
probability 
\begin{equation}
P(\hat{s}_{i}^{l}|\hat{\boldsymbol{w}}^{l},\hat{\boldsymbol{s}}^{l-1})=\sqrt{\frac{\beta}{2\pi}}\exp\left\{ -\frac{\beta}{2}\left[\hat{s}_{i}^{l}-\phi\left(\hat{h}_{i}^{l}(\hat{\boldsymbol{w}}^{l},\hat{\boldsymbol{s}}^{l-1})\right)\right]^{2}\right\} ,
\end{equation}
where the temperature $\beta$ quantifies the strength of the additive
white noise. In the noiseless limit $\beta\to\infty$, the system
corresponds to a deterministic neural network with ReLU activation
function, which is the limit that we will consider in the end, to
reduce the number of free variables. The perturbed network operates
in the same manner. We further consider the Gaussian weight-disorder
and perturbation to be the same as in the case of Sec.~A1, namely
$\hat{w}_{ij}^{l}\sim\mathcal{N}(0,\sigma_{w}^{2})$; the perturbed
weight has the form $w_{ij}^{l}=\sqrt{1-(\eta^{l})^{2}}\hat{w}_{ij}^{l}+\eta^{l}\delta w_{ij}^{l}$
where $\delta w_{ij}^{l}\sim\mathcal{N}(0,\sigma_{w}^{2})$ but is
independent of $\hat{w}_{ij}^{l}$. The input is assumed to follow
the normal distribution 
\begin{equation}
P(\hat{\boldsymbol{s}}^{0})=\prod_{i}P(\hat{s}_{i}^{0})=\prod_{i}\frac{1}{\sqrt{2\pi}}\exp\left\{ -\frac{1}{2}\left(\hat{s}_{i}^{0}\right)^{2}\right\} .\label{eq:relu_gauss_initial}
\end{equation}

The generating functional analysis follows the same procedure as in
the binary spin variable case, except for the addition of two new
order parameters $q_{11}^{l}:=\frac{1}{N}\sum_{i}(\hat{s}_{i}^{l})^{2}$
and $q_{22}^{l}:=\frac{1}{N}\sum_{i}(s_{i}^{l})^{2}$ in addition
to the overlap $q^{l}=\frac{1}{N}\sum_{i}\hat{s}_{i}^{l}s_{i}^{l}$
to characterize the macroscopic dynamics. Essentially $q_{11}^{l}$
and $q_{22}^{l}$ measure the scales of the spin variables at layer
$l$. Expressing the order parameters through the integral representation
of the $\delta$-function 
\begin{align}
1 & =\int\frac{d\mathcal{Q}^{l}dq^{l}}{2\pi/N}e^{\mathrm{i}N\mathcal{Q}^{l}\left[q^{l}-\frac{1}{N}\sum_{i}s_{i}^{l}\hat{s}_{i}^{l}\right]},\quad1=\int\frac{d\mathcal{Q}_{11}^{l}dq_{11}^{l}}{2\pi/N}e^{\mathrm{i}N\mathcal{Q}_{11}^{l}\left[q_{11}^{l}-\frac{1}{N}\sum_{i}(\hat{s}_{i}^{l})^{2}\right]},\nonumber \\
 & \qquad\qquad\qquad1=\int\frac{d\mathcal{Q}_{22}^{l}dq_{22}^{l}}{2\pi/N}e^{\mathrm{i}N\mathcal{Q}_{22}^{l}\left[q_{22}^{l}-\frac{1}{N}\sum_{i}(s_{i}^{l})^{2}\right]},
\end{align}
the disorder-averaged generating functional $\overline{\Gamma}$ can
be factorized over sites (assuming site-independent conjugate field
$\hat{\psi}_{i}^{l}=\hat{\psi}^{l}$ and $\psi_{i}^{l}=\psi^{l}$)
and expressed as 
\begin{equation}
\overline{\Gamma}=\int\prod_{l=0}^{L}\frac{d\mathcal{Q}^{l}dq^{l}}{2\pi/N}\frac{d\mathcal{Q}_{11}^{l}dq_{11}^{l}}{2\pi/N}\frac{d\mathcal{Q}_{22}^{l}dq_{22}^{l}}{2\pi/N}e^{N\Psi[\boldsymbol{q},\boldsymbol{\mathcal{Q}},\boldsymbol{q}_{11},\boldsymbol{\mathcal{Q}}_{11},\boldsymbol{q}_{22},\boldsymbol{\mathcal{Q}}_{22}]},
\end{equation}
where the potential function $\Psi[...]$ in the exponent assumes
the form 
\begin{equation}
\Psi[\boldsymbol{q},\boldsymbol{\mathcal{Q}},\boldsymbol{q}_{11},\boldsymbol{\mathcal{Q}}_{11},\boldsymbol{q}_{22},\boldsymbol{\mathcal{Q}}_{22}]=\mathrm{i}\sum_{l=0}^{L}\left(\mathcal{Q}^{l}q^{l}+\mathcal{Q}_{11}^{l}q_{11}^{l}+\mathcal{Q}_{22}^{l}q_{22}^{l}\right)+\log\int\prod_{l=1}^{L}d\hat{h}^{l}dh^{l}\prod_{l=0}^{L}d\hat{s}^{l}ds^{l}M[\hat{s},s,\hat{h},h],
\end{equation}
with the effective single site measure $M[...]$ in the form of 
\begin{align}
M\left[\{\hat{s}^{l},s^{l},\hat{h}^{l},h^{l}\}\right] & =P(\hat{s}^{0})\delta(\hat{s}^{0}-s^{0})e^{-\mathrm{i}\sum_{l}\left[\mathcal{Q}^{l}\hat{s}^{l}s^{l}+\mathcal{Q}_{11}^{l}(\hat{s}^{l})^{2}+\mathcal{Q}_{22}^{l}(s^{l})^{2}\right]}e^{-\mathrm{i}\sum_{l}(\hat{\psi}^{l}\hat{s}^{l}+\psi^{l}s^{l})}\nonumber \\
 & \quad\times\prod_{l=1}^{L}\left\{ \frac{\exp\left[-\frac{\beta}{2}(\hat{s}^{l}-\phi(\hat{h}^{l}))^{2}-\frac{\beta}{2}(s^{l}-\phi(h^{l}))^{2}\right]}{2\pi/\beta}\frac{\exp\left[-\frac{1}{2}(H^{l})^{T}\cdot\Sigma_{l}(q^{l-1})^{-1}\cdot H^{l}\right]}{\sqrt{(2\pi)^{2}|\Sigma_{l}(q^{l-1})|}}\right\} ~.
\end{align}
The covariance matrix between the local field $\hat{h}^{l}$ and $h^{l}$
takes the form of 
\begin{equation}
\Sigma_{l}(\eta^{l},q^{l-1})=\sigma_{w}^{2}\begin{bmatrix}q_{11}^{l-1} & \sqrt{1-(\eta^{l})^{2}}q^{l-1}\\
\sqrt{1-(\eta^{l})^{2}}q^{l-1} & q_{22}^{l-1}
\end{bmatrix}.\label{eq:Sigma_relu_uncorrelated_w}
\end{equation}

In the limit $N\to\infty$, $\overline{\Gamma}$ is dominated by the
saddle point of $\Psi[\boldsymbol{q},\boldsymbol{\mathcal{Q}},\boldsymbol{q}_{11},\boldsymbol{\mathcal{Q}}_{11},\boldsymbol{q}_{22},\boldsymbol{\mathcal{Q}}_{22}]$.
Similar to Sec. A4, the conjugate order parameter can be shown to
vanish identically at the saddle point 
\begin{equation}
\mathcal{Q}^{l}=\mathcal{Q}_{11}^{l}=\mathcal{Q}_{22}^{l}=0,\quad\forall l.\label{eq:saddle_point_three_Q}
\end{equation}
It gives rise to evolution of the order parameters with vanishing
conjugate field $\{\hat{\psi}^{l},\psi^{l}\}$ 
\begin{align}
q_{11}^{l} & =\langle(\hat{s}^{l})^{2}\rangle_{M[...]}=\frac{1}{\beta}+\int d\hat{h}^{l}dh^{l}\left(\phi(\hat{h}^{l})\right)^{2}\frac{e^{-\frac{1}{2}(H^{l})^{T}\cdot\Sigma_{l}^{-1}\cdot H^{l}}}{\sqrt{(2\pi)^{2}|\Sigma_{l}|}},\\
q_{22}^{l} & =\langle(s^{l})^{2}\rangle_{M[...]}=\frac{1}{\beta}+\int d\hat{h}^{l}dh^{l}\left(\phi(h^{l})\right)^{2}\frac{e^{-\frac{1}{2}(H^{l})^{T}\cdot\Sigma_{l}^{-1}\cdot H^{l}}}{\sqrt{(2\pi)^{2}|\Sigma_{l}|}},\\
q^{l} & =\langle\hat{s}^{l}s^{l}\rangle_{M[...]}=\int d\hat{h}^{l}dh^{l}\phi(\hat{h}^{l})\phi(h^{l})\frac{e^{-\frac{1}{2}(H^{l})^{T}\cdot\Sigma_{l}^{-1}\cdot H^{l}}}{\sqrt{(2\pi)^{2}|\Sigma_{l}|}}.
\end{align}

Under the ReLU non-linearity $\phi(x)=\max(0,x)$, the above integrals
can be carried out analytically, yielding 
\begin{align}
q_{11}^{l} & =\frac{1}{2}\Sigma_{l,11}+\frac{1}{\beta},\qquad\qquad q_{22}^{l}=\frac{1}{2}\Sigma_{l,22}+\frac{1}{\beta},\label{eq:saddle_point_q11_q22}\\
q^{l} & =\frac{1}{2\pi}\left[\sqrt{|\Sigma_{l}|}+\frac{\pi}{2}\Sigma_{l,12}+\Sigma_{l,12}\tan^{-1}\left(\frac{\Sigma_{l,12}}{\sqrt{|\Sigma_{l}|}}\right)\right],\label{eq:saddle_point_q_relu}
\end{align}
with initial condition 
\begin{equation}
q_{11}^{0}=q_{22}^{0}=q^{0}=1.
\end{equation}
In the noiseless limit $\beta\to\infty,$ one can show that $q_{11}^{l}=q_{22}^{l}=1\,\,\forall l$
if and only if $\sigma_{w}=\sqrt{2}$. In contrast, $q_{11}^{l}$
and $q_{22}^{l}$ grow indefinitely as $l$ increases if $\sigma_{w}>\sqrt{2}$
or decay to zero as the $l$ increases if $\sigma_{w}<\sqrt{2}$.

Other macroscopic quantities such as the mean activations $m_{1}^{l}:=1/N\sum_{i}\langle\hat{s}_{i}^{l}\rangle$
and $m_{2}^{l}:=1/N\sum_{i}\langle s_{i}^{l}\rangle$ can be calculated
by resorting to the generating functional evaluated at the saddle
point following Eqs. (\ref{eq:saddle_point_three_Q}) (\ref{eq:saddle_point_q11_q22})
and (\ref{eq:saddle_point_q_relu})

\begin{align}
\overline{\Gamma[\{\hat{\psi}^{l},\psi^{l}\}]} & =\int\prod_{l=0}^{L}\frac{d\mathcal{Q}^{l}dq^{l}}{2\pi/N}\frac{d\mathcal{Q}_{11}^{l}dq_{11}^{l}}{2\pi/N}\frac{d\mathcal{Q}_{22}^{l}dq_{22}^{l}}{2\pi/N}e^{N\Psi[\boldsymbol{q},\boldsymbol{\mathcal{Q}},\boldsymbol{q}_{11},\boldsymbol{\mathcal{Q}}_{11},\boldsymbol{q}_{22},\boldsymbol{\mathcal{Q}}_{22}]}\nonumber \\
 & \approx\text{const}\times\exp\left\{ N\Psi_{\text{saddle}}[\boldsymbol{q},\boldsymbol{\mathcal{Q}},\boldsymbol{q}_{11},\boldsymbol{\mathcal{Q}}_{11},\boldsymbol{q}_{22},\boldsymbol{\mathcal{Q}}_{22}]\right\} ,
\end{align}
which can be differentiated with respect to $\hat{\psi}^{l}$ and
$\psi^{l}$ to obtain other moments.

The relevant similarity measure in this case is the correlation coefficient
\begin{equation}
\rho^{l}:=\frac{q^{l}-m_{1}^{l}m_{2}^{l}}{\sqrt{q_{11}^{l}-(m_{1}^{l})^{2}}\sqrt{q_{22}^{l}-(m_{2}^{l})^{2}}}.\label{eq:relu_rho_l}
\end{equation}
where mean activations are given by 
\begin{equation}
m_{1}^{l}=\sqrt{\frac{\Sigma_{l,11}}{2\pi}},\qquad m_{2}^{l}=\sqrt{\frac{\Sigma_{l,22}}{2\pi}},\qquad\forall l>0.
\end{equation}

\begin{figure}
\includegraphics[scale=0.42]{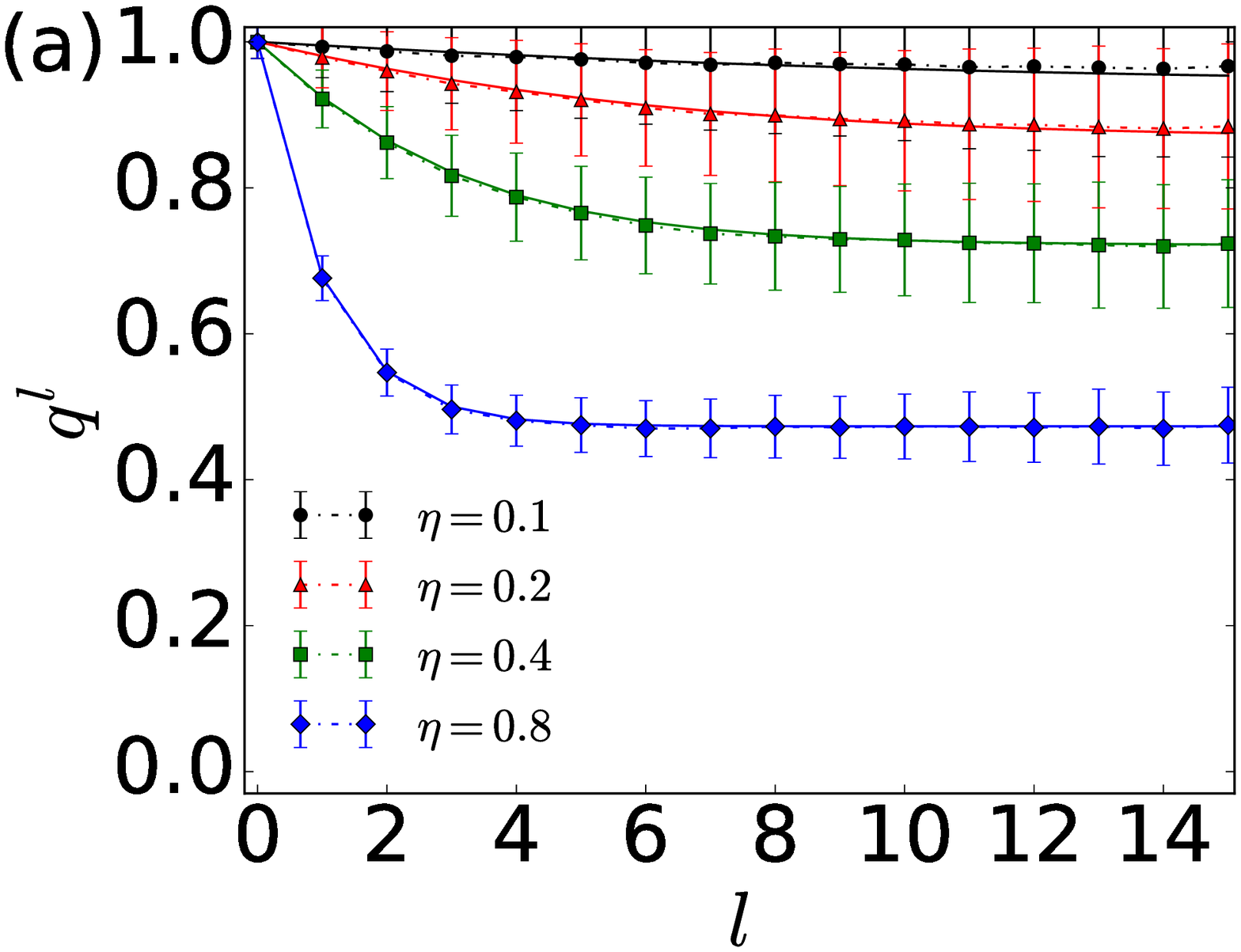}\includegraphics[scale=0.42]{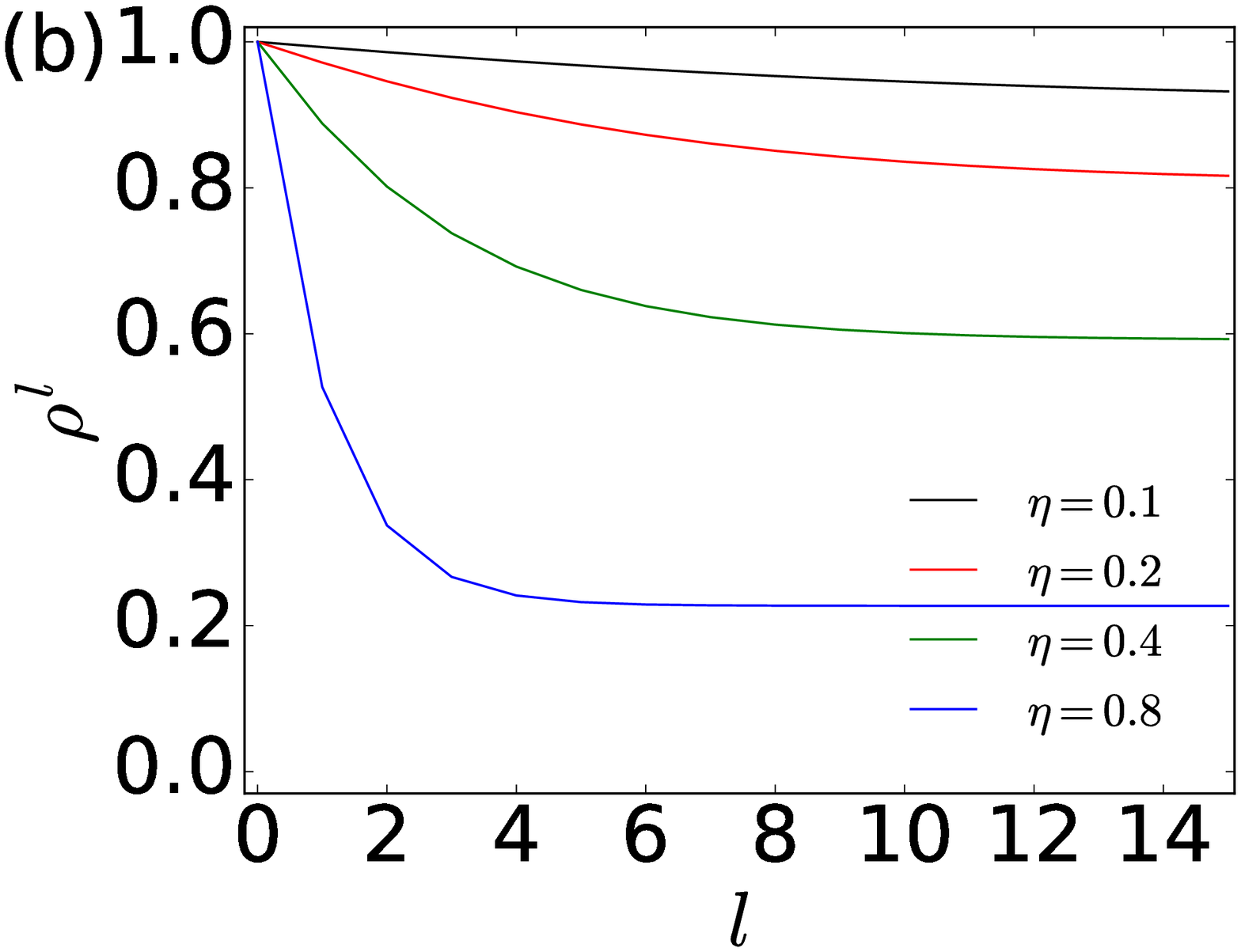}

\caption{(a) Comparison between analytical and simulation results of the overlap
evolution in layers, in networks with ReLU activation function at
$\sigma_{w}=\sqrt{2}$ and $\beta\to\infty$. Analytical results based
on Eq.~(\ref{eq:saddle_point_q_relu}) (solid lines) and Monte Carlo
simulations (dashed-dotted lines with markers) shows a perfect match.
The perturbations are uniform in all layers, i.e., $\eta^{l}=\eta$.
The systems are of size $N=4000$ and the results are averaged over
$100$ disorder realizations. (b) Evolution of the correlation coefficient
$\rho^{l}$ from Eq.~(\ref{eq:relu_rho_l}) using the same parameters
as in (a). \label{fig:relu_dynamics} }
\end{figure}

The mean field overlap evolution in layers of Eq.~(\ref{eq:saddle_point_q_relu})
at $\sigma_{w}=\sqrt{2}$ and $\beta\to\infty$ is compared to Monte
Carlo simulations results in Fig.~\ref{fig:relu_dynamics}(a). Differently
from results obtained for networks with sign activation function,
the two systems exhibit non-zero overlap in the deep network regime;
similar behavior is also found in the evolution of correlation coefficient
$\rho^{l}$ as shown in Fig.~\ref{fig:relu_dynamics}(b). That
is, there exist residual correlations in deep ReLU networks even
under very large weight perturbations, indicating that a large number
of networks in the weight space correspond to similar functions. It
implies that deep ReLU networks with uncorrelated Gaussian weight-disorder
are typically not very expressive, but should be easier to train given the corresponding function landscape. Another approach to understand
such behavior is through mapping DLM in the thermodynamic
limit $N\to\infty$ to Gaussian process and analyzing the resulting
kernel~\cite{Lee2018_sm}. Note also that deep ReLU networks used
in practice have finite depth, finite width and the weight distribution
may differ from uncorrelated Gaussian distribution that is examined
here.

\subsection{ReLU Activation Functions with Correlated Weight Variables\label{subsec:relu_correlated_w}}

In order to investigate the effect of weight correlation on
deep ReLU networks, we consider fully-connected networks with reference function weights
drawn from the distribution 
\begin{align}
P(\{\hat{w}_{ij}^{l}\}) & =\prod_{i}P(\hat{\boldsymbol{w}}_{i}^{l})=\prod_{i}\frac{\exp\left\{ -\frac{1}{2}(\hat{\boldsymbol{w}}_{i}^{l})^{T}\cdot A^{-1}\cdot(\hat{\boldsymbol{w}}_{i}^{l})\right\} }{\sqrt{(2\pi)^{N}\det A}},\label{eq:correlated_weight_distribution}
\end{align}
which accommodates possible correlation between weights of each sub-perceptron.

The previous section represents a special case where the precision matrix $A^{-1}$ is taken to
be $I/\sigma_{w}^{2}$ and $I$ is the identity matrix, such that
the no correlation between components of $\hat{\boldsymbol{w}}_{i}^{l}$
exist. Here we consider one simple extension, of correlation matrices that assume the form $A^{-1}=(I+\frac{b}{N}\boldsymbol{v}\boldsymbol{v}^{T})/\sigma_{w}^{2}$,
where $\boldsymbol{v}$ is a rank-one vector and $b\sim O(1)$ is
a scalar. For simplicity, we further consider $\boldsymbol{v}=[1,1,...1]^{T}$
such that $\boldsymbol{v}\boldsymbol{v}^{T}=J$ is the all-one matrix.
Note that 
\begin{equation}
A=\sigma_{w}^{2}\left(I-\frac{b}{(1+b)N}J\right),\label{eq:covariance_matrix_A}
\end{equation}
thus the components of $\hat{\boldsymbol{w}}_{i}^{l}$ are weakly
correlated, depending on the value of $b$. For a valid model, the precision matrix is required to
be semi-positive definite $A^{-1}\succeq0$, imposing the constraint
$b\geq-1$. Note that the components of $\hat{\boldsymbol{w}}_{i}^{l}$
are positively correlated for $-1<b<0$, while they are negatively
correlated for $b>0$.
As before, we assume independent and identical distribution of perturbation $w_{ij}^{l}=\sqrt{1-(\eta^{l})^{2}}\hat{w}_{ij}^{l}+\eta^{l}\delta w_{ij}^{l}$
with $\delta w_{ij}^{l}\sim\mathcal{N}(0,\sigma_{w}^{2})$. The derivation
of the GF follows the same lines as in the uncorrelated case, where the average over the weight-disorder
start to diverge, e.g., from the first line of Eq.~(\ref{eq:x_density_from_int})

\begin{align}
 & \int P(\hat{\boldsymbol{w}})P(\delta\boldsymbol{w})d\hat{\boldsymbol{w}}d\delta\hat{\boldsymbol{w}}\exp\Biggr\{\frac{-\mathrm{i}}{\sqrt{N}}\sum_{l=1}^{L}\Biggr[\sum_{ij}\hat{w}_{ij}^{l}\left(\hat{x}_{i}^{l}\hat{s}_{j}^{l-1}+\sqrt{1-(\eta^{l})^{2}}x_{i}^{l}\hat{s}_{j}^{l-1}\right)+\eta^{l}\sum_{ij}\delta w_{ij}^{l}x_{i}^{l}s_{j}^{l-1}\Biggr]\Biggr\}\nonumber \\
&= \exp\Biggr\{-\frac{\sigma_{w}^{2}}{2}\sum_{l=1}^{L}\sum_{i}\Biggr[(\hat{x}_{i}^{l})^{2}\Biggr(\frac{1}{N}\sum_{j}(\hat{s}_{j}^{l-1})^{2}-\frac{b}{1+b}\left(\frac{1}{N}\sum_{j}\hat{s}_{j}^{l-1}\right)^{2}\Biggr)\nonumber \\
&+ 2\sqrt{1-(\eta^{l})^{2}}\hat{x}_{i}^{l}x_{i}^{l}\Biggr(\frac{1}{N}\sum_{j}\hat{s}_{j}^{l-1}s_{j}^{l-1}-\frac{b}{1+b}\left(\frac{1}{N}\sum_{j}\hat{s}_{j}^{l-1}\right)^{2}\left(\frac{1}{N}\sum_{j}s_{j}^{l-1}\right)^{2}\Biggr)\nonumber \\
&+
(x_{i}^{l})^{2}\Biggr(\frac{1}{N}\sum_{j}(s_{j}^{l-1})^{2}-(1-(\eta^{l})^{2})\left(\frac{b}{1+b}\right)\left(\frac{1}{N}\sum_{j}s_{j}^{l-1}\right)^{2}\Biggr)\Biggr]\Biggr\}.
\end{align}

Defining the order parameters $q_{11}^{l},q_{22}^{l},m_{1}^{l},m_{2}^{l},q^{l}$
by

\begin{align}
1 & =\int\frac{d\mathcal{Q}_{11}^{l}dq_{11}^{l}}{2\pi/N}e^{\mathrm{i}N\mathcal{Q}_{11}^{l}\left[q_{11}^{l}-\frac{1}{N}\sum_{j}(\hat{s}_{j}^{l})^{2}\right]},\quad1=\int\frac{d\mathcal{Q}_{22}^{l}dq_{22}^{l}}{2\pi/N}e^{\mathrm{i}N\mathcal{Q}_{22}^{l}\left[q_{22}^{l}-\frac{1}{N}\sum_{j}(s_{j}^{l})^{2}\right]},\nonumber \\
1 & =\int\frac{d\mathcal{M}_{1}^{l}dm_{1}^{l}}{2\pi/N}e^{\mathrm{i}N\mathcal{M}_{1}^{l}\left[m_{1}^{l}-\frac{1}{N}\sum_{j}\hat{s}_{j}^{l}\right]},\qquad1=\int\frac{d\mathcal{M}_{2}^{l}dm_{2}^{l}}{2\pi/N}e^{\mathrm{i}N\mathcal{M}_{2}^{l}\left[m_{2}^{l}-\frac{1}{N}\sum_{j}s_{j}^{l}\right]},\nonumber \\
 & \qquad\qquad\qquad1=\int\frac{d\mathcal{Q}^{l}dq^{l}}{2\pi/N}e^{\mathrm{i}N\mathcal{Q}^{l}\left[q^{l}-\frac{1}{N}\sum_{i}\hat{s}_{i}^{l}s_{i}^{l}\right]},
\end{align}
the disorder-averaged generating functional $\overline{\Gamma}$ can
be factorized over sites and eventually leads to the effective single
site measure $M[...]$ of the form  
\begin{align}
M\left[\{\hat{s}^{l},s^{l},\hat{h}^{l},h^{l}\}\right] & =P(\hat{s}^{0})\delta(\hat{s}^{0}-s^{0})e^{-i\sum_{l}\left[\mathcal{Q}_{11}^{l}(\hat{s}^{l})^{2}+\mathcal{Q}_{22}^{l}(s^{l})^{2}+\mathcal{Q}^{l}\hat{s}^{l}s^{l}+\mathcal{M}_{1}^{l}\hat{s}^{l}+\mathcal{M}_{2}^{l}s^{l}\right]}e^{-\mathrm{i}\sum_{l}(\hat{\psi}^{l}\hat{s}^{l}+\psi^{l}s^{l})}\nonumber \\
 & \quad\times\prod_{l=1}^{L}\left\{ \frac{\exp\left[-\frac{\beta}{2}(\hat{s}^{l}-\phi(\hat{h}^{l}))^{2}-\frac{\beta}{2}(s^{l}-\phi(h^{l}))^{2}\right]}{2\pi/\beta}\frac{\exp\left[-\frac{1}{2}(H^{l})^{T}\cdot\Sigma_{l}(\eta^{l},q^{l-1},m^{l-1})^{-1}\cdot H^{l}\right]}{\sqrt{(2\pi)^{2}|\Sigma_{l}(\eta^{l},q^{l-1},m^{l-1})|}}\right\} ,
\end{align}
where the covariance matrix between the local field $\hat{h}^{l}$
and $h^{l}$ takes the form of 
\begin{equation}
\Sigma_{l}(\eta^{l},q^{l-1},m^{l-1})=\sigma_{w}^{2}\begin{bmatrix}q_{11}^{l-1}-\frac{b}{1+b}(m_{1}^{l-1})^{2} & \sqrt{1-(\eta^{l})^{2}}\left(q^{l-1}-\frac{b}{1+b}m_{1}^{l-1}m_{2}^{l-1}\right)\\
\sqrt{1-(\eta^{l})^{2}}\left(q^{l-1}-\frac{b}{1+b}m_{1}^{l-1}m_{2}^{l-1}\right) & q_{22}^{l-1}-\frac{b}{1+b}(1-(\eta^{l})^{2})(m_{2}^{l-1})^{2}
\end{bmatrix}.\label{eq:Sigma_relu_correlated_w}
\end{equation}

Consider the ReLU activation function $\phi(h)=\max(0,h)$ and the deterministic
limit $\beta\to\infty$. In the limit $N\to\infty$, $\overline{\Gamma}$
is dominated by the saddle point of the potential function $\Psi[...]$,
giving rise to the evolution of order parameters with vanishing conjugate
field $\{\hat{\psi}^{l},\psi^{l}\}$,

\begin{align}
q_{11}^{l} & =\frac{1}{2}\Sigma_{l,11},\quad q_{22}^{l}=\frac{1}{2}\Sigma_{l,22},\quad m_{1}^{l}=\sqrt{\frac{\Sigma_{l,11}}{2\pi}},\quad m_{2}^{l}=\sqrt{\frac{\Sigma_{l,22}}{2\pi},}\\
 & \qquad q^{l}=\frac{1}{2\pi}\left[\sqrt{|\Sigma_{l}|}+\frac{\pi}{2}\Sigma_{l,12}+\Sigma_{l,12}\tan^{-1}\left(\frac{\Sigma_{l,12}}{\sqrt{|\Sigma_{l}|}}\right)\right],\label{eq:saddle_point_relu_correlated_w}
\end{align}
which is similar to the evolution of systems with uncorrelated weights
in Sec.~\ref{subsec:relu_activiation}. However, in Sec.~\ref{subsec:relu_activiation}
the mean activations $m_{1}^{l}$ and $m_{2}^{l}$ do not directly
determine the layer evolution (as seen in Eq.~(\ref{eq:Sigma_relu_uncorrelated_w})),
unlike the results in this section (as seen in Eq.~(\ref{eq:Sigma_relu_correlated_w})).
The relevant similarity measure in this case is again the correlation
coefficient 
\begin{equation}
\rho^{l}:=\frac{q^{l}-m_{1}^{l}m_{2}^{l}}{\sqrt{q_{11}^{l}-(m_{1}^{l})^{2}}\sqrt{q_{22}^{l}-(m_{2}^{l})^{2}}}.
\end{equation}

\begin{figure}
\includegraphics[scale=0.5]{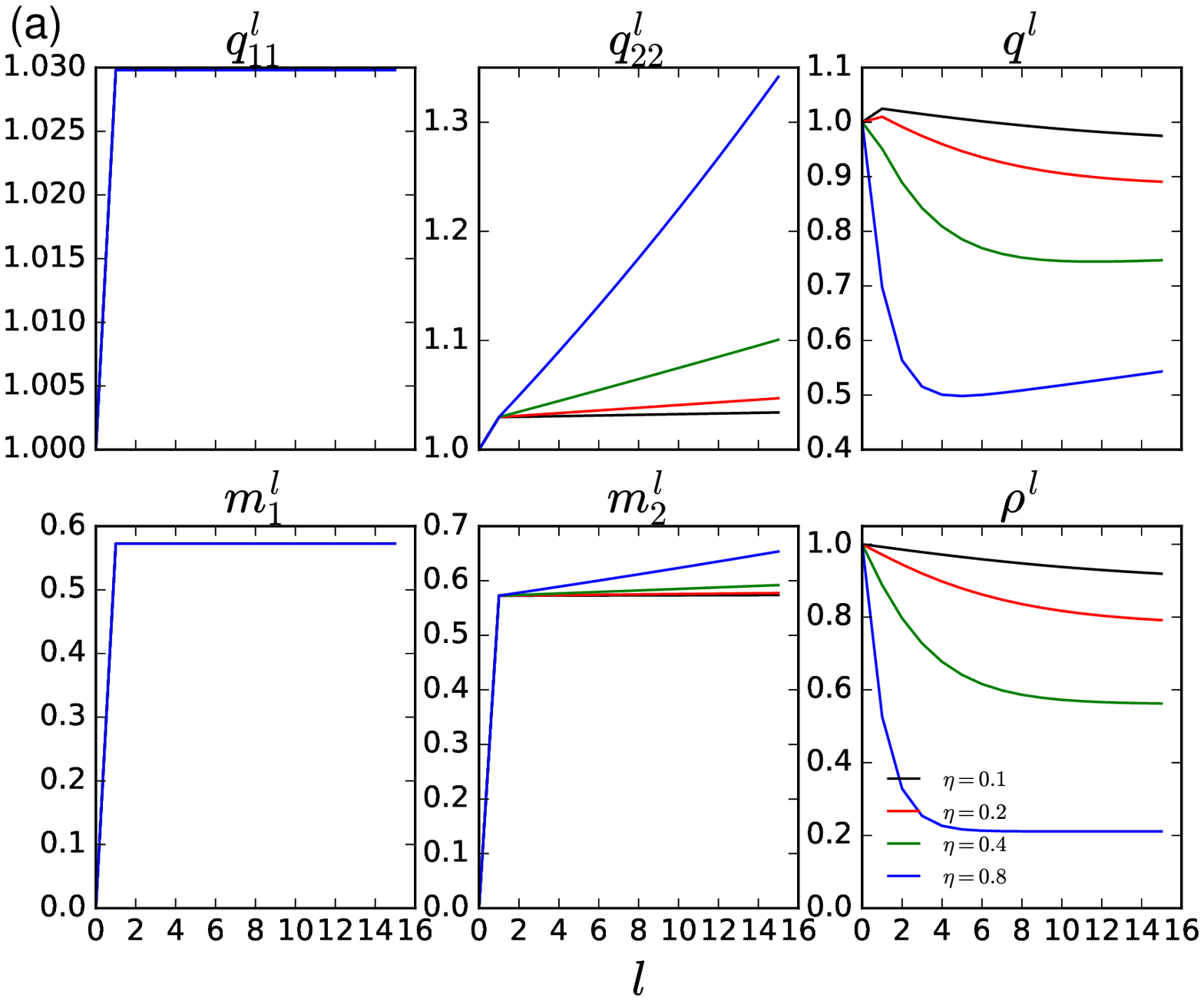}\includegraphics[scale=0.5]{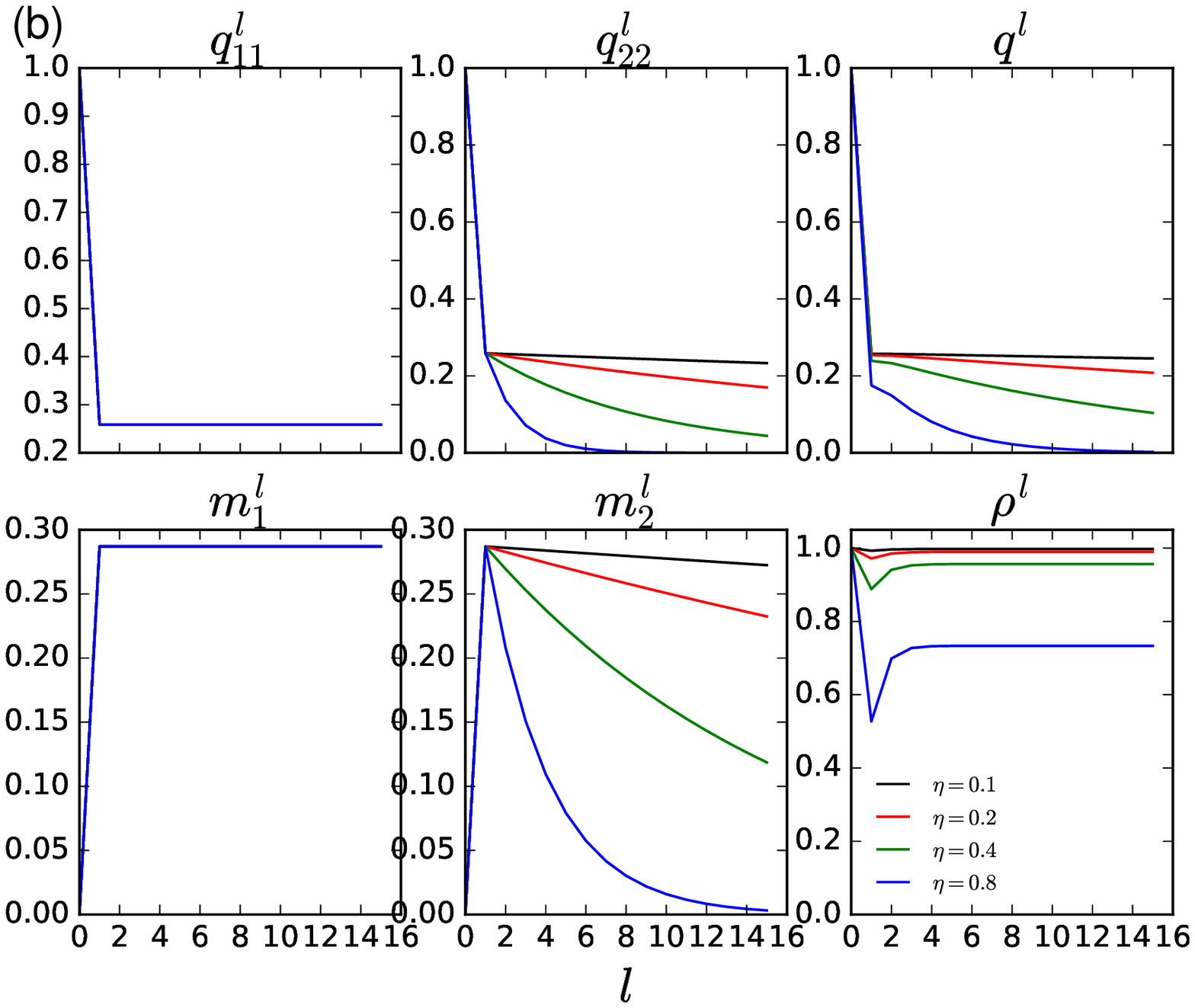}

\caption{Evolution of all the order parameters $q_{11}^{l},q_{22}^{l},q^{l},m_{1}^{l},m_{2}^{l}$
and the correlation coefficient $\rho^{l}$ in layers, in DLM
with ReLU activation function with correlated weight distribution
given by Eq.~(\ref{eq:correlated_weight_distribution}) and weight
covariance given by Eq.~(\ref{eq:covariance_matrix_A}). The parameter
$\sigma_{w}$ is chosen such that $q_{11}^{l}$ and $m_{1}^{l}$ can
attain stationary points, i.e., $\sigma_{w}=1/\sqrt{\frac{1}{2}-\frac{1}{2\pi}\frac{b}{1+b}}$.
(a) For negatively correlated weights $b=0.1$. (b) For positively correlated weights $b=-0.9$.\label{fig:relu_all_order_parameters_correlated_w}}
\end{figure}

Fig.~\ref{fig:relu_all_order_parameters_correlated_w} shows the evolution
of all the order parameters $q_{11}^{l},q_{22}^{l},q^{l},m_{1}^{l},m_{2}^{l}$
and the correlation coefficient $\rho^{l}$ for two different level
of weight correlation $b=0.1$ and $b=-0.9$. The initial input $\hat{\boldsymbol{s}}^{0}$
follows the distribution Eq.~(\ref{eq:relu_gauss_initial}), such
that 
\begin{equation}
q_{11}^{0}=q_{22}^{0}=q^{0}=1,\qquad m_{1}^{0}=m_{2}^{0}=0.
\end{equation}
It is observed that though $q_{11}^{l}$ and $m_{1}^{l}$ can
remain stationary for the particular choice of the weight standard deviation $\sigma_{w}=1/\sqrt{\frac{1}{2}-\frac{1}{2\pi}\frac{b}{1+b}}$,
$q_{22}^{l}$ and $m_{2}^{l}$ vary with the number of layers $l$. For negatively correlated
$\hat{\boldsymbol{w}}_{i}^{l}$ with $b=0.1>0$, $q_{22}^{l}$ and
$m_{2}^{l}$ increase with $l$, while the correlation coefficient
$\rho^{l}$ decreases with $l$. For positively correlated $\hat{\boldsymbol{w}}_{i}^{l}$
with $b=-0.9<0$, $q_{22}^{l}$ and $m_{2}^{l}$ decrease with $l$,
while interestingly, $\rho^{l}$  increases after a drop in the
first layer, suppressing the effect of weight perturbations in  subsequent
layers. 

\begin{figure}
\includegraphics[scale=0.5]{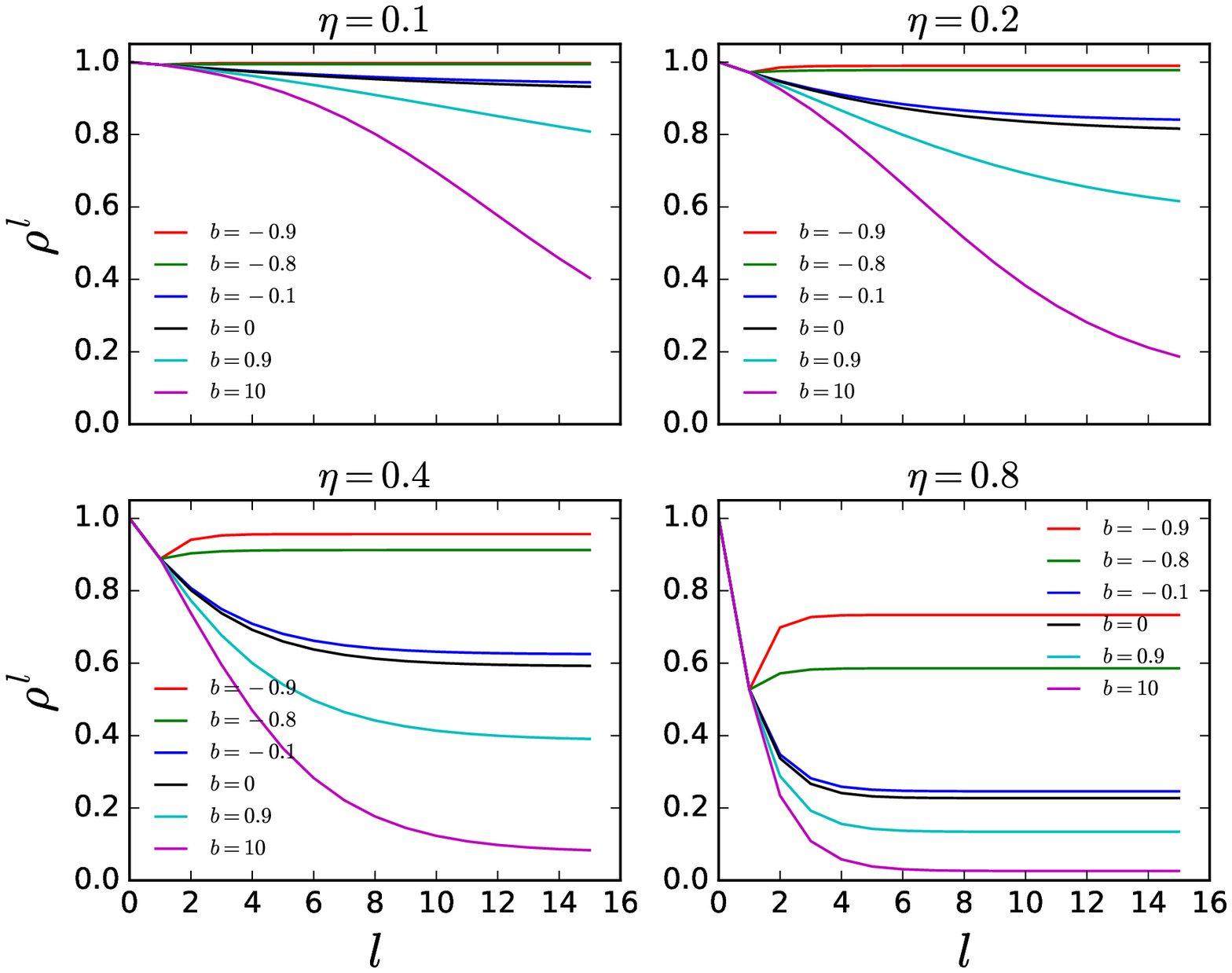}\caption{Evolution of the correlation coefficient $\rho^{l}$ in layers, in
networks with ReLU activation function and correlated weight distribution
(\ref{eq:correlated_weight_distribution}) for different
strengths of weight perturbation $\eta$ and weight correlation $b$.\label{fig:relu_rho_vary_b}}
\end{figure}

In Fig.~\ref{fig:relu_rho_vary_b}, we further investigate
the effect of weight correlation on $\rho^{l}$. Since $m_{1}^{0}=m_{2}^{0}=0$,
the evolutions from $\rho^{0}$ to $\rho^{1}$ are the same for different
values of $b$ which can be seen from Eq.~(\ref{eq:Sigma_relu_correlated_w}).
It is observed that $\rho^{l}$ attains a lower value for larger $b$,
indicating that functions with more negatively correlated weight
variables are more sensitive to weight perturbations. Sweeping the
parameter space through varying $\eta$, the perturbed systems with
larger $b$ (negatively correlated) explore further away the function space compared to the ones with smaller $b$ values (positively correlated), which indicates a potential effect of negative weight correlations on the expressive power of DLM with ReLU activation function~\cite{Poole2016_sm}. 
On the other hand, DLM with positively correlated weights show robustness to perturbation and hence contribute towards broader regions in function space and a more effective training.  Interestingly, it is observed in a recent study that negative weight correlation emerges in networks with ReLU activation function after training~\cite{Shang2016_sm}.

\subsection{Convolutional Neural Network}

\subsubsection{The Model}

\begin{figure}
\includegraphics[scale=0.6]{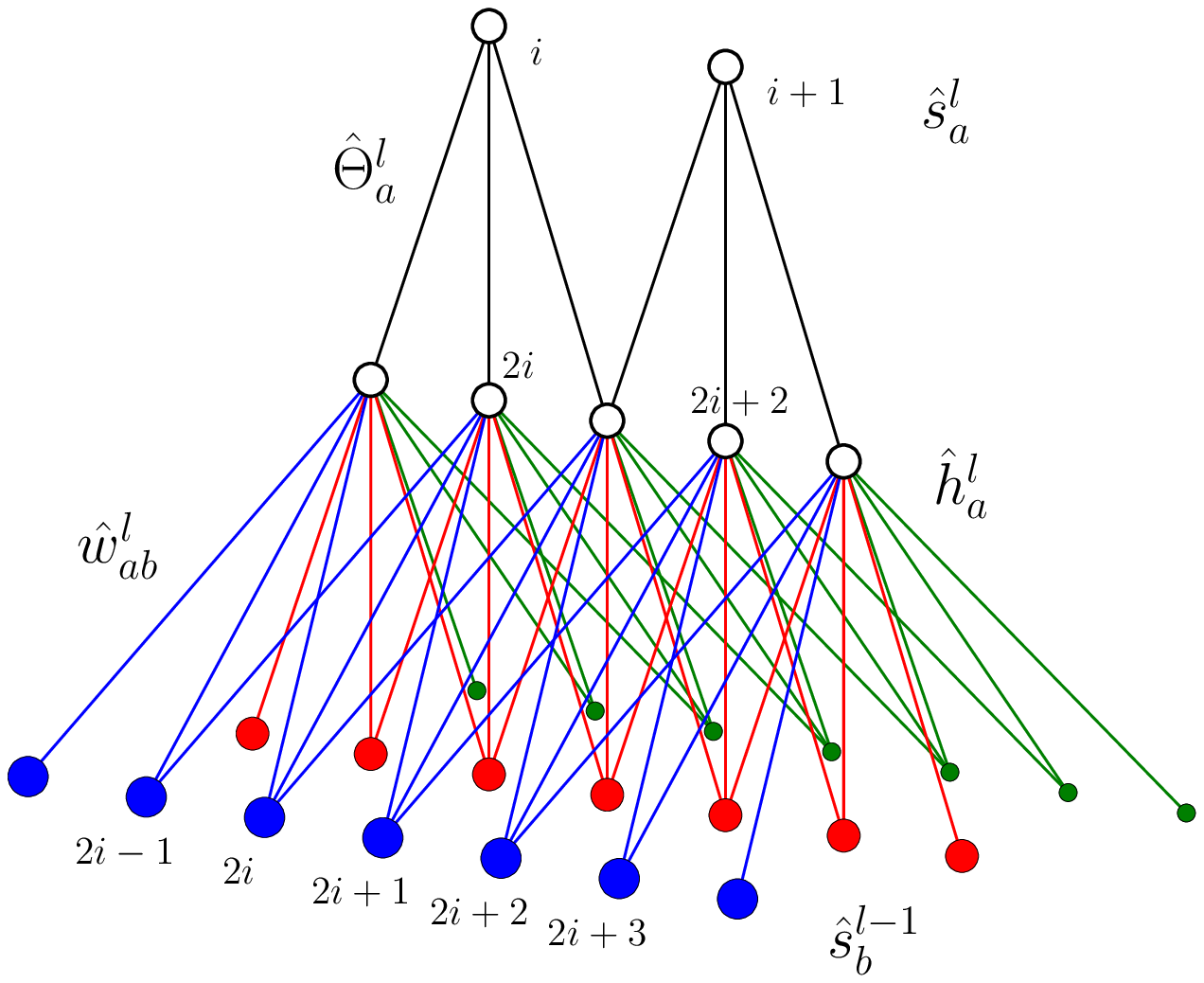}

\caption{A convolution block composed a normal convolution layer and a stride-two
convolution layer followed by element-wise non-linearity. Different
colors are used to denote different channels. The weight variables
within the same channel are shared such that they are translation
invariant and there are only $|\Delta|$ degrees of freedom for each
filter $\hat{w}_{ab}^{l}$ and $\hat{\Theta}_{a}^{l}$. \label{fig:conv_block}}
\end{figure}

In this section, we consider the convolutional neural networks (CNN)
stacked by multiple convolution blocks, each of which typically contains
a convolution layer, a pooling layer and a non-linearity layer. We
consider 1D sequential data and omit the bias variables for simplicity.
Let $a,b,c=1,...,N_{c}$ denote the channel indices and $i,j=1,...,N_{l}$
denote the pixel/site indices, where $N_{c}$ is the number of channels
(assumed to be the same for all layers including the input layer)
and $N_{l}$ is the number of sites of each channel on the $l$-th
block. To ease the computation, we replace the popular max-pooling
layer with a stride-two convolution as suggested in Ref. \cite{Springenberg2015_sm},
which also reduces the number of sites of each channel as max-pooling
(e.g., $N_{l}=N_{l-1}/2$ with appropriate padding). Each convolution
block has the following structure (see also Fig.~\ref{fig:conv_block}
for illustration),

\begin{align}
\hat{h}_{a}^{l}(j) &=\frac{1}{\sqrt{N_{c}}}\sum_{b}\left(\hat{w}_{ab}^{l}*\hat{s}_{b}^{l-1}\right)(j)  :=\frac{1}{\sqrt{N_{c}}}\sum_{b=1}^{N_{c}}\sum_{k\in\Delta}\hat{w}_{ab}^{l}(k)\hat{s}_{b}^{l-1}(j-k),\qquad j=1,...,N_{l-1},\nonumber \\
\hat{\tau}_{a}^{l}(i) & =\sum_{k\in\Delta}\hat{\Theta}_{a}^{l}(k)\hat{h}_{a}^{l}(2i-k),\qquad i=1,...,\frac{N_{l-1}}{2},\\
\hat{s}_{a}^{l}(i) & =\phi\left(\hat{\tau}_{a}^{l}(i)\right),\qquad i=1,...,\frac{N_{l-1}}{2},\nonumber 
\end{align}
where $*$ denotes the normal convolution operation and $\Delta$
is the set of possible indices for the filter (in the following we
consider $\Delta=\{-1,0,1\}$ for a kernel of size 3). The variables
$\{\hat{\Theta}_{a}^{l}(k)\}$ are the weights of stride-two filters.
The representation of a channel $s_{a}^{l}$ is also called a feature
map.

Different channels are coupled in a fully-connected manner, where
it is tempted to derive the mean field relation in the feature map
level in the limit $N_{c}\to\infty$. We assume that weight variables
of filters $\hat{w}_{ab}^{l}(k)$ are independent and follow a Gaussian
density $\hat{w}_{ab}^{l}(k)\sim\mathcal{N}(0,\sigma_{w}^{2})$, and
that the perturbed weight has the form $w_{ab}^{l}(k)=\sqrt{1-(\eta_{w}^{l})^{2}}\hat{w}_{ab}^{l}(k)+\eta_{w}^{l}\delta w_{ab}^{l}(k)$
where $\delta w_{ab}^{l}(k)$ are drawn from the Gaussian density $\mathcal{N}(0,\sigma_{w}^{2})$
independently of $\hat{w}_{ab}^{l}(k)$. We assume a similar form for the density
and perturbations for the stride-two filters, $\Theta_{a}^{l}(k)=\sqrt{1-(\eta_{\theta}^{l})^{2}}\hat{\Theta}_{a}^{l}(k)+\eta_{\theta}^{l}\delta\Theta_{a}^{l}(k)$,
which both $\Theta_{a}^{l}(k)$ and $\delta\Theta_{a}^{l}(k)$ follow
the Gaussian density $\mathcal{N}(0,\sigma_{\theta}^{2})$. In the
following, we focus on ReLU and sign activation
functions.

\subsubsection{GF Analysis}

The GF analysis for CNN follows a similar derivation as before.
Introducing auxiliary fields through the integral representation of
$\delta$-function

\begin{align}
1 & =\int_{-\infty}^{\infty}\frac{d\hat{h}_{a}^{l}(i)d\hat{x}_{a}^{l}(i)}{2\pi}e^{\mathrm{i}\hat{x}_{a}^{l}(i)\left(\hat{h}_{a}^{l}(i)-\frac{1}{\sqrt{N_{c}}}\sum_{b}\sum_{k\in\Delta}\hat{w}_{ab}^{l}(k)\hat{s}_{b}^{l-1}(i-k)\right)},\nonumber \\
1 & =\int_{-\infty}^{\infty}\frac{dh_{a}^{l}(i)dx_{a}^{l}(i)}{2\pi}e^{\mathrm{i}x_{a}^{l}(i)\left(h_{a}^{l}(i)-\frac{1}{\sqrt{N_{c}}}\sum_{b}\sum_{k\in\Delta}w_{ab}^{l}(k)s_{b}^{l-1}(i-k)\right)},\\
1 & =\int_{-\infty}^{\infty}\frac{d\hat{\tau}_{a}^{l}(i)d\hat{y}_{a}^{l}(i)}{2\pi}e^{\mathrm{i}\hat{y}_{a}^{l}(i)\left(\hat{\tau}_{a}^{l}(i)-\sum_{k\in\Delta}\hat{\Theta}_{a}^{l}(k)\hat{h}_{a}^{l}(2i-k)\right)},\quad1=\int_{-\infty}^{\infty}\frac{d\tau_{a}^{l}(i)dy_{a}^{l}(i)}{2\pi}e^{\mathrm{i}y_{a}^{l}(i)\left(\tau_{a}^{l}(i)-\sum_{k\in\Delta}\hat{\Theta}_{a}^{l}(k)h_{a}^{l}(2i-k)\right),}\nonumber 
\end{align}

we can express the weight variables linearly in the exponential of
the GF; averaging the term involving $\{\hat{w}_{ab}^{l}(k)\}$, $\{\delta w_{ab}^{l}(k)\}$
and $\{\hat{\Theta}_{a}^{l}(k)\}$ requires to compute the integrals
\small{
		\begin{align}
I_{w} & :=\int P(\hat{\boldsymbol{w}}^{l})P(\delta\boldsymbol{w}^{l})d\hat{\boldsymbol{w}}^{l}d\delta\boldsymbol{w}^{l}\exp\biggr\{\frac{-\mathrm{i}}{\sqrt{N_{c}}}\sum_{a,b=1}^{N_{c}}\sum_{k\in\Delta}\nonumber \\
 & \phantom{=} \biggr[\hat{w}_{ab}^{l}(k)\sum_{i=1}^{N_{l-1}}\left(\hat{x}_{a}^{l}(i)\hat{s}_{b}^{l-1}(i-k)+\sqrt{1-(\eta_{w}^{l})^{2}}x_{a}^{l}(i)\hat{s}_{b}^{l-1}(i-k)\right)+\eta_{w}^{l}\delta w_{ab}^{l}(k)\sum_{i=1}^{N_{l-1}}x_{a}^{l}(i)s_{b}^{l-1}(i-k)\biggr]\biggr\}\nonumber \\
& =  \exp\biggr\{\frac{-\sigma_{w}^{2}}{2}\sum_{i,j=1}^{N_{l-1}}\sum_{a=1}^{N_{c}}\biggr[\hat{x}_{a}^{l}(i)\hat{x}_{a}^{l}(j)\biggr(\frac{1}{N_{c}}\sum_{k\in\Delta}\sum_{b=1}^{N_{c}}\hat{s}_{b}^{l-1}(i\!-\!k)\hat{s}_{b}^{l-1}(j\!-\!k)\biggr)\!+\!x_{a}^{l}(i)x_{a}^{l}(j)\biggr(\frac{1}{N_{c}}\sum_{k\in\Delta}\sum_{b=1}^{N_{c}}s_{b}^{l-1}(i\!-\!k)s_{b}^{l-1}(j\!-\!k)\biggr)\nonumber \\
 & +2\sqrt{1-(\eta_{w}^{l})^{2}}\hat{x}_{a}^{l}(i)x_{a}^{l}(j)\biggr(\frac{1}{N_{c}}\sum_{k\in\Delta}\sum_{b=1}^{N_{c}}\hat{s}_{b}^{l-1}(i-k)s_{b}^{l-1}(j-k)\biggr)\biggr]\biggr\},\label{eq:Iw_real_space}
\end{align}
}
and 
\begin{align}
I_{\Theta} & =\int P(\hat{\boldsymbol{\Theta}}^{l})d\hat{\boldsymbol{\Theta}}^{l}\int P(\delta\hat{\boldsymbol{\Theta}}^{l})d\delta\hat{\boldsymbol{\Theta}}^{l}\exp\biggr\{-\mathrm{i}\sum_{a=1}^{N_{c}}\sum_{k\in\Delta}\nonumber \\
 & \phantom{=}\biggr[\hat{\Theta}_{a}^{l}(k)\sum_{i=1}^{N_{l}}\biggr(\hat{y}_{a}^{l}(i)\hat{h}_{a}^{l}(2i-k)+\sqrt{1-(\eta_{\theta}^{l})^{2}}y_{a}^{l}(i)h_{a}^{l}(2i-k)\biggr)+\eta_{\theta}^{l}\delta\hat{\Theta}_{a}^{l}(k)\sum_{i=1}^{N_{l}}y_{a}^{l}(i)h_{a}^{l}(2i-k)\biggr]\biggr\}\nonumber \\
 & =\exp\biggr\{\frac{-\sigma_{\theta}^{2}}{2}\sum_{i,j=1}^{N_{l}}\sum_{a=1}^{N_{c}}\biggr[\hat{y}_{a}^{l}(i)\hat{y}_{a}^{l}(j)\biggr(\sum_{k\in\Delta}\hat{h}_{a}^{l}(2i-k)\hat{h}_{a}^{l}(2j-k)\biggr)+y_{a}^{l}(i)y_{a}^{l}(j)\biggr(\sum_{k\in\Delta}h_{a}^{l}(2i-k)h_{a}^{l}(2j-k)\biggr)\nonumber \\
 & \qquad\qquad\qquad\qquad\qquad+2\sqrt{1-(\eta_{\theta}^{l})^{2}}\hat{y}_{a}^{l}(i)y_{a}^{l}(j)\biggr(\sum_{k\in\Delta}\hat{h}_{a}^{l}(2i-k)h_{a}^{l}(2j-k)\biggr)\biggr]\biggr\}.
\end{align}
While the pixels/variables (with site indices $i,j$) are still coupled, the
feature maps (with channel indices $a,b$) can be decoupled if we
define the order parameters $q_{11}^{l-1}(i,j),q_{22}^{l-1}(i,j),q_{12}^{l-1}(i,j)$
through 
\begin{align}
1 & =\int\frac{d\mathcal{Q}_{11}^{l-1}(i,j)dq_{11}^{l-1}(i,j)}{2\pi/N_{c}}e^{\mathrm{i}N_{c}\mathcal{Q}_{11}^{l-1}(i,j)\left[q_{11}^{l-1}(i,j)-\frac{1}{N_{c}}\sum_{b}\hat{s}_{b}^{l-1}(i)\hat{s}_{b}^{l-1}(j)\right]},\\
1 & =\int\frac{d\mathcal{Q}_{22}^{l-1}(i,j)dq_{22}^{l-1}(i,j)}{2\pi/N_{c}}e^{\mathrm{i}N_{c}\mathcal{Q}_{22}^{l-1}(i,j)\left[q_{22}^{l-1}(i,j)-\frac{1}{N_{c}}\sum_{b}s_{b}^{l-1}(i)s_{b}^{l-1}(j)\right]},\\
1 & =\int\frac{d\mathcal{Q}_{12}^{l-1}(i,j)dq_{12}^{l-1}(i,j)}{2\pi/N_{c}}e^{\mathrm{i}N_{c}\mathcal{Q}_{12}^{l-1}(i,j)\left[q_{12}^{l-1}(i,j)-\frac{1}{N_{c}}\sum_{b}\hat{s}_{b}^{l-1}(i)s_{b}^{l-1}(j)\right]}.
\end{align}
To simplify the analysis in the following, we assume a distance-based structure for the order parameters, termed here the stationary
assumption for the order parameters, e.g., 
\begin{equation}
q_{11}^{l-1}(i,j)=q_{11}^{l-1}(|i-j|)~.\label{eq:stationary_assumption}
\end{equation}
This is a reasonable assumption for structured data, of the type CNN are used for. Moreover, it is satisfied throughout the DLM if the initial data $q_{11}^{0}(i,j)$ have this property and the dynamical evolution across layers preserves it (see the following section for an example). It leads to 
\begin{equation}
\sum_{k\in\Delta}q_{11}^{l-1}(i-k,j-k)=\sum_{k\in\Delta}q_{11}^{l-1}(|(i-k)-(j-k)|)=|\Delta|q_{11}^{l-1}(|i-j|),
\end{equation}
and similarly for $q_{22}^{l-1}(i,j)$ and $q_{12}^{l-1}(i,j)$.

Since the different channels are decoupled, we drop the channel indices
of $\hat{x}_{a}^{l}(i),x_{a}^{l}(i),\hat{h}_{a}^{l}(i),h_{a}^{l}(i)$, and consider them as $N_{c}$ factorized copies (similar to the
case of fully-connected networks where the site indices of $\hat{x}_{i}^{l},x_{i}^{l},\hat{h}_{i}^{l},h_{i}^{l}$
are dropped in the previous sections). Defining the field vector $H^{l}$
of size $2N_{l-1}$ as $H^{l}:=[\hat{h}^{l}(1),...,\hat{h}^{l}(N_{l-1}),h^{l}(1),...,h^{l}(N_{l-1})]$,
its covariance matrix (with shape $2N_{l-1}\times2N_{l-1}$) has the
element in the form of 
\begin{equation}
\left[\Sigma_{h,l}(\eta_{w}^{l},q^{l-1})\right]_{ij}=\begin{cases}
\sigma_{w}^{2}|\Delta|q_{11}^{l-1}(|i-j|) & 1\leq i,j\leq N_{l-1}\\
\sigma_{w}^{2}|\Delta|q_{22}^{l-1}(|i-j|) & N_{l-1}<i,j\leq2N_{l-1}\\
\sigma_{w}^{2}|\Delta|\sqrt{1-(\eta_{w}^{l})^{2}}q_{12}^{l-1}(|i-(j-N_{l-1})|), & 1\leq i\leq N_{l-1},N_{l-1}<j\leq2N_{l-1}
\end{cases}
\end{equation}
Similarly, defining the field vector $\mathcal{T}^{l}$ of size $2N_{l}$
as $\mathcal{T}^{l}:=[\hat{\tau}^{l}(1),...,\hat{\tau}^{l}(N_{l}),\tau^{l}(1),...,\tau^{l}(N_{l})]$,
its covariance matrix (with shape $2N_{l}\times2N_{l}$) has the element
in the form of 
\begin{equation}
\left[\Sigma_{\tau,l}(\eta_{\theta}^{l},H^{l})\right]_{ij}=\begin{cases}
\sigma_{\theta}^{2}\sum_{k\in\Delta}\hat{h}^{l}(2i-k)\hat{h}^{l}(2j-k), & 1\leq i,j\leq N_{l}\\
\sigma_{\theta}^{2}\sum_{k\in\Delta}h^{l}(2(i-N_{l})-k)h^{l}(2(j-N_{l})-k), & N_{l}<i,j\leq2N_{l}\\
\sigma_{\theta}^{2}\sqrt{1-(\eta_{\theta}^{l})^{2}}\sum_{k\in\Delta}\hat{h}^{l}(2i-k)h^{l}(2(j-N_{l})-k), & 1\leq i\leq N_{l},N_{l}<j\leq2N_{l}
\end{cases}
\end{equation}

Integrating out the auxiliary fields $\{\hat{x}^{l}(i),x^{l}(i),\hat{y}^{l}(i),y^{l}(i)\}$,
the GF has the form 
\begin{align}
\overline{\Gamma} & =\int\prod_{l'}\prod_{i,j=1}^{N_{l'}}\frac{d\mathcal{Q}_{11}^{l'}(i,j)dq_{11}^{l'}(i,j)}{2\pi/N_{c}}\frac{d\mathcal{Q}_{22}^{l'}(i,j)dq_{22}^{l'}(i,j)}{2\pi/N_{c}}\frac{d\mathcal{Q}^{l'}(i,j)dq^{l'}(i,j)}{2\pi/N_{c}}\nonumber \\
 & \qquad\times\exp\biggr\{\mathrm{i}N_{c}\sum_{l'}\sum_{i,j=1}^{N_{l'}}\left(\mathcal{Q}_{11}^{l'}(i,j)q_{11}^{l'}(i,j)+\mathcal{Q}_{22}^{l'}(i,j)q_{22}^{l'}(i,j)+\mathcal{Q}^{l'}(i,j)q^{l'}(i,j)\right)\biggr\}\nonumber \\
 & \qquad\times\biggr[\int dH^{l}\int d\mathcal{T}^{l}\int\prod_{i=1}^{N_{l}}d\hat{s}^{l}(i)ds^{l}(i)\frac{1}{\sqrt{(2\pi)^{2N_{l-1}}|\Sigma_{h,l}|}}\exp\biggr\{-\frac{1}{2}(H^{l})^{T}\cdot\Sigma_{h,l}(\eta_{w}^{l},q^{l-1})^{-1}\cdot H^{l}\biggr\}\nonumber \\
 & \qquad\quad\times\frac{1}{\sqrt{(2\pi)^{2N_{l}}|\Sigma_{\tau,l}|}}\exp\biggr\{-\frac{1}{2}(\mathcal{T}^{l})^{T}\cdot\Sigma_{\tau,l}(\eta_{\theta}^{l},H^{l})^{-1}\cdot\mathcal{T}^{l}\biggr\}\nonumber \\
 & \qquad\quad\times\exp\biggr\{\sum_{i=1}^{N_{l}}\biggr[-\frac{\beta}{2}\left(\hat{s}^{l}(i)-\phi(\hat{\tau}^{l}(i))\right)^{2}-\frac{\beta}{2}\left(s^{l}(i)-\phi(\tau^{l}(i))\right)^{2}-\log(2\pi/\beta)\biggr]\biggr\}\nonumber \\
 & \qquad\quad\times\exp\biggr\{-\mathrm{i}\sum_{i,j=1}^{N_{l}}\left(\mathcal{Q}_{11}^{l}(i,j)\hat{s}^{l}(i)\hat{s}^{l}(j)+\mathcal{Q}_{22}^{l}(i,j)s^{l}(i)s^{l}(j)+\mathcal{Q}^{l}(i,j)\hat{s}^{l}(i)s^{l}(j)\right)\biggr\}\biggr]^{N_{c}}\nonumber \\
 & \qquad\times\text{terms of remaining blocks}.
\end{align}
In the limit $N_{c}\to\infty,\beta\to\infty$, the saddle point approximation
can be applied and the auxiliary fields $\{\mathcal{Q}_{11}^{l}(i,j),\dots\}$
can be shown to vanish, leading to a single-channel effective measure
\begin{align}
M[\cdots] & =\prod_{i=1}^{N_{l}}\delta\big(\hat{s}^{l}(i)-\phi(\hat{\tau}^{l}(i))\big)\delta\big(s^{l}(i)-\phi(\tau^{l}(i))\big)\frac{e^{-\frac{1}{2}(H^{l})^{T}\cdot\Sigma_{h,l}^{-1}(\eta_{w}^{l},q^{l-1})\cdot H^{l}}}{\sqrt{(2\pi)^{2N_{l-1}}|\Sigma_{h,l}|}}\frac{e^{-\frac{1}{2}(\mathcal{T}^{l})^{T}\cdot\Sigma_{\tau,l}^{-1}(\eta_{\theta}^{l},H^{l})\cdot\mathcal{T}^{l}}}{\sqrt{(2\pi)^{2N_{l}}|\Sigma_{\tau,l}|}}\\
 & \qquad\times\text{terms of remaining blocks}.\nonumber 
\end{align}

\subsubsection{Order Parameters of Networks with ReLU Activation Function}

We first consider ReLU activation function $\phi(x)=\max(0,x)$. The
order parameters $q_{11}^{l}(i,i)$ can be computed as

\begin{align}
q_{11}^{l}(i,i) & =\big<\big(\hat{s}^{l}(i)\big)^{2}\big>_{M}=\int\frac{dH^{l}e^{-\frac{1}{2}(H^{l})^{T}\cdot\Sigma_{h,l}^{-1}\cdot H^{l}}}{\sqrt{(2\pi)^{2N_{l-1}}|\Sigma_{h,l}|}}\int\frac{d\mathcal{T}^{l}e^{-\frac{1}{2}(\mathcal{T}^{l})^{T}\cdot\Sigma_{\tau,l}^{-1}\cdot\mathcal{T}^{l}}}{\sqrt{(2\pi)^{2N_{l}}|\Sigma_{\tau,l}|}}\phi(\hat{\tau}^{l}(i))^{2}\nonumber \\
 & =\int\frac{dH^{l}e^{-\frac{1}{2}(H^{l})^{T}\cdot\Sigma_{h,l}^{-1}\cdot H^{l}}}{\sqrt{(2\pi)^{2N_{l-1}}|\Sigma_{h,l}|}}\frac{\sigma_{\theta}^{2}}{2}\sum_{k\in\Delta}\bigr(\hat{h}^{l}(2i-k)\bigr)^{2}\nonumber \\
 & =\frac{\sigma_{\theta}^{2}\sigma_{w}^{2}|\Delta|^{2}}{2}q_{11}^{l-1}(0)=:q_{11}^{l}(0),
\end{align}
where we have applied the stationary assumption Eq.~(\ref{eq:stationary_assumption})
of layer $l-1$. The same relation holds for $q_{22}^{l}(i,i)$. By
choosing $q_{11}^{0}(0)=1$, $\sigma_{w}=\sqrt{1/3}$, $\sigma_{\theta}=\sqrt{2/3}$,
we can ensure $q_{11}^{l}(0)=1$ and $\left[\Sigma_{h,l}(\eta_{w}^{l},q^{l-1})\right]_{ii}=1,\forall i$.

The order parameters $q_{11}^{l}(i,j)$ can be computed as

\begin{align}
 & q_{11}^{l}(i,j)=\langle\hat{s}^{l}(i)\hat{s}^{l}(j)\rangle_{M}\nonumber \\
= & \int\frac{dH^{l}e^{-\frac{1}{2}(H^{l})^{T}\cdot\Sigma_{h,l}^{-1}\cdot H^{l}}}{\sqrt{(2\pi)^{2N_{l-1}}|\Sigma_{h,l}|}}\int\frac{d\mathcal{T}^{l}e^{-\frac{1}{2}(\mathcal{T}^{l})^{T}\cdot\Sigma_{\tau,l}^{-1}\cdot\mathcal{T}^{l}}}{\sqrt{(2\pi)^{2N_{l}}|\Sigma_{\tau,l}|}}\phi(\hat{\tau}^{l}(i))\phi(\hat{\tau}^{l}(j))\nonumber \\
= & \int\frac{dH^{l}e^{-\frac{1}{2}(H^{l})^{T}\cdot\Sigma_{h,l}^{-1}\cdot H^{l}}}{\sqrt{(2\pi)^{2N_{l-1}}|\Sigma_{h,l}|}}\frac{1}{2\pi}\bigg[\sqrt{\left[\Sigma_{\tau,l}\right]_{ii}\left[\Sigma_{\tau,l}\right]_{jj}-\left[\Sigma_{\tau,l}\right]_{ij}^{2}}+\frac{\pi}{2}\left[\Sigma_{\tau,l}\right]_{ij}+\left[\Sigma_{\tau,l}\right]_{ij}\tan^{-1}\frac{\left[\Sigma_{\tau,l}\right]_{ij}}{\sqrt{\left[\Sigma_{\tau,l}\right]_{ii}\left[\Sigma_{\tau,l}\right]_{jj}-\left[\Sigma_{\tau,l}\right]_{ij}^{2}}}\bigg],\label{eq:q11ij_relu}
\end{align}
which can be integrated numerically (remind that $\Sigma_{\tau,l}$
depends on $H^{l}$). Note that Eq.~(\ref{eq:q11ij_relu}) preserves
the stationary property, i.e., if $q_{11}^{l-1}(i,j)=q_{11}^{l-1}(|i-j|)$,
then $q_{11}^{l}(i,j)=q_{11}^{l}(|i-j|)$. Other order parameters
$q_{12}^{l}(i,j),q_{22}^{l}(i,j)$ and the mean activations $m_{1}^{l}(i)=\langle\hat{s}^{l}(i)\rangle,m_{2}^{l}(i)=\langle s^{l}(i)\rangle$
can be computed and shown to preserve the stationary property similarly.
We assume the input signal also has $N_{c}$ channels, all of which
have the same statistical properties; we also assume that they are
stationary $\langle\hat{s}^{0}(i)\hat{s}^{0}(j)\rangle=q_{11}^{0}(|i-j|)$.
It is then directly justified by induction that the stationary property
$q_{11}^{l}(i,j)=q_{11}^{l}(|i-j|)$ holds for all layers.

\subsubsection{Order Parameters of Networks with Sign Activation Function}

The derivation for the  sign activation function $\phi(x)=\text{sgn}(x)$ follows a similar form as that of the ReLU-activation networks with a different form of the order parameters,
\begin{align}
 & q_{11}^{l}(i,j)=\langle\hat{s}^{l}(i)\hat{s}^{l}(j)\rangle_{M}\nonumber \\
= & \int\frac{dH^{l}e^{-\frac{1}{2}(H^{l})^{T}\cdot\Sigma_{h,l}^{-1}\cdot H^{l}}}{\sqrt{(2\pi)^{2N_{l-1}}|\Sigma_{h,l}|}}\int\frac{d\mathcal{T}^{l}e^{-\frac{1}{2}(\mathcal{T}^{l})^{T}\cdot\Sigma_{\tau,l}^{-1}\cdot\mathcal{T}^{l}}}{\sqrt{(2\pi)^{2N_{l}}|\Sigma_{\tau,l}|}}\phi(\hat{\tau}^{l}(i))\phi(\hat{\tau}^{l}(j))\nonumber \\
= & \int\frac{dH^{l}e^{-\frac{1}{2}(H^{l})^{T}\cdot\Sigma_{h,l}^{-1}\cdot H^{l}}}{\sqrt{(2\pi)^{2N_{l-1}}|\Sigma_{h,l}|}}\frac{2}{\pi}\tan^{-1}\frac{\left[\Sigma_{\tau,l}\right]_{ij}}{\sqrt{\left[\Sigma_{\tau,l}\right]_{ii}\left[\Sigma_{\tau,l}\right]_{jj}-\left[\Sigma_{\tau,l}\right]_{ij}^{2}}},
\end{align}
which can also be integrated numerically. The stationary property
is also preserved by the layer evolution. Note that for sign activation,
$q_{11}^{l}(i,i)=1,$ $m_{1}^{l}(i)=0$ for $l>0$.

\subsubsection{Results}

As an example we specify input data (with $N_{c}$ channels) that obeys the stationarity properties and is represented for convenience in the form (other input representations may be considered as well with minor modifications)
\begin{align}
q_{11}^{0}(i,j) & =\exp\left(-|i-j|/\ell\right)\label{eq:input_OU_kernel}\\
m_{1}^{0}(i) & =0,
\end{align}
where $\ell$ is a parameter governing the length scale of the correlation.
It satisfies $q_{11}^{0}(i,j)=q_{11}^{0}(|i-j|)$ and $q_{11}^{0}(0)=1$;
such one dimensional data can be generated by Ornstein\textendash Uhlenbeck
(OU) processes. Since we restrict $s^{0}(i)=\hat{s}^{0}(i)$ for the
perturbed system, we have $q_{22}^{0}(i,j)=q_{12}^{0}(i,j)=q_{11}^{0}(i,j)$.

\begin{figure}
\includegraphics[scale=0.4]{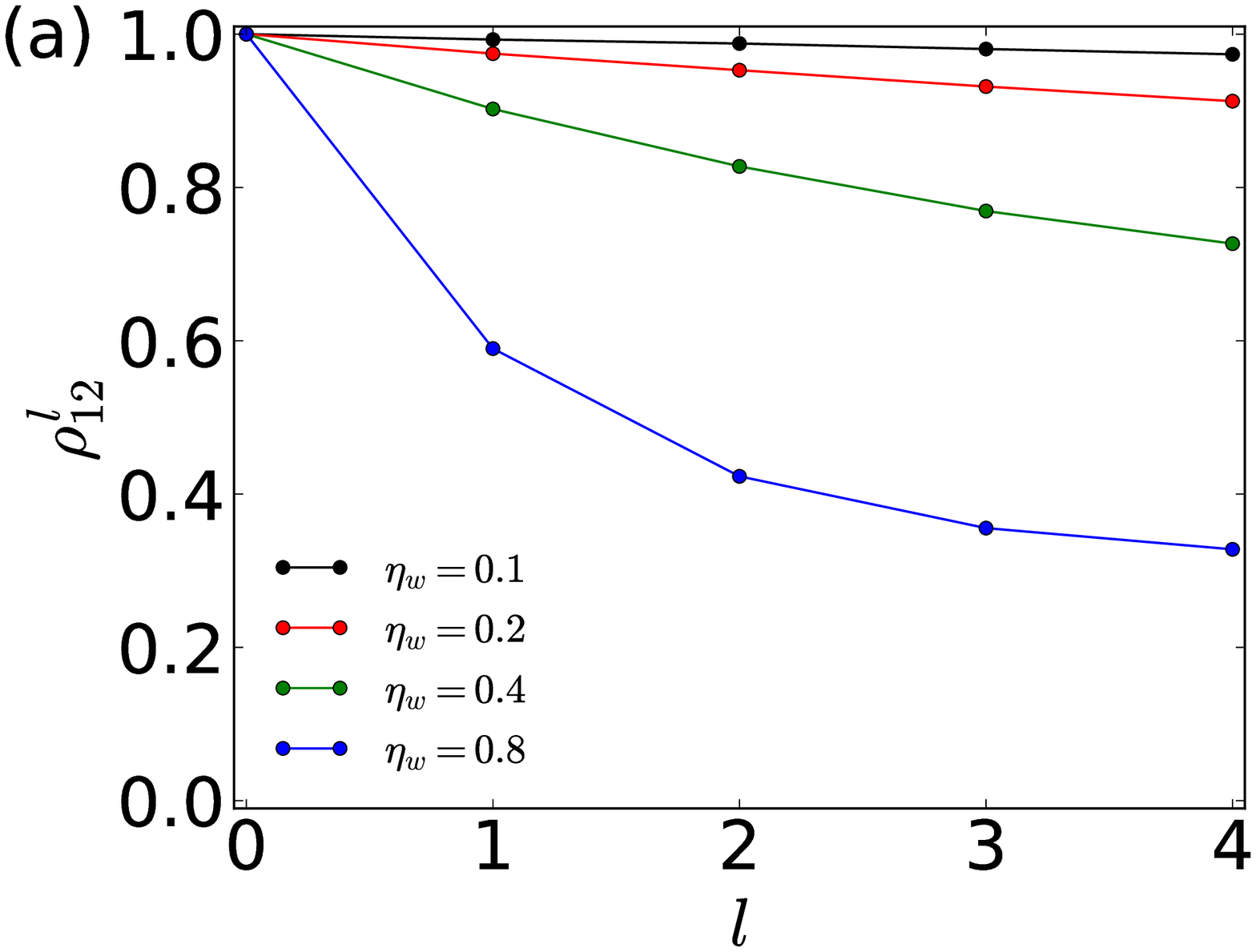}\includegraphics[scale=0.4]{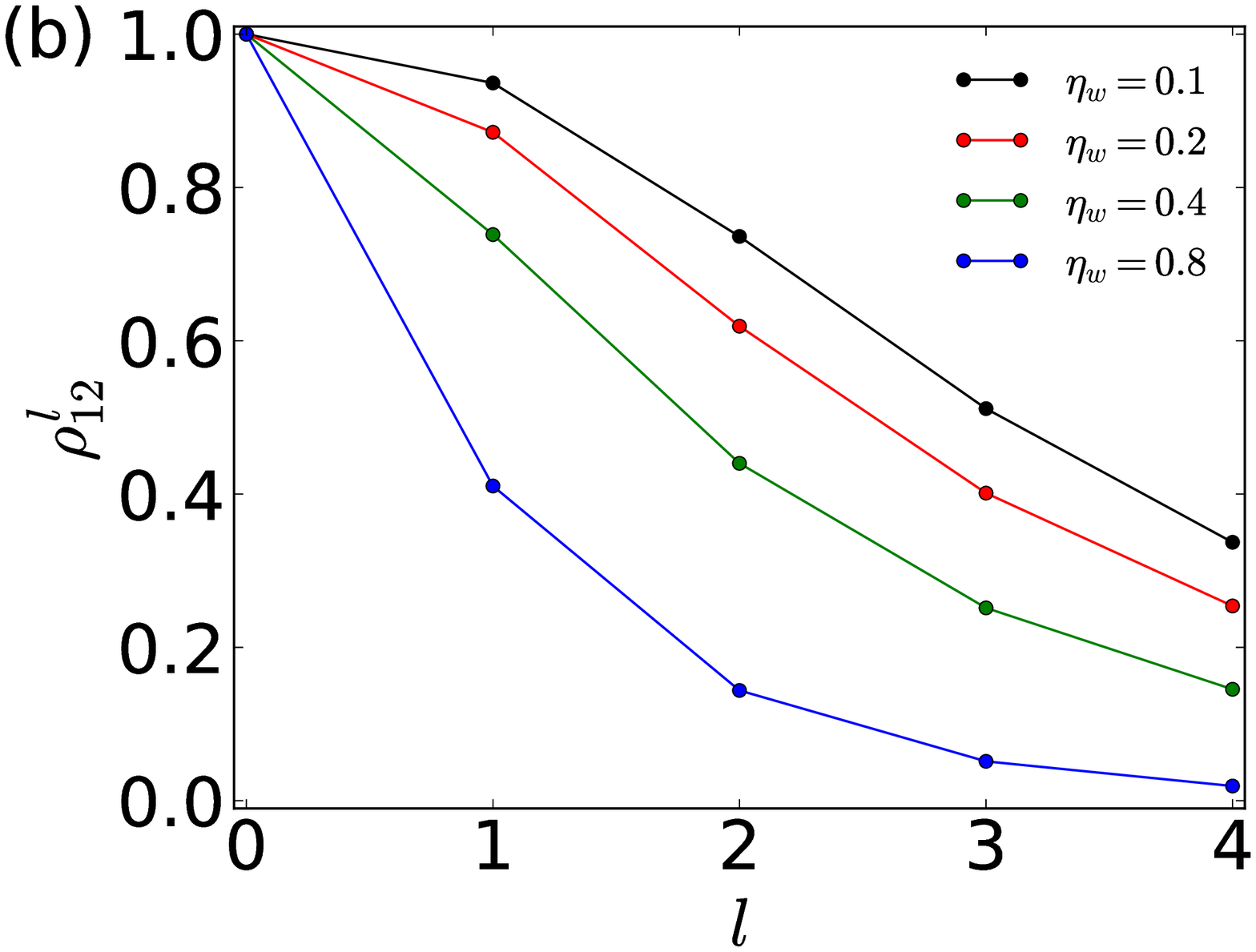}
\caption{Evolution of $\rho_{12}^{l}$ over layers in convolutional neural networks.
The perturbation strength $\eta_{\theta}$ on $\{ \hat{\Theta}^{l}(k) \}$
is set to zero, while perturbation strength $\eta_{w}$ on $\{\hat{w}_{ab}^{l}(k)\}$ varies. The input data is specified by Eq.~(\ref{eq:input_OU_kernel}),
with length scale $\ell=32$. (a) Networks with ReLU activation function.
(b) Networks with sign activation function. \label{fig:rhol_12}}

\end{figure}

The order parameters $q_{11}^{l}(|i-j|),q_{22}^{l}(|i-j|),q_{12}^{l}(|i-j|)$
can be computed numerically as described above. To gain insight into
the corresponding networks under weight perturbations, we consider
the average correlation coefficient as a similarity measure between
the two systems
\begin{align}
\rho_{12}^{l} & =\frac{\sum_{i}\big<\left(\hat{s}^{l}(i)-m_{1}^{l}(i)\right)\left(s^{l}(i)-m_{2}^{l}(i)\right)\big>}{\sqrt{\sum_{i}\left[\big<\left(\hat{s}^{l}(i)\right)^{2}\big>-\left(m_{1}^{l}(i)\right)^{2}\right]}\sqrt{\sum_{i}\left[\big<\left(s^{l}(i)\right)^{2}\big>-\left(m_{2}^{l}(i)\right)^{2}\right]}}\nonumber \\
 & =\frac{q_{12}^{l}(0)-m_{1}^{l}m_{2}^{l}}{\sqrt{q_{11}^{l}(0)-(m_{1}^{l})^{2}}\sqrt{q_{22}^{l}(0)-(m_{2}^{l})^{2}}},
\end{align}
where we have applied the stationary property, e.g., $q_{12}^{l}(i,i)=q_{12}^{l}(|i-i|)=q_{12}^{l}(0)$,
$m_{1}^{l}(i)=m_{1}^{l}$ and so on. In Fig.~\ref{fig:rhol_12},
we sketch two examples of how $\rho_{12}^{l}$ evolves over layers for both ReLU and sign activation functions. Compared to Fig.~\ref{fig:relu_dynamics}(b)
and Fig.~\ref{fig:compare_to_MC_dense}(a), the qualitative behaviors
of $\rho_{12}^{l}$ are similar to the fully-connected networks with
the same activation function, which is expected since the
features maps are fully coupled.

\begin{figure}
\includegraphics[scale=0.4]{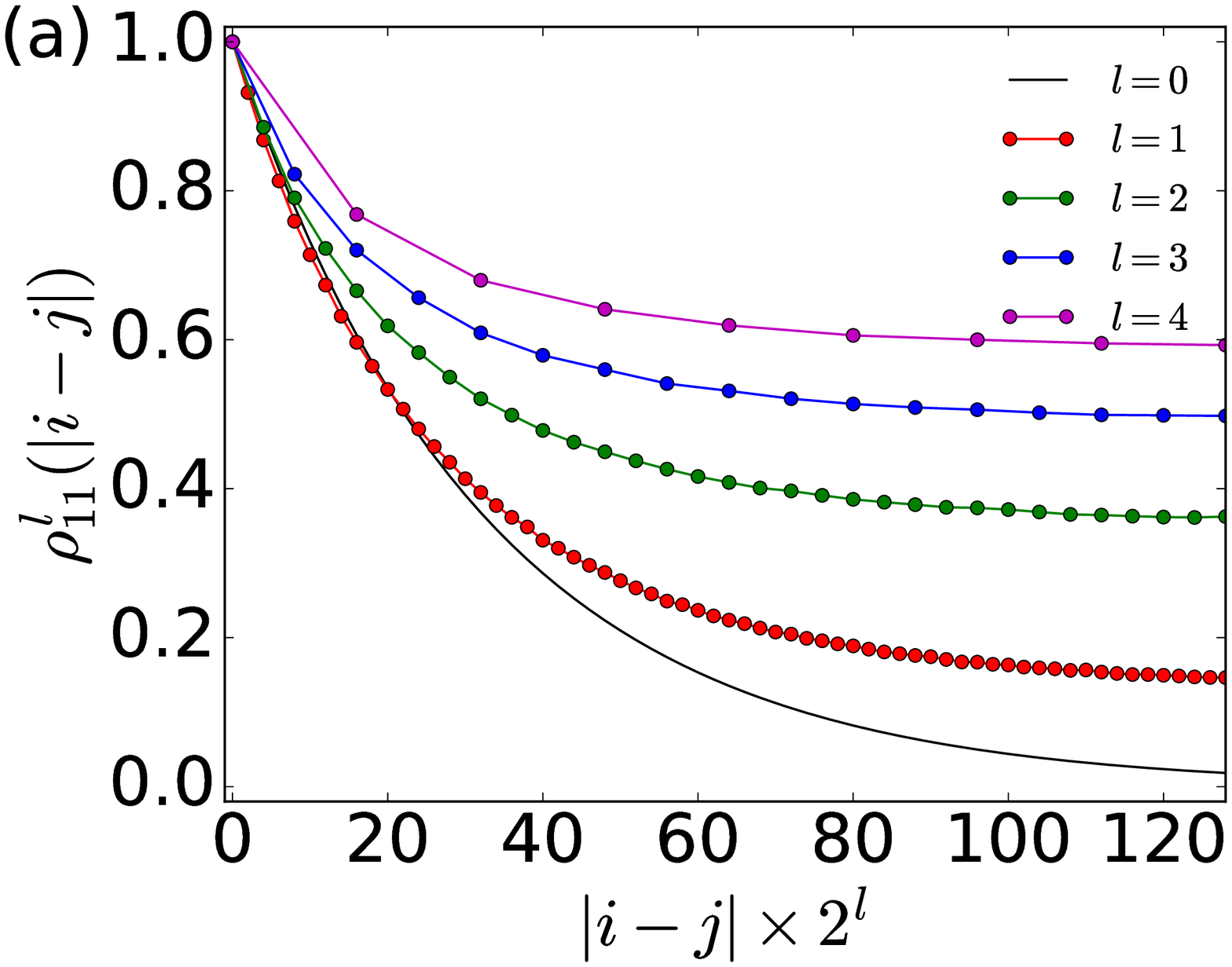}\includegraphics[scale=0.4]{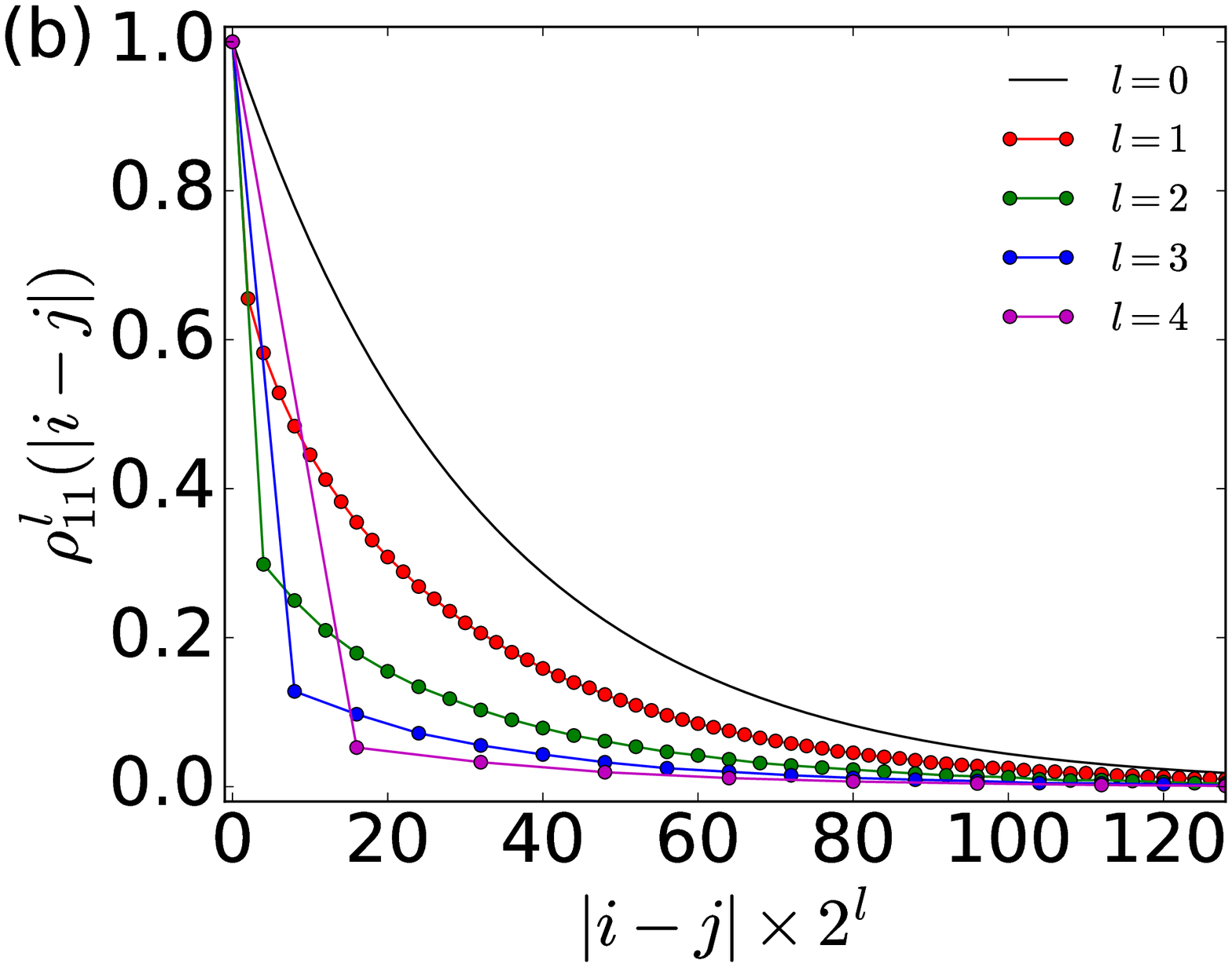}
\caption{Between-site correlation coefficient of the reference convolutional network $\rho_{11}^{l}(|i-j|)$
vs the rescaled distance $|i-j|\times2^{l}$. The input data is specified
by Eq.~(\ref{eq:input_OU_kernel}), with length scale $\ell=32$.
(a) Convolutional networks with ReLU activation function. (b) Convolutional networks with sign
activation function. \label{fig:rholij_11}}
\end{figure}

We further investigate how spatial correlations evolve in layers,
focusing on the between-site correlation coefficient of the reference
network
\begin{align}
\rho_{11}^{l}(i,j) & =\frac{q_{11}^{l}(i,j)-m_{1}^{l}(i)m_{1}^{l}(j)}{\sqrt{q_{11}^{l}(i,i)-(m_{1}^{l}(i))^{2}}\sqrt{q_{11}^{l}(j,j)-(m_{1}^{l}(j))^{2}}}\nonumber \\
 & =\frac{q_{11}^{l}(|i-j|)-(m_{1}^{l})^{2}}{q_{11}^{l}(0)-(m_{1}^{l})^{2}}=:\rho_{11}^{l}(|i-j|) \label{eq:rho_11ij_def}
\end{align}
In Fig.~\ref{fig:rholij_11}, we sketch two examples of how
$\rho_{11}^{l}(|i-j|)$ behaves, one for ReLU activation and the other
for sign activation. Every time the system propagates through a convolutional
block, the lattice spacing between neighboring sites doubles (see
Fig.~\ref{fig:conv_block}); we therefore rescaled the distance $|i-j|$
at layer $l$ by a factor $2^{l}$ for comparison across layers. It
is observed that networks with ReLU activation tend to \emph{correlate
sites} as evolving in layers, while the networks with sign activation
tend to \emph{decorrelate sites} as evolving in layers. Based on the observations and insights into
networks with ReLU activation and correlated weight variables in
Sec.~\ref{subsec:relu_correlated_w}, we also hypothesize that introducing
negatively correlated filters can counteract the correlation inducing
behavior of CNN with ReLU activation (as in Fig.~\ref{fig:rholij_11}(a)),
leading to more expressive learning machines.

\begin{figure}
\includegraphics[scale=0.5]{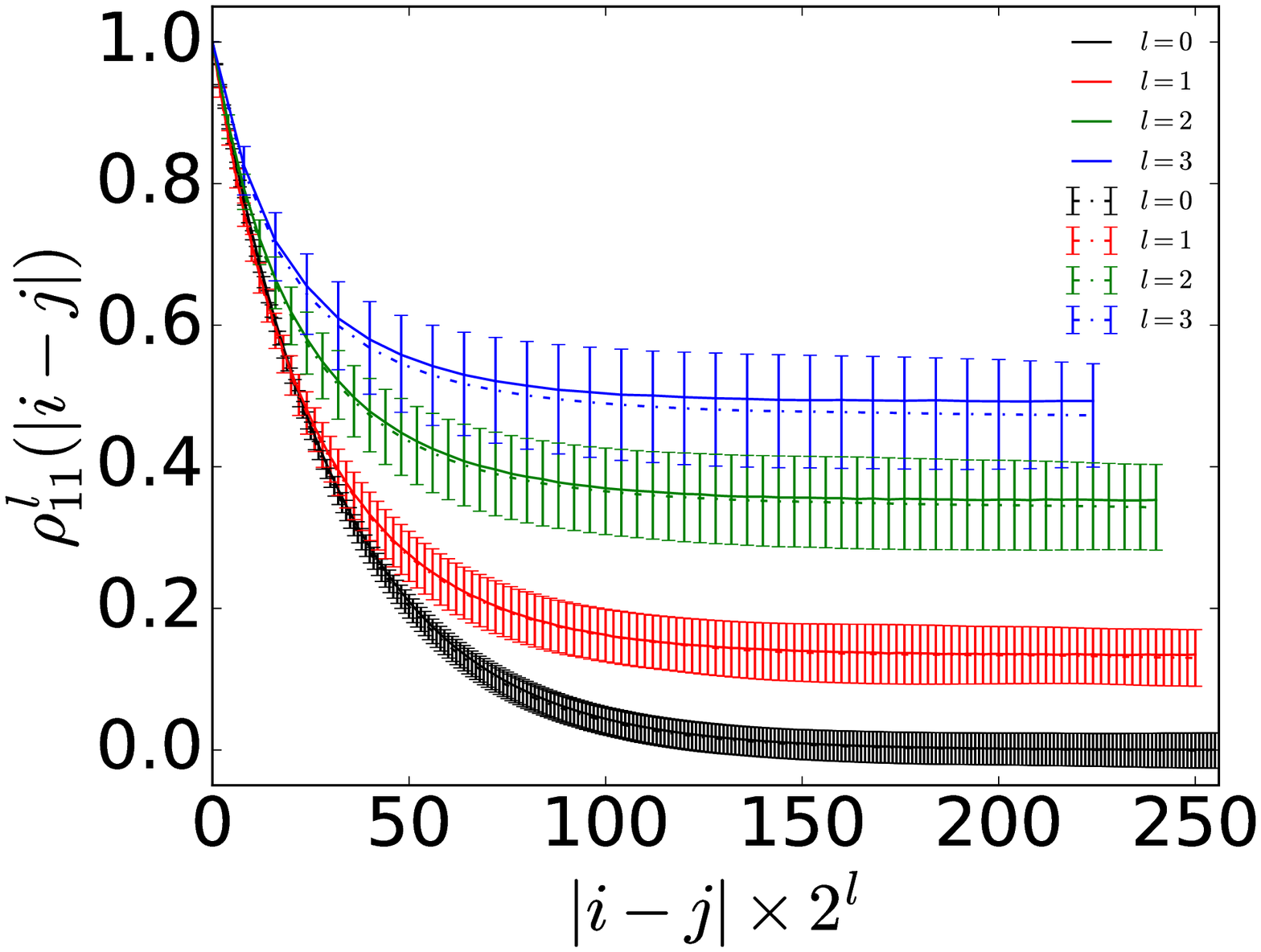}
\caption{Comparison between analytical and simulation results of the correaltion coefficient $\rho_{11}^{l}(|i-j|)$ of the reference convolutional networks with ReLU activation function. The parameters are the same as those in Fig.~\ref{fig:rholij_11}(a). The analytical results (solid lines) are based on saddle point approximation in the limit $N_{c} \to \infty$. The Monte Carlo simulations (dashed-dotted lines) are performed on systems with $N_{c}=200$; in simulations, the correlation coefficient $\rho_{11}^{l}(|i-j|)$ is first calculated according to Eq.~(\ref{eq:rho_11ij_def}) for each realization of the disorder, after which the disorder average is performed. \label{fig:rholij_11_saddle_vs_MC}}
\end{figure}

Finally, we remark that though taking the limit $N_{c}\to\infty$
seems quite unrealistic, our Monte-Carlo simulation results (see Fig.~\ref{fig:rholij_11_saddle_vs_MC}) suggest that the average behavior of finite-size system of
moderate size $N_{c}\sim O(100)$ is already well captured by the theoretical
predictions. However, the assumptions that there are also $N_{c}$
channels in the input signal and that they are stationary may be
restricted. We defer to future works to relax these assumptions, investigate
other weight-disorders, and study the learning processes.

\end{document}